\documentclass[11pt]{article}
\usepackage{jcapmod}

\usepackage{booktabs}
\usepackage[english]{babel}
\usepackage{amsmath,amssymb,amsbsy,amstext, amsthm, simplewick}
\usepackage{graphicx}
\usepackage{mathtools}
\usepackage{amsfonts}
\usepackage{amssymb}
\usepackage[small]{caption}
\usepackage{upgreek}
  \usepackage{framed}
\usepackage{enumitem}
\usepackage[normalem]{ulem}
  
  \usepackage{jcapmod}
\usepackage{booktabs}
\usepackage[english]{babel}
\usepackage{amsmath,amssymb,amsbsy,amstext, amsthm, simplewick}
\usepackage{wrapfig}
\usepackage{hyperref}
\usepackage{graphicx}
\usepackage{amsfonts}
\usepackage{amssymb}
\usepackage{subfig}
\usepackage{upgreek}
 \usepackage{exscale,relsize}
 \usepackage[makeroom]{cancel}
\usepackage{soul}
\usepackage{bbold}
\usepackage{lipsum}
\usepackage{mdframed}
\usepackage{mathtools}
\usepackage[dvipsnames]{xcolor}
\usepackage{tikz}
\usepackage{tikz-cd}
\usepackage[export]{adjustbox}

 \newcommand{\circled}[2][]{%
  \tikz[baseline=(char.base)]{%
    \node[shape = circle, draw, inner sep = 1pt]
    (char) {\phantom{\ifblank{#1}{#2}{#1}}};%
    \node at (char.center) {\makebox[0pt][c]{#2}};}}
\robustify{\circled}

\usepackage{colortbl}
\definecolor{lightgreen}{cmyk}{0.2, 0, 0.2, 0.2}
\definecolor{lightgray}{cmyk}{0.1,0.2,0,0.1}
\definecolor{lightgray2}{cmyk}{0.1,0.1,0,0.1}

\makeatletter
\newlength{\apb@width}
\newcommand{\autoparbox}[2][c]{\settowidth{\apb@width}{#2}\parbox[#1]{\apb@width}{#2}}

\newcommand{\Cen}[2]{%
  \ifmeasuring@
    #2%
  \else
    \makebox[\ifcase\expandafter #1\maxcolumn@widths\fi]{$\displaystyle#2$}%
  \fi
}
\makeatother

\newcommand{\beq}{\begin{equation}\begin{aligned}}
\newcommand{\eeq}{\end{aligned}\end{equation}}

\numberwithin{equation}{section}

\def\beq{\begin{equation}}
\def\eeq{\end{equation}}

\def\Beq{\begin{equation}\begin{aligned}}
\def\Eeq{\end{aligned}\end{equation}}

\def\bea{\begin{eqnarray}}
\def\eea{\end{eqnarray}}

\def\beq{\begin{equation}}
\def\eeq{\end{equation}}
\def\bea{\begin{eqnarray}}
\def\eea{\end{eqnarray}}

\def\bp{\boldsymbol{p}}
\def\bP{\boldsymbol{P}}

\def\bx{\boldsymbol{x}}

\def\trh{T_{\rm RH}}
\def\tmax{T_{\rm max}}

\newcommand{\lslashslash}{%

  \raisebox{0.8ex}{%
    \scalebox{.7}{%
      \rotatebox[origin=c]{18}{$-$}%
    }%
  }%
}

\newcommand{\lslash}{%
  {%
   \vphantom{d}%
   \ooalign{\kern-.1em\smash{\lslashslash}\hidewidth\cr${\rm l}$\cr}%
   \kern.05em
  }%
}

\DeclareRobustCommand{\SkipTocEntry}[4]{}
\DeclareSymbolFont{extraup}{U}{zavm}{m}{n}
\DeclareMathSymbol{\varheart}{\mathalpha}{extraup}{86}
\DeclareMathSymbol{\vardiamond}{\mathalpha}{extraup}{87}
\begin{document}

\hypersetup{pageanchor=false, citecolor={blue}}

\begin{titlepage}
\begin{flushright}
UMN--TH--4101/21, FTPI--MINN--21/19 \\
CERN-TH-2021-121
\end{flushright}

\setcounter{page}{1} \baselineskip=15.5pt \thispagestyle{empty}

\bigskip\

\vspace{1cm}
\begin{center}

{\fontsize{20.74}{24}\selectfont  \sffamily \bfseries  Freeze-in from Preheating}

\end{center}

\vspace{0.2cm}

%~~~~~~~~~~~~~~~~~~~~~~~~~~~~~
\begin{center}
{\fontsize{12}{30}\selectfont  Marcos A.~G.~Garcia$^{a,b,c,d}$,
%\footnote{marcosa.garcia@uam.es}, 
Kunio Kaneta$^{e}$, Yann Mambrini$^{f,g}$\\ Keith A.~Olive$^{h}$, Sarunas Verner$^{h}$}
\end{center}
%~~~~~~~~~~~~~~~~~~~~~~~~~~~~~
\begin{center}

\vskip 7pt

\textsl{$^{a}$ Instituto de F\'isica Te\'orica (IFT) UAM-CSIC, Campus de Cantoblanco, 28049, Madrid, Spain}\\
\textsl{$^{b}$ Departamento de F\'isica Te\'orica, Universidad Aut\'onoma de Madrid (UAM), Campus de Cantoblanco, 28049 Madrid, Spain}\\
\textsl{$^{c}$ Dipartimento di Fisica e Astronomia ``Galileo Galilei'', Universit\`{a} degli Studi di Padova, 35131 Padova, Italy}\\
\textsl{$^{d}$ Istituto Nazionale di Fisica Nucleare (INFN), Sezione di Padova, 35131 Padova, Italy}\\
\textsl{$^{e}$ Department of Mathematics, Tokyo Woman’s Christian University, Tokyo 167-8585, Japan}\\
\textsl{$^{f}$ Universit\'e Paris-Saclay, CNRS/IN2P3, IJCLab, 91405 Orsay, France}\\
\textsl{$^{g}$ 
CERN, Theoretical Physics Department, Geneva, Switzerland}
\\
\textsl{$^{h}$ William I. Fine Theoretical Physics Institute, School of Physics and Astronomy, University of Minnesota, Minneapolis, MN 55455, U.S.A.}
\vskip 7pt

\end{center}

\vspace{0.3cm}
\centerline{\bf ABSTRACT}
\vspace{0.1cm}

%\vspace{1.2cm}
%\hrule \vspace{0.3cm}
%\begin{center}
%\noindent {\sffamily \bfseries Abstract} \\[0.1cm]
%\end{center}
%~~~~~~~~~~~~~~~~~~~~~~~~~~~~~~~~~~~~~
We consider the production of dark matter during the process of reheating after inflation.  The relic density of dark matter from freeze-in depends on both the energy density and energy distribution of the inflaton scattering or decay products composing the radiation bath. We compare the perturbative and non-perturbative calculations of the energy density in radiation. We also consider the (likely) possibility that the final state scalar products are unstable. 
Assuming either thermal or non-thermal 
energy distribution functions, we compare
the resulting relic density based on these different approaches. We show that the present-day cold dark matter density can be obtained through freeze-in from preheating for a large range of dark matter masses. 
%~~~~~~~~~~~~~~~~~~~~~~~~~~~~~~~~~~~~~

\vspace{0.2in}

\begin{flushleft}
{September} 2021
\end{flushleft}
\medskip
\noindent

\newpage
\vskip 10pt
%\hrule
\vskip 10pt

\vspace{0.6cm}
 \end{titlepage}

\hypersetup{pageanchor=true}

\tableofcontents

\newpage

%%%%%%%%%%%%%%%%%%%%%%%%%%%%%%%%%%%
\section{Introduction}
\label{sec:intro}
%%%%%%%%%%%%%%%%%%%%%%%%%%%%%%%%%%%

Reheating and a graceful exit from inflationary expansion is
a necessary component of any model of inflation. 
For most applications, it is sufficient that the radiation dominated epoch of the Universe begins at a suitably high temperature
to allow for standard Big Bang nucleosynthesis and baryogenesis.
While the former requires only that the reheating temperature, $\trh$,
is greater that $\sim 1$ MeV, the latter typically requires $\trh > \mathcal{O}$(TeV). Similarly, the production of weakly-interacting 
dark matter only requires that the reheating temperature was sufficiently
high so that the dark matter candidate was brought into thermal equilibrium with the radiation bath. However, in some cases,
the production and abundance of very weakly-interacting massive particles (WIMPs) such as the gravitino depends explicitly on the value of the reheating temperature \cite{nos,ehnos,kl,oss}. More generally, the abundance of any `feebly' interacting massive particle (FIMP)
will also depend on $\trh$ \cite{fimp,brv,Bernal:2017kxu,Bernal:2019mhf,Bernal:2020gzm}.

In these cases, it is usually sufficient to assume that
reheating occurs instantaneously.  For example, 
one can often assume that once the period of exponential expansion has ceased, and the inflaton begins 
oscillating about its minimum, inflaton decay (which occurs roughly when the effective inflaton decay rate $\Gamma_\phi$ is equal to the Hubble expansion rate, $H$)
will produce radiation~\cite{dg,nos} which eventually thermalizes~\cite{Davidson:2000er,Harigaya:2013vwa,Mukaida:2015ria,Garcia:2018wtq}. Here, we define the `moment' of reheating when the energy density stored in the inflaton, $\rho_\phi$, is equal to the energy density of the newly produced radiation, denoted by $\rho_R$. In this case, 
\beq
\rho_R (\trh) \;=\; \frac{\pi^2 }{30}g_{\rm RH} T_{\rm RH}^4 \, ,
\eeq
where $g_{\rm RH}$ is the number of relativistic degrees of freedom at $\trh$. 

More accurately, reheating is not an instantaneous process, and if the radiation thermalizes rapidly as it is produced, the temperature of the 
dilute bath may far exceed that of the reheating temperature when $\rho_\phi = \rho_R$ \cite{Giudice:2000ex,egnop,Garcia:2017tuj,Chen:2017kvz,Garcia:2020eof,Bernal:2020gzm,Co:2020xaf,Garcia:2020wiy}.
In the case of the weak scale gravitino, the thermal production cross section is independent of the temperature and therefore constant in time, implying a 
gravitino abundance proportional to $\trh$. However, when the particle production cross section depends strongly on temperature, $\sigma \sim T^n$ with $n \geq 6$, the abundance depends on the maximum temperature attained during the reheating process \cite{ Turner:1983he,Garcia:2017tuj,Chen:2017kvz,Kaneta:2019zgw,Garcia:2020eof,Garcia:2020hyo,Anastasopoulos:2020gbu,Brax:2020gqg,Garcia:2020wiy,Mambrini:2021zpp,Bernal:2021kaj,Garcia:2021gsy,Kaneta:2021pyx,Bhattacharyya:2018evo,Chowdhury:2018tzw}
(see also \cite{book} for a pedagogical introduction on the subject). Thus, the details of the reheating process between $\tmax$ and $\trh$ become important. Relaxing the assumption of instantaneous thermalization, the dark matter abundance could be produced in-between the end of inflation and the equilibration of the primordial plasma, well before the end of reheating~\cite{Garcia:2018wtq}.

As noted above, the radiation bath is produced as the inflaton decays. For an inflaton potential, which is dominated by a quadratic term near its minimum, reheating always occurs as the inflaton energy density is depleted. A thermal bath may also be produced by scatterings such as $\phi \phi \to \chi \chi$, where $\chi$ is another scalar field. However, in this case, the produced radiation never comes to dominate the total energy density, as the radiation density falls off as $a^{-4}$ and the inflaton energy density scales as $a^{-3}$, where $a$ is the cosmological scale factor \cite{Garcia:2020wiy}. In contrast, when the inflaton potential is dominated by a quartic (or higher-order) term, with $V(\phi) \sim \phi^k$, the produced radiation density scales as 
$a^{-\frac{18}{k+2}}$
whereas the inflaton density scales as $a^{-\frac{6k}{k+2}}$, which means that reheating occurs for $\frac{6 k}{k+2}>\frac{18}{k+2}$, or $k > 3$ \cite{Garcia:2020wiy}. Nevertheless,  quadratic 
potential scatterings that do not lead to reheating without additional decay channels that drain the energy density of the inflaton,
can have an important effect on particle production in the early stages of reheating. 

In this work, we consider an inflaton potential dominated by a quadratic term after
inflation. We approximate 
\beq
V(\phi) \;\simeq\; \frac12 m_\phi^2 \phi^2 \;\equiv\; \lambda \phi^2 M_P^2\, ,  \quad \phi\ll M_P \, ,
\label{infmass}
\eeq
where $M_P = \sqrt{8 \pi/G_N}  \simeq 2.4 \times 10^{18}~\rm{GeV}$  is the reduced Planck mass.
We assume that the scale of $\lambda$ is set by the normalization of the cosmic microwave background (CMB) anisotropies \cite{Planck}, which typically gives $\lambda \sim \mathcal{O}(10^{-11})$.
The full Lagrangian should also include couplings of the inflaton to fermions, $\psi$, and bosons, $\chi$, of the form
\begin{equation}
\mathcal{L} \supset\left\{\begin{array}{ll}
-y \phi \bar{\psi} \psi & \qquad \phi \rightarrow \bar{\psi} \psi \, , \\[10pt]
-\dfrac{\sigma}{2} \phi^{2} \chi^{2} & \qquad \phi \phi \rightarrow \chi \chi \, ,
\end{array}\right.
\label{inter}
\end{equation}
where $y$ is a Yukawa-like coupling and $\sigma$ is a four-point coupling. It should be noted that we do not consider interactions of the form $\mathcal{L} \supset - \mu \frac{\phi^n \chi^2}{M_P^{n-2}}$, with $n=1$, or the non-renormalizable interactions with $n \geq 3$, and a detailed discussion of preheating with such couplings is presented in~\cite{Dufaux:2006ee}.
During reheating, the inflaton oscillates about the origin, and for oscillations about a quadratic minimum, we can write $\phi(t) = \phi_0 (t) \cdot \cos(\sqrt{2 \lambda}M_P t)$, where we refer to $\phi_0(t) \sim a^{-3/2}$ as the time-dependent amplitude that includes the effects of redshift.

Na\"ively, one might expect that initially, for $\phi_0 \sim M_P$, 
decays and scatterings are kinematically forbidden when 
$y \phi_0 > \sqrt{\lambda/2} M_P$ and $\sigma \phi_0^2 > 2\lambda M_P^2$, respectively.
However, during each oscillation, some decays and scatterings occur as $\phi(t)$ passes through the origin. 
It was found in \cite{Garcia:2020wiy} that the na\"ive perturbative processes
can be corrected by defining effective couplings, $y_{\rm eff}$ and $\sigma_{\rm eff}$, both proportional to $\mathcal{R}^{-1/2}$ for $\mathcal{R} > 1$, where the induced mass parameter is given by
\beq\label{eq:kincond}
\mathcal{R}(t) \;\equiv\; \frac{2}{\lambda} \times \begin{cases}
y^2 \left(\dfrac{\phi_0(t)}{M_P}\right)^{2}\,, & \phi\rightarrow \bar{\psi} \psi\,,\\[9pt]
\sigma \left(\dfrac{\phi_0(t)}{M_P}\right)^{2}\,, & \phi\phi\rightarrow \chi\chi\,,
\end{cases}
\eeq
and $y_{\rm eff}=y$ and $\sigma_{\rm eff}=\sigma$ for ${\cal R} < 1$.
Reheating processes and $\rho_R$ are then reduced by a factor ${\cal R}^{-1/2}=\left(\frac{m_\phi}{2 m_{\psi/\chi}(t)}\right)_{\phi\rightarrow \phi_0}$.
However, it was also recognized in \cite{Garcia:2020wiy} that $\mathcal{R} > 1$
signified the potential importance of non-perturbative effects also playing a role.\footnote{Na\"ively, one may expect that the final state suppression is exponential, $e^{-m_{\psi/\chi}/m_\phi}$. However, once inflaton oscillations commence, during some fraction of each oscillatory period, the final state masses become small enough to allow the scattering to proceed. This results in a power-law suppression $\propto {\mathcal{R}}^{-1/2}$ rather than an exponential suppression \cite{Garcia:2020wiy}.} These effects are
among the primary subjects of this work. 

The perturbative approximation provides an adequate description of inflationary reheating and relic production for sufficiently small couplings. However, above this model-dependent threshold the perturbative description breaks down, on one part due to the relevance of quantum-mechanical enhancement/blocking effects dependent on the spin of the inflaton decay products, and on the other hand on the relevance of inverse processes. The effect on the background dynamics ({\em back-reaction}) due to particle production is not limited to a depletion of the energy density of $\phi$. When quantum fluctuations are sufficiently amplified, mode-mode couplings and non-linear interactions will mediate in the conversion of the energy density of the $\phi$ condensate into gradients, potentially leading to the fragmentation of the field. In the following sections we study the transition between the perturbative, the linear non-perturbative, and the non-linear non-perturbative regimes in terms of the energy densities of the inflaton and its decay products, and subsequently apply these results to dark relic production during
reheating. 

We will show that even though the non-perturbative effects themselves 
do not lead to successful reheating and a radiation-dominated era, they can play a role in particle production in the early stages of reheating and delay the reheating process. 
The paper is organized as follows:
In the next section, we consider
the perturbative production of 
a radiation bath composed of either
bosons (in the case $\phi \phi \to \chi \chi$) or fermions (in the case $\phi \to {\bar \psi} \psi$). In the former we review the effect of the effective boson mass which is induced by its coupling to the inflaton. In Section~\ref{sec:nonpert}, we review
the methods used for our numerical calculation of the energy density in radiation. More details are given in Appendix~\ref{sec:appA}. We concentrate primarily on boson production in Section~\ref{sec:nonpert-bos} and we explicitly include the effects of preheating in  Section~\ref{sec:nonpert-pre},
the effects of back-reactions in Section~\ref{sec:nonpert-back},
and the effects of boson decay in Section~\ref{sec:unstablechi}.
Fermion production is discussed in Section~\ref{Sec:fermions}. Further details on our non-perturbative calculation involving fermions is given in Appendix~\ref{sec:appB}.
We apply these results to the production of dark matter in Section~\ref{sec:DM}. We start by introducing the dark matter production rates and particle distribution functions in Section~\ref{sec:rates} which are applied to the case of bosonic production of dark matter in Section~\ref{sec:ndmb} and fermionic production in Section~\ref{sec:ndmf}.
Our summary is given in Section~\ref{sec:summary}.

%%%%%%%%%%%%%%%%%%%%%%%%%%%%%%%%%%%
\section{Perturbative particle production after inflation}
%%%%%%%%%%%%%%%%%%%%%%%%%%%%%%%%%%%

The end of the inflationary epoch of the early Universe in most scenarios is not only marked by the transition to a decelerated expansion, $\ddot{a}<0$,
but also by the beginning of a stage of coherent oscillations of the spatially homogeneous inflaton field $\phi$. As stated in the Introduction, in this work we assume that the inflaton is massive (i.e., dominated by a quadratic potential near the origin) and that its potential near its minimum can be approximated as in Eq.~(\ref{infmass}).
The oscillations of the inflaton are accompanied by particle production processes which originate from the coupling of the inflaton field, $\phi$, to the Standard Model and/or dark sector (\ref{inter}). At the perturbative level, these processes are described by the following Friedmann-Boltzmann set of equations,
\begin{align} \label{eq:FBeqs1}
\dot{\rho}_{\phi} + 3H\rho_{\phi} \;&=\; -\Gamma_{\phi}\rho_{\phi}\,,\\ \label{eq:FBeqs2}
\dot{\rho}_R + 4H\rho_R \;&=\; \Gamma_{\phi}\rho_{\phi}\,,\\ \label{eq:FBeqs3}
\rho_{\phi} + \rho_R \;&=\; 3H^2M_P^2\,,
\end{align}
where 
$H = \frac{\dot{a}}{a}$ is the Hubble parameter and $\Gamma_\phi$ is
the decay/scattering rate of $\phi$.
Note that neither the nature of the radiation bath nor of the decay rate are explicit in these expressions; the inflaton could in general have multiple decay channels to bosonic and fermionic matter, or to gauge bosons. Moreover, the individual rates need not be constant in time, as is the case of inflaton annihilation processes, or due to a time-dependent kinematic suppression of the rate, described by the parameter ${\cal R}(t)$ (see~\cite{Kainulainen:2016vzv,Ichikawa:2008ne,Garcia:2020wiy} and below). \par\medskip

Although our analysis is largely independent of specific form of $V(\phi)$, to illustrate our point we will consider an inflationary potential
of the form
motivated by T-models~\cite{Kallosh:2013hoa} which can be derived in the context of no-scale supergravity~\cite{Garcia:2020eof},
\beq
\label{tmodelpot}
V(\phi) \;=\; \lambda M_P^4 \left[ \sqrt{6} \tanh \left( \frac{\phi}{\sqrt{6}M_P} \right) \right]^2 \, ,
\eeq
which when expanded about the minimum takes the form in Eq.~(\ref{infmass}). This model is easily generalized to obtain
quartic or higher-order polynomial expansions about the origin.
As noted above, $\lambda$ is set by the normalization of the CMB anisotropies, and we take $\lambda = 2.05 \times 10^{-11}$, corresponding to 55 $e$-folds of inflation for the Planck pivot scale.
At large values of $\phi$, the Universe undergoes a period of exponential expansion which ends when $\ddot{a} = 0$, at $a(t_{\rm end}) \equiv a_{\rm end}$ when $\phi(t_{\rm end}) \equiv \phi_{\rm end} \simeq 0.84 \, M_P$, corresponding to 
$\rho_\phi(t_{\rm end}) \equiv \rho_{\rm end} \simeq 2 \times 10^{-11} M_P^4$. 
For $a > a_{\rm end}$, the inflaton begins to oscillate about the origin
and its energy density scales like non-relativistic matter,
\beq
\rho_\phi(a) = \rho_{\rm end} \left(\frac{a}{a_{\rm end}} \right)^{-3} \, ,
\label{Eq:rhophi}
\eeq
so long as $\Gamma_\phi \ll 3 H \simeq \sqrt{3} \rho_\phi^{1/2}/M_P$.
Subsequently, when $\Gamma_\phi$ remains constant, the energy density of the inflaton begins to decrease. 

For the processes in Eq.~(\ref{inter}), the decay/scattering rate is given by
\beq\label{eq:rates}
\Gamma_\phi \;=\; \frac{1}{8\pi} \begin{cases}
y_{\rm eff}^2  m_{\phi}\,, & \phi\rightarrow \bar{\psi} \psi\,,\\[10pt]
\dfrac{\sigma_{\rm eff}^2}{4}
\dfrac{\rho_\phi(t)}{ m^3_\phi}\,, & \phi\phi\rightarrow \chi \chi\, ,
\end{cases}
\eeq
where the calculations of the effective couplings, $y_{\rm eff}$ and $\sigma_{\rm eff}$ can be found in \cite{Garcia:2020wiy}, with ${\mathcal{R}}$ given in Eq.~(\ref{eq:kincond}). Here, $y_{\rm eff}^2 \simeq 0.38 y^2/{\mathcal{R}}^{1/2}$ and $\sigma_{\rm eff}^2 \simeq \sigma^2/{\mathcal{R}}^{1/2}$ for $\mathcal{R}\gg 1$, while $y_{\rm eff}\simeq y$ and $\sigma_{\rm eff}\simeq \sigma$ if $\mathcal{R}\ll 1$. When $\mathcal{R}\ll 1$,  as the radiation bath begins to appear, and the energy density of radiation
evolves as 
\beq\label{eq:rho}
\rho_R \;=\; 2 \rho_{\rm end} \frac{\Gamma_\phi}{H_{\rm end}}  \begin{cases}
\dfrac15 \left( \dfrac{a_{\rm end}}{a} \right)^4 \left[ \left( \dfrac{a}{a_{\rm end}}\right)^{\frac52} -1\right] \,, &  \phi\rightarrow \bar{\psi}\psi\,,\\[10pt]
\dfrac{\rho_{\rm end}}{\rho_{\phi}} \left( \dfrac{a_{\rm end}}{a} \right)^4 \left[1- \left(  \dfrac{a_{\rm end}}{a}\right)^{\frac12}\right] \,, & \phi\phi\rightarrow \chi \chi\, ,
\end{cases}
\eeq
where $H_{\rm end} = \rho_{\rm end}/\sqrt{3}M_P$.

As one can see, for $a \gg a_{\rm end}$, decays to fermions lead to a temperature dependence $T \propto a^{-3/8}$, while scatterings to bosons leads to $T \propto a^{-1}$, making reheating impossible without decays. From Eq.~(\ref{eq:rho}), we see that as $\rho_R$ begins to increase, it reaches a maximum at $a_{\rm{max}}/a_{\rm end} = \left(\frac{8}{3}\right)^{2/5}$
and $a_{\rm{max}}/a_{\rm end}= \frac{81}{64}$, respectively, which gives
\beq\label{eq:rhomax}
\rho_R^{\rm max} \;=\; 2 \rho_{\rm end} \dfrac{\Gamma_\phi}{H_{\rm end}}  \begin{cases}
\dfrac13 \left( \dfrac38 \right)^{8/5}  \,, & \phi\rightarrow \bar{\psi} \psi\,,\\[10pt]
\frac19 \left( \dfrac89 \right)^{8} \dfrac{\rho_{\rm end}}{\rho_\phi}   \,, &  \phi\phi\rightarrow \chi \chi\, .
\end{cases}
\eeq

\begin{figure}[!t]
\centering
    \includegraphics[width=0.83\textwidth]{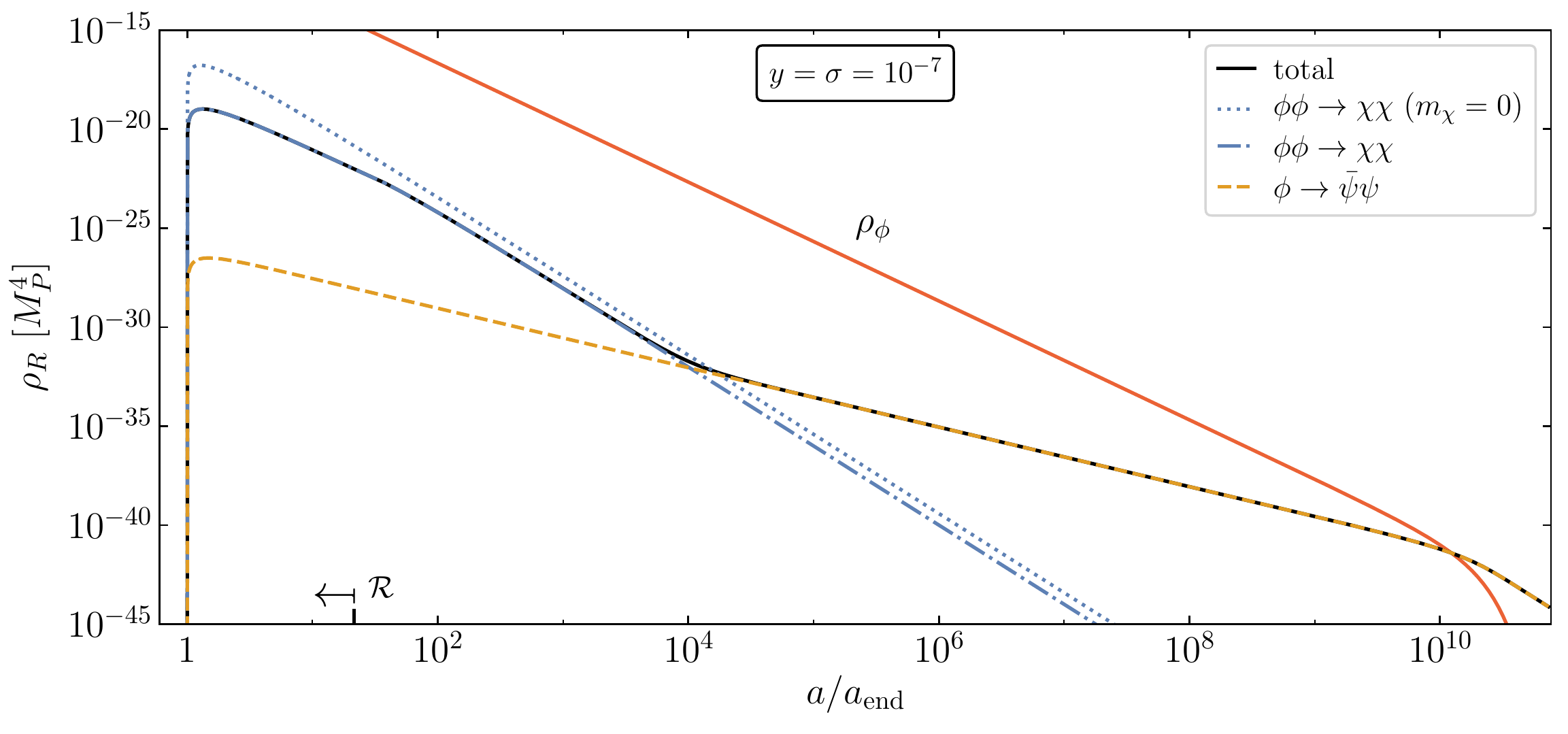}
    \caption{\it Evolution of the energy densities during reheating of the inflaton (solid red) and radiation produced by inflaton decays to fermions (dashed, orange) and annihilations to bosons (blue). In the latter, we show separately the case when the effective masses of the decay products are ignored (dotted) and included (dashed-dotted). The total energy density of the decay/annihilation products is also shown (black solid). We set $y = \sigma = 10^{-7}$, $\rho_{\rm end}=2\times 10^{-11}\, M_P^4$ and $\lambda=2.05\times 10^{-11}$, assuming T-attractor inflation boundary conditions.
    The arrow points toward the region where $\mathcal{R}>1$, indicating the region where non-perturbative effects are expected to affect the evolution of $\rho_R$. The `moment' of reheating is defined when $\rho_\phi = \rho_R$, which occurs at $a/a_{\rm end} \simeq 2.4 \times 10^{10}$.}
    \label{fig:mixedtempsk2}
\end{figure}

The evolution of the radiation and inflaton energy densities are shown in Fig.~\ref{fig:mixedtempsk2} for $y=\sigma=10^{-7}$. The inflaton energy density is shown by the red curve. The energy density falls off as $a^{-3}$ as given in Eq.~(\ref{Eq:rhophi}) until $a/a_{\rm end} \sim 10^{10}$, where the decay rate to fermions becomes comparable to the Hubble rate, and the density begins to fall off exponentially.  For this value of $y$ (and $\lambda$), the ratio ${\cal R}^{1/2} = \frac{2 m_{\psi}}{m_\phi}\ll1$ even for $\phi_0(t) \simeq M_P$ 
and the kinematic effects do not affect the decay to fermions. The energy density in fermions is shown by the orange dashed curve.
The energy density in radiation due to bosons is shown by the blue dotted and dashed-dotted curves when kinematic suppression is ignored (labelled as $m_{\chi} = 0$) and included, respectively.
In the case of annihilation to bosons, the suppression factor
\beq
{\cal R}^{-1/2}=\frac{m_\phi}{2 m_\chi} \; = \; \sqrt{\frac{\lambda}{2\sigma}} \; \simeq \; 0.01 \eeq
when $\phi_0(t)\simeq M_P$ and
$\rho_R$ near its maximum is reduced relative to the case $m_{\chi}=0$ where the suppression is ignored ($\mathcal{R} = 1$). We also find that, since ${\cal R}$ evolves as $a^{-3}$ with ${\cal R}_{\rm end}=2 \frac{\sigma}{\lambda}\simeq 10^4$, ${\cal R}(t)\simeq 1$ for $\frac{a}{a_{\rm end}}\simeq 20$, which is clearly
seen in the figure by the change in slope of the solid black curve, which represents the sum of the radiation contributions from bosons and fermions. When $a \lesssim (2\frac{\sigma}{\lambda})^{1/3}~a_{\rm end}$,
non-perturbative effects are expected to be important. At larger $a$, the two blue curves run parallel to each other and scale as $a^{-4}$. Thus in the absence of decays to fermions, reheating does not occur. The `moment' of reheating is shown in the figure when $\rho_\phi = \rho_R$ at 
$\frac{a}{a_{\rm end}} 
\simeq 2.4 \times 10^{10}$~\cite{Garcia:2020wiy}.   A more precise treatment of the non-perturbative effects is given in the next section.

%%%%%%%%%%%%%%%%%%%%%%%%%%%%%%%%%%%
\section{Non-perturbative particle production}
\label{sec:nonpert}
%%%%%%%%%%%%%%%%%%%%%%%%%%%%%%%%%%%

We now proceed to study the excitation of scalar and fermion fields coupled to the inflaton using non-perturbative tools. Unlike the perturbative approximation, which relies on particle production rates computed averaging over the oscillations of $\phi$, we now take into account the short-time non-adiabaticity encoded in the time-dependence of the effective masses of $\chi$ and $\psi$. In doing so we recover well-known results while exploring in detail others that are not as widely known, and which are crucial for the correct determination of the present dark matter relic abundance. In the following subsection we discuss spin-0 preheating, while in Section~\ref{Sec:fermions} we present spin-1/2 preheating.

%%%%%%%%%%%%%%%%%%%%%%%%%%%%%%%%%%%
\subsection{Production of spin-0 bosons}
\label{sec:nonpert-bos}
%%%%%%%%%%%%%%%%%%%%%%%%%%%%%%%%%%%

%%%%%%%%%%%%%%%%%%%%%%%%%%%%%%%%%%%
\subsubsection{Preheating}
\label{sec:nonpert-pre}
%%%%%%%%%%%%%%%%%%%%%%%%%%%%%%%%%%%
In this section we consider the excitation of a spin-0 field due to its direct coupling to the inflaton during reheating. For such a field, the relevant action is given by\footnote{In this work, we use the metric $g_{\mu \nu} = {\rm{diag}}(1, -1, -1, -1)$.}
\beq
\mathcal{S}_{\chi} \;=\; \frac{1}{2} \int d^4x \,\sqrt{-g}\,\left[ \partial_{\mu}\chi \partial^{\mu}\chi - m^2_{\chi}(t)\chi^2 + \cdots \right]\,,
\label{Eq:action}
\eeq
 where $m_{\chi}(t)$ is the time-varying mass of the spin-0 boson, and we have omitted the self-interaction terms and couplings that are not related to the spatially homogeneous inflaton. It should also be noted that we disregard the inflaton coupling to the curvature because the gravitational particle production~\cite{gravprod,ema} is subdominant compared to non-perturbative particle production in the range of couplings that we will consider.

The equation of motion for the field $\chi$ is given by
\beq
\label{boson:eom1}
\left(\frac{d^2}{dt^2} - \frac{\nabla^2}{a^2} + 3H\frac{d}{dt} + m^2_{\chi}(t) \right)\chi \;=\; 0\,.
\eeq
To proceed further, we switch for convenience to conformal time, $\tau$, related to cosmic time $t$ via $dt/d\tau=a$, and introduce the re-scaled field $X\equiv a\chi$, which leads to
\beq
\label{boson:eombis}
\left[ \frac{d^2}{d\tau^2} - \nabla^2 - \frac{a''}{a} + a^2m^2_{\chi}(\tau) \right] X \;=\; 0\, ,
\eeq
where we introduce the notation $' \equiv d/d \tau$.

We expand the quantum field $X$ in terms of its Fourier components,
\beq
X(\tau,\bx) \;=\; \int \frac{d^3 \bp}{(2\pi)^{3/2}} e^{-i\bp\cdot \bx} \left[ X_p(\tau) \hat{a}_{\bp} + X_p^*(\tau) \hat{a}_{-\bp}^{\dagger} \right]\,,
\label{Eq:X}
\eeq
where $\hat{a}_p^{\dagger}$ and $\hat{a}_{p}$ are the creation and annihilation operators, respectively, that obey the commutation relations $[\hat{a}_{\bp},\hat{a}_{\bp'}^{\dagger}]=\delta(\bp-\bp')$ and $[\hat{a}_{\bp},\hat{a}_{\bp'}]=[\hat{a}_{\bp}^{\dagger},\hat{a}_{\bp'}^{\dagger}]=0$. From Eq.~(\ref{boson:eombis}), we find that the equation of motion satisfied by the mode functions is 
\beq\label{eq:Xktau}
X_p'' + \left[ p^2 - \frac{a''}{a} + a^2m^2_{\chi} \right] X_p \;=\; 0\,.
\eeq
To ensure that the canonical commutation relation between the field, $X$, and its conjugate momentum 
is satisfied, the Wronskian constraint, $X_p X_p^{*\prime}  -  X_p^*  X_p'  \;=\; i$ is imposed.
The zero-particle initial condition is taken as the positive frequency Bunch-Davies vacuum,
\beq\label{eq:icondX}
X_p(\tau_0) \;=\; \frac{1}{\sqrt{2\omega_p}} \,,\qquad X_p'(\tau_0) \;=\; -\frac{i\omega_p}{\sqrt{2\omega_p}}\,,
\eeq
where the angular frequency is in this case defined as 
\beq\label{eq:omS}
\omega_p^2 \;\equiv\; p^2 - \frac{a''}{a} + a^2 m_{\chi}^2\,,
\eeq
and the mode equation of motion~(\ref{eq:Xktau}) can be recast as $X_p'' + \omega_p^2 X_p = 0$.

The energy density in the field $\chi$ can be readily determined from its energy-momentum tensor, $T^{\mu \nu}$, with
\beq
T^{00}\;=\; \frac{1}{2}\dot{\chi}^2 + \frac{1}{2a^2}\left(\nabla\chi\right)^2 + \frac{1}{2}m_{\chi}^2 \chi^2\,,
\eeq
and the energy of each mode $X_p$ is given by 
$E_p = \omega_p (n_p+\frac{1}{2}) = \frac{1}{2}|X_p'|^2 + \frac{1}{2}\omega_p^2 |X_p|^2$. We obtain for the normal-ordered, UV convergent expectation value of the energy density the following expression~\cite{preheating2},
\beq\label{eq:rhochi}
\rho_{\chi} \;=\; \frac{1}{(2\pi)^3 a^4}\int d^3\bp \,\omega_p n_p\,,
\eeq
where in this case the particle occupation number is given by 
\beq\label{eq:nchi}
n_p \;=\; \frac{1}{2\omega_p} \left[\left|X_p'  \right|^2 + \omega_p^2 \left|X_p\right|^2 \right] -\frac{1}{2}\,.
\eeq

\begin{figure}[!t]
\centering
    \includegraphics[width=.785\textwidth]{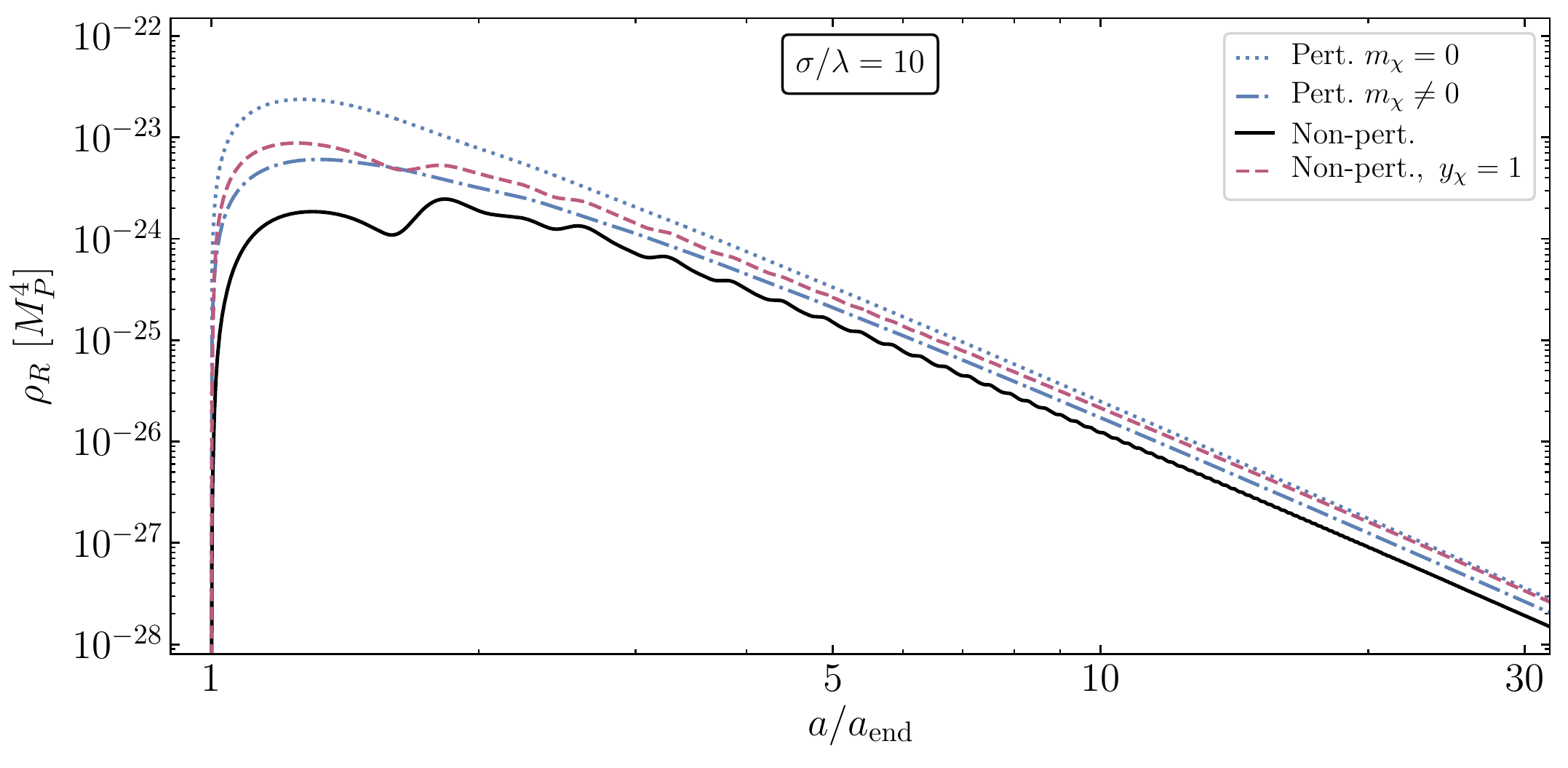}
    \includegraphics[width=.785\textwidth]{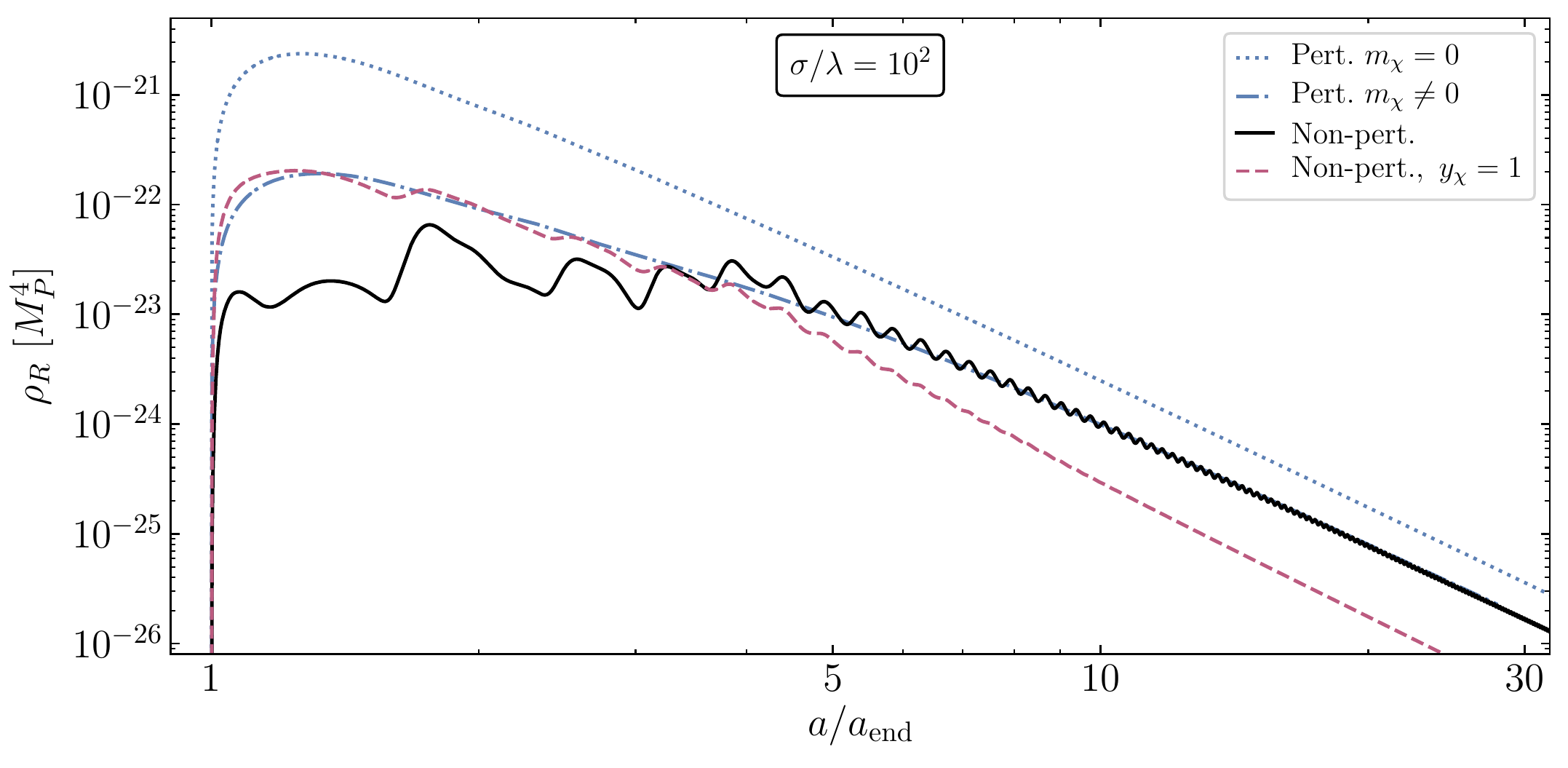}
    \includegraphics[width=.785\textwidth]{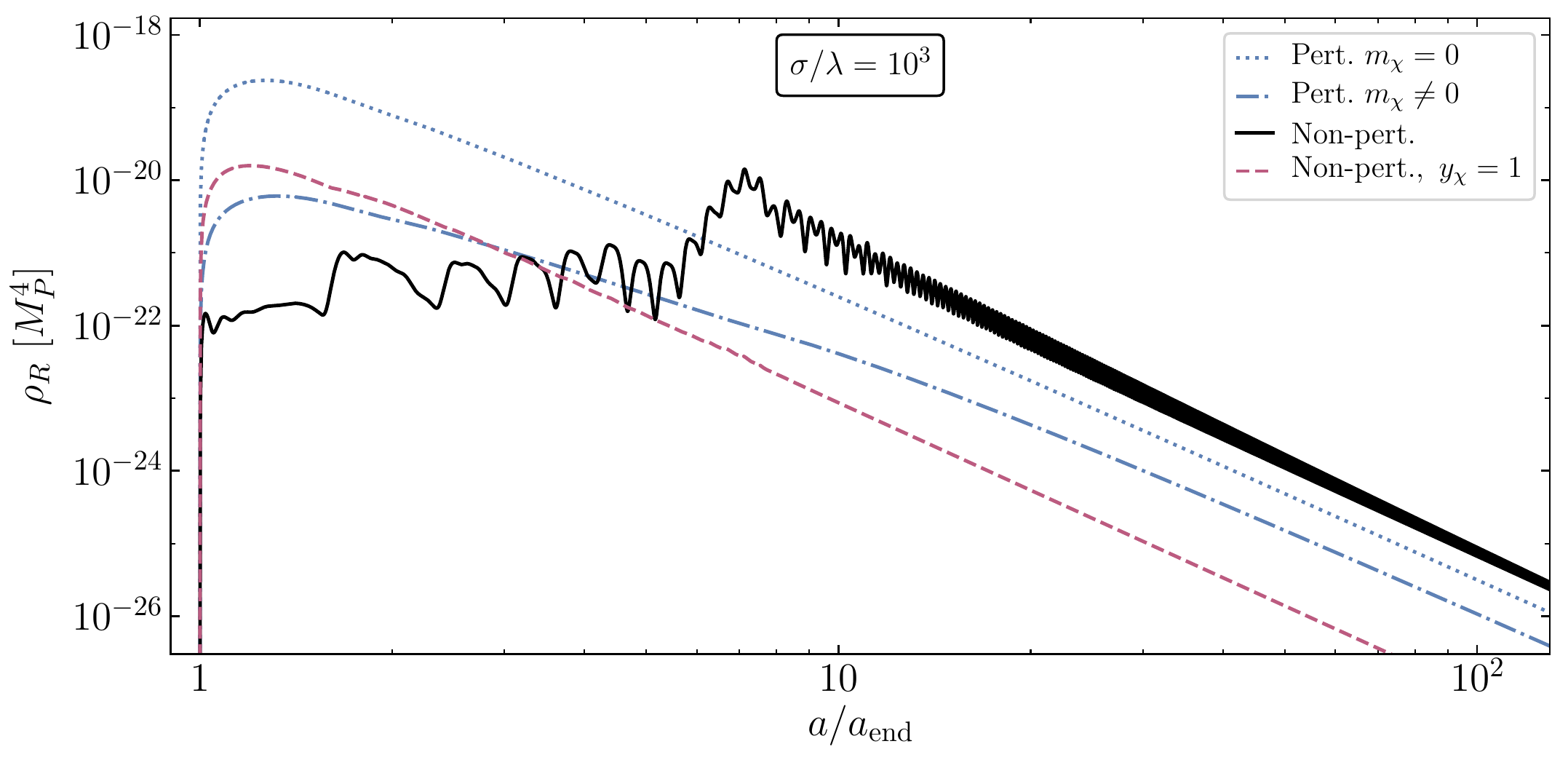}
    \caption{\it Instantaneous energy density in relativistic bosons during reheating sourced via the coupling $\frac{\sigma}{2}\phi^2\chi^2$ for values of $\sigma/\lambda = 10, \, 100$, and $1000$. We show the energy density $\rho_R=\rho_{\chi}$ determined perturbatively when the $\phi$-induced mass of $\chi$ is ignored (dotted blue), accounting for this induced mass (dashed-dotted blue), and computed non-perturbatively (solid black) as a function of the scale factor.
    We also show the energy density $\rho_R=\rho_{\chi}+\rho_f$ when $\chi$ decays rapidly assuming the coupling $y_\chi = 1$ (dashed purple).  
    }
    \label{fig:rhochi}\vspace{-20pt}
\end{figure}

\begin{figure}[!t]
\centering
    \includegraphics[width=.785\textwidth]{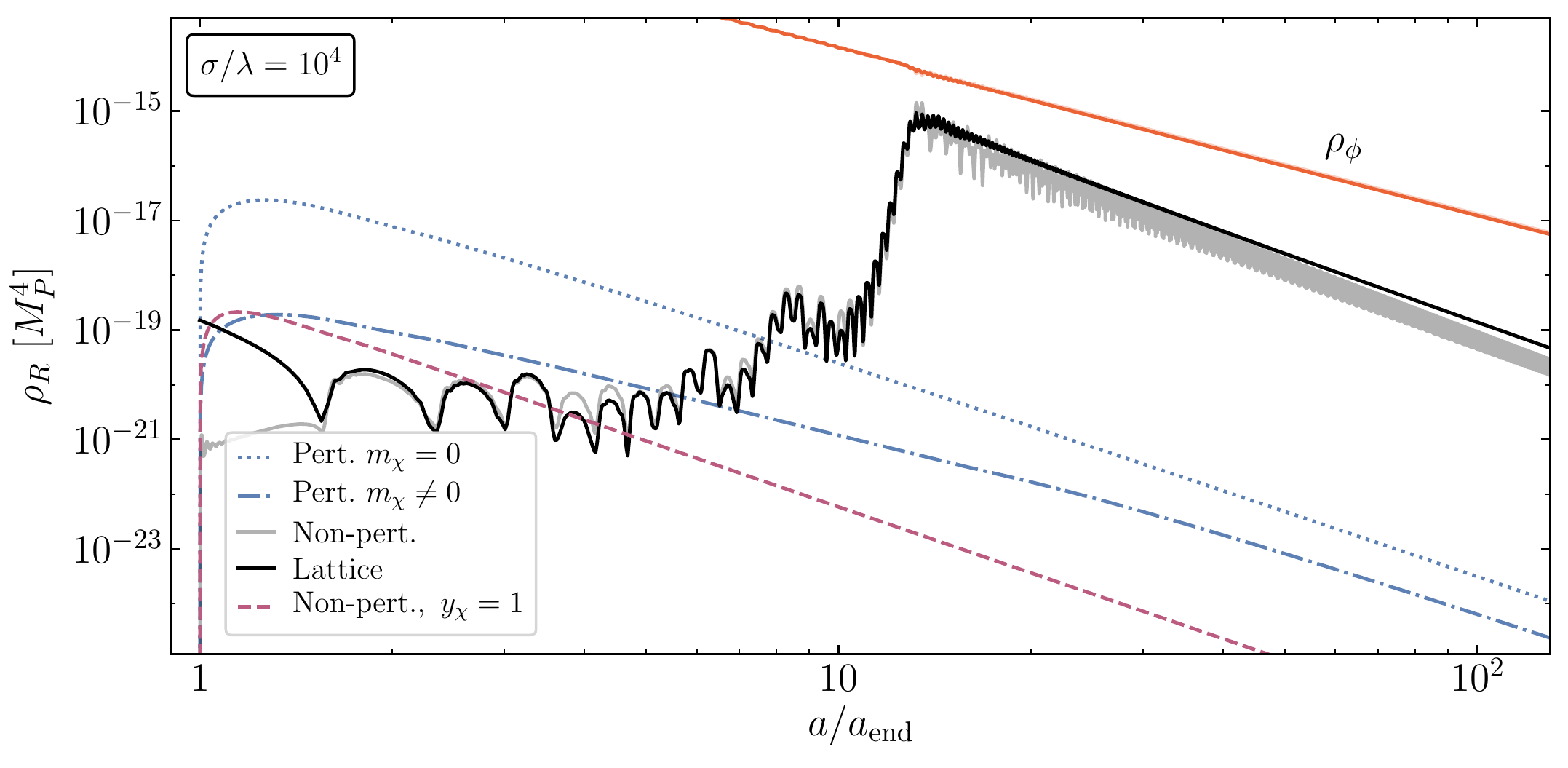}
    \includegraphics[width=.785\textwidth]{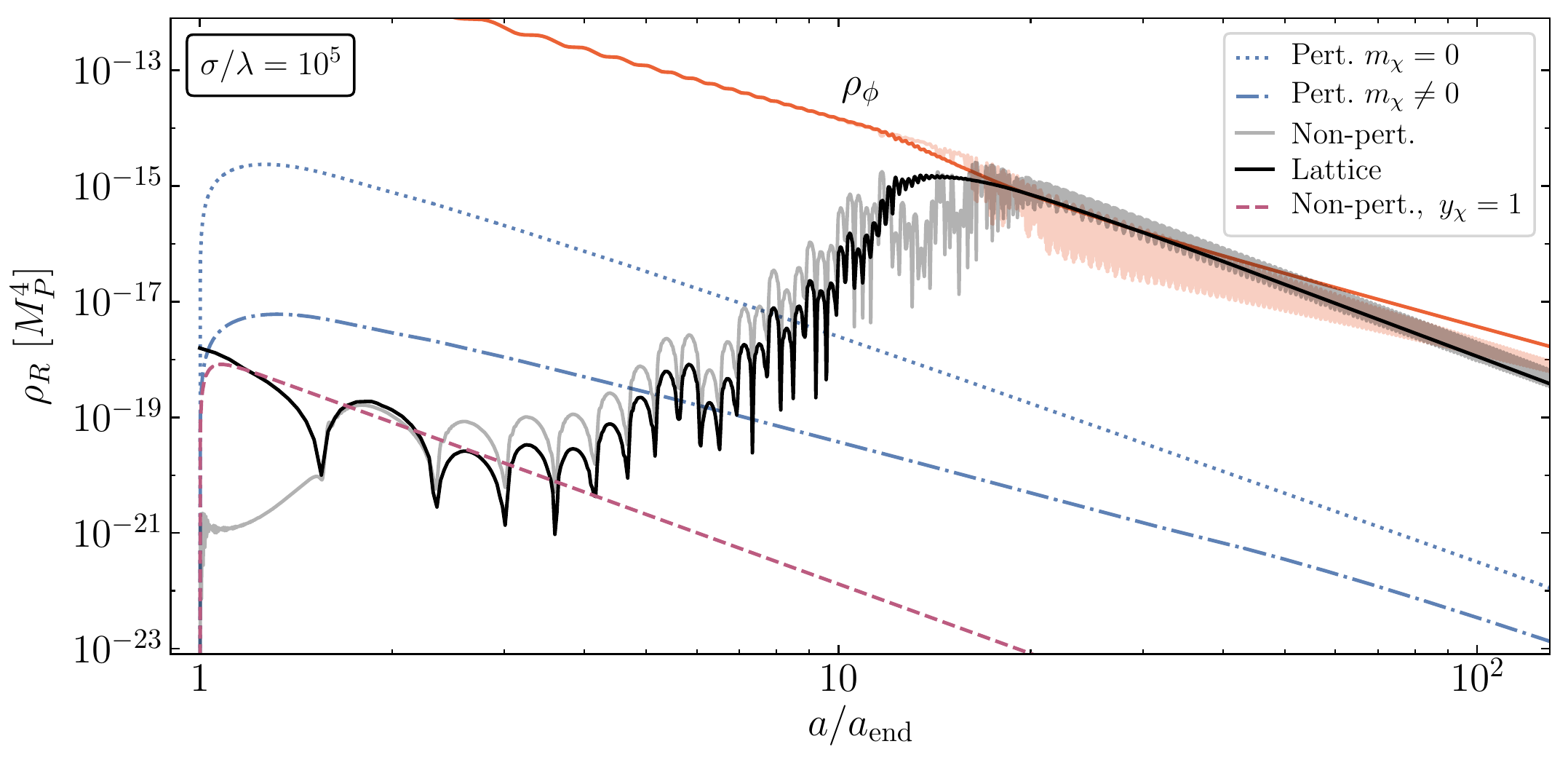}
    \includegraphics[width=.785\textwidth]{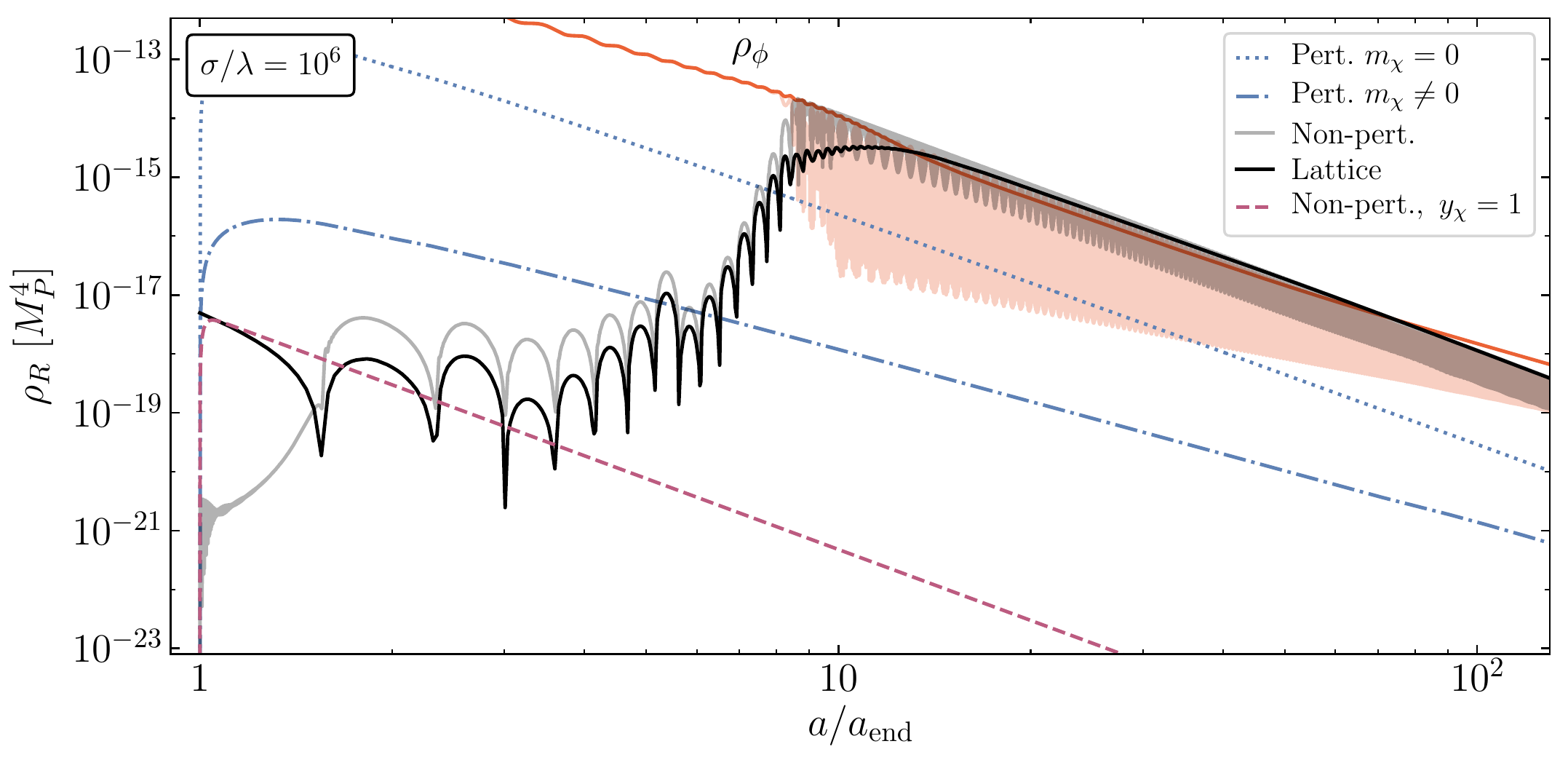}
        \caption{\it Instantaneous energy density in relativistic bosons during reheating sourced via the coupling $\frac{\sigma}{2}\phi^2\chi^2$ for values of $\sigma/\lambda = 10^4, \, 10^5$, and $10^6$. We show the energy density $\rho_R=\rho_{\chi}$ determined perturbatively when the $\phi$-induced mass of $\chi$ is ignored (dotted blue), accounting for this induced mass (dashed-dotted blue), and computed non-perturbatively in the Hartree approximation (transparent black) as a function of the scale factor. The solid black curves are computed using {\tt CosmoLattice}. We also show the energy density $\rho_R=\rho_{\chi}+\rho_f$ when $\chi$ decays rapidly assuming the coupling $y_\chi = 1$ (dashed purple).}
    \label{fig:rhochi2}\vspace{-20pt}
\end{figure}

In Figs.~\ref{fig:rhochi} and \ref{fig:rhochi2} we show the scale factor dependence of the energy density in the scalar field $\chi$ for a selection of couplings $\sigma$ with $10 \lesssim \sigma/\lambda \lesssim 10^6$.
In each panel, $\rho_{\chi}$ is shown as a function of the scale factor in the early stages of reheating. 
For the larger values of $10^4 \lesssim \sigma/\lambda \lesssim 10^6$ in Fig.~\ref{fig:rhochi2}, the inflaton energy density $\rho_\phi$ is also shown by the solid red curve.  
The dotted blue lines corresponds to the ``na\"ive'' result, $m_{\chi} = 0$, which ignores the instantaneous induced mass of $\chi$ due to the non-vanishing VEV of the inflaton, and
the dashed-dotted blue lines correspond the solution of the perturbative set keeping the kinematic effect induced by the oscillation of $\phi$ when $m_{\chi} \neq 0$, as discussed in the previous section. These results are computed by solving the perturbative Eqs.~(\ref{eq:FBeqs1})-(\ref{eq:FBeqs3}).
It can be seen clearly from the figures that the processes become suppressed due to $\mathcal{R} > 1$. Indeed, one can see a subtle change in slope of the blue dashed curves when $\mathcal{R} \gtrsim 1$ at progressively larger values of $a/a_{\rm end} \sim \left(2{\sigma}/{\lambda}\right)^{1/3}$, which increase with $\sigma/\lambda$.

The solid black lines in Figs.~\ref{fig:rhochi} and \ref{fig:rhochi2} correspond to the non-perturbative production of bosons $\chi$, $\rho_R=\rho_{\chi}$. Initially, for any arbitrary value of $\sigma/\lambda$, we see a delay in the production of bosons
due to the fact that the exponential production of final state bosons has not had sufficient time to produce particles as estimated in the perturbative analysis. As discussed in Appendix~\ref{sec:appA}, 
for small values of $\sigma/\lambda$ the exponential growth of particle production ends when
\begin{equation}
    q^2 m_{\phi} H^{-1} \lesssim 1~~~\Rightarrow ~~~ \sigma^2 \phi_0^3 \lesssim \frac{16 m_{\phi}^4}{\sqrt{6}M_P} \, ,
    \label{endexp}
\end{equation}
where $q=\frac{\sigma \phi_0^2}{4 m_{\phi}^2}$ is the resonance parameter and we assumed that at the end of inflation $H \simeq \sqrt{\lambda}\phi_0/\sqrt{3}$ for $\phi \lesssim 1$, which can be calculated from Eq.~(\ref{tmodelpot}). 
From this bound we can estimate that the explosive particle production approximately ends at $a/a_{\rm{end}} = 1.6$ (4.5) for $\sigma / \lambda =10$ (100) respectively, which is in agreement with Fig.~\ref{fig:rhochi}.
Since the rate of particle production scales with $\sigma/\lambda$, for larger values of this ratio the mode functions cross through a larger number of resonance bands. However, as the resonance parameter $q$ decreases due to expansion of the Universe, the resonance bands become narrower, thereby 
delaying the explosive particle production accordingly. This delay is a characteristic of the Mathieu equation that we describe in Appendix~\ref{sec:appA}.

As the scale factor $a$ increases, the amplitude of oscillations decreases sufficiently, $\phi_0 \sim a^{-3/2}$,  so that the induced mass $m_{\chi}$ becomes significantly smaller than $m_\phi$ causing the energy density $\rho_\chi$ to scale as radiation with $\rho_{\chi} \sim a^{-4}$. At this point, we recover the same perturbative slope, as in the case where the suppression effect induced by ${\cal R}$ is neglected (dotted blue lines). 

For $\sigma/\lambda\lesssim 10$, the energy density of $\chi$ is smaller than that obtained in the perturbative case, even after accounting for the kinematic factor $\mathcal{R}$. The relatively weak non-perturbative production of bosons is dominated by the first inflaton oscillations, with $n_p\lesssim 1$ (see Fig.~\ref{fig:npS} in Appendix~\ref{sec:appB}). Adiabaticity is violated more weakly as the amplitude of $\phi$ decreases. 
For such weak couplings, the effect of the initial Hubble friction, with a value of $H_{\rm end}$ inherited from inflation, leads to the observed suppression in the rate of particle production. More precisely, we can re-write Eq.~(\ref{eq:omS}) as 
\beq
\omega_p^2 \;=\; p^2 + a^2 \left[ m_{\chi}^2 - \frac{1-3w}{2}H^2\right]\,,
\eeq
where $w$ denotes the equation of state parameter. As $H^2/m_{\chi}^2 \simeq 4(\lambda/\sigma)$ at the end of inflation, and $-1/3<w<1$ during the first oscillation of $\phi$, the result is a suppression in the particle production rate of $\chi$. Only for $\sigma/\lambda \gg 1$ can the $a''/a$ term in $\omega_p^2$ be ignored and the usual Floquet analysis in the absence of expansion is valid during the first zero-crossings of $\phi$ (see Appendix~\ref{sec:appA} for more details). 

For larger values of the inflaton-matter coupling, the resonant production of $\chi$ is capable of overcoming the Hubble damping. The maximum of $\rho_{\chi}$, which is reached during the first oscillation of $\phi$ for $10\lesssim \sigma/\lambda \lesssim 300$, can be easily estimated. This is the case for the middle panel in Fig.~\ref{fig:rhochi}. The occupation number for each $X_p$ can be approximated as 
\beq
\label{numbdens}
n_{p} \;\simeq\; \exp\left(-\frac{\pi (p^2 + m_{\chi}^2)}{|\dot{m}_{\chi}|}\right) \, ,
\eeq
after the first zero crossing of $\phi$~\cite{Felder:1998vq,Felder:1999pv,Chung:1999ve}. At the subsequent maximum of the $\phi$ oscillation, integration gives $\rho_{\chi} \propto \sigma^{5/4}$ by straightforward substitution of (\ref{numbdens}) into (\ref{eq:rhochi}) with $\omega_p \simeq \sqrt{p^2 + \sigma\phi^2}$.\footnote{More precisely, $\rho_{\chi}\propto \sigma^{5/4}$ for $\sigma/\lambda \gtrsim 20$.} It can also be verified that at this maximum, the ratio of the perturbative to the non-perturbative results, including the $\mathcal{R}$-suppression, is $\rho_{\chi,{\rm pert.}}/\rho_{\chi,{\rm non\mbox{-}pert.}} \simeq (\sigma/\lambda)^{1/4}$.

For larger values of $\sigma/\lambda$, we see that the resonant production leads to a strong enhancement in the energy density, which occurs when the bound Eq.~(\ref{endexp}) is violated. Occupation numbers grow exponentially with cosmic time, $n_p\gg 1$ (see Fig.~\ref{fig:npL} in Appendix~\ref{sec:appB}). This enhancement continues to grow until $\sigma/\lambda \sim 10^4$, when the back-reaction effects become important,  discussed in more detail in the following subsections. Non-perturbative effects only become important and influence the evolution of $\rho_\phi$ when $\sigma/\lambda \gtrsim 10^4$. Below this value, classical perturbative analysis for $\rho_\phi$ remains valid.

\begin{figure}[!ht]
\centering
    \includegraphics[width=\textwidth]{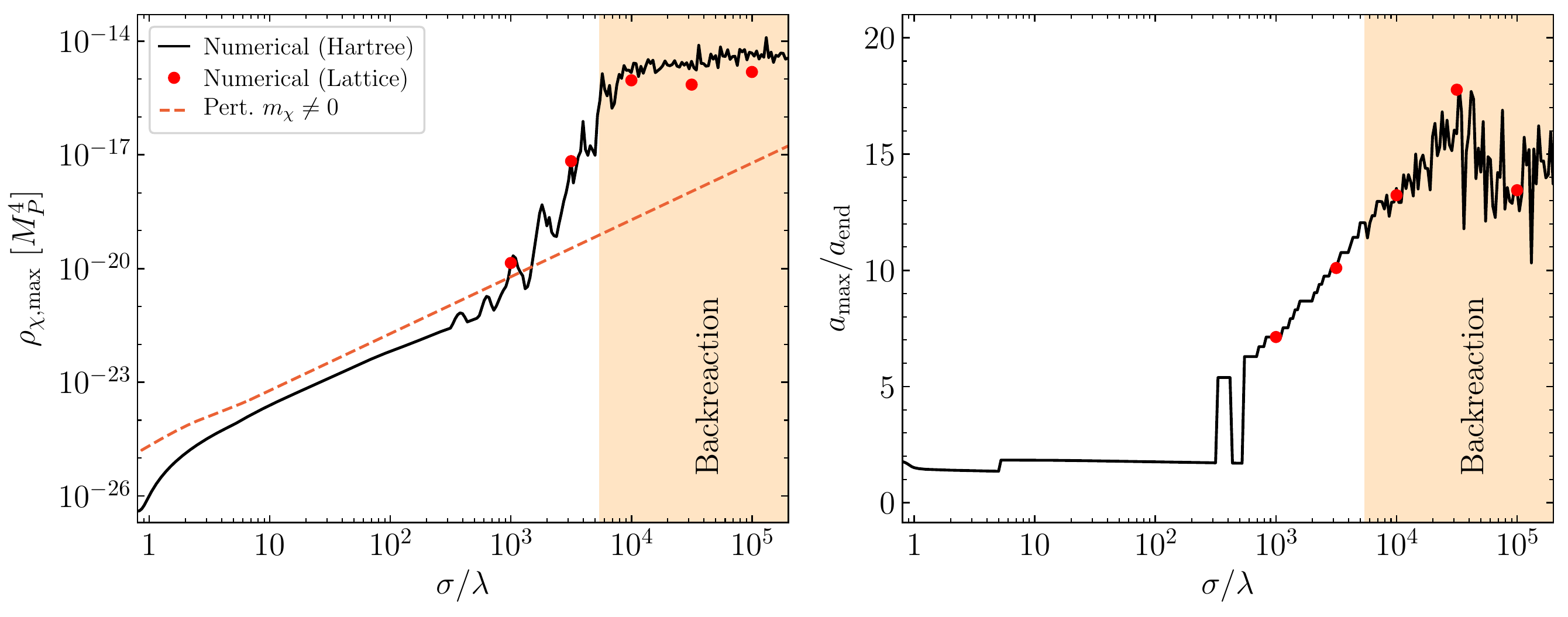}
    \caption{\it Maximum energy density in $\chi$, and the scale factor at which this maximum is reached, as a function of $\sigma/\lambda$. The backreaction band corresponds to the region for which $\rho_{\chi}\geq 0.1 \rho_{\phi}$ at $a=a_{\rm max}$. The solid curves are computed in the Hartree approximation. The red dots were
    computed using {\tt CosmoLattice}. For comparison we also show in the left panel the perturbative approximation (dashed).}
    \label{boseback}%\vspace{-20pt}
\end{figure}

Fig.~\ref{boseback} summarizes our results for the numerically computed maximum energy density in $\chi$, $\rho_{\chi,{\rm max}}$, shown as a function of the coupling $\sigma/\lambda$ (left panel, solid black line). In the right panel, we show the corresponding scale factor. In the left panel, we further show the perturbative estimate for the maximum of $\rho_{\chi}$ (red dashed line). For $\sigma/\lambda \lesssim 300$, $\rho_{\chi}$ is maximized at the first $\phi$ oscillation and grows monotonically with $\sigma$. As discussed above, this monotonic growth does not scale linearly with $\sigma$ either because of the suppression by expansion, for $\sigma/\lambda \lesssim 20$, or due to the $\rho_{\chi} \propto \sigma^{5/4}$ behavior for $20 \lesssim \sigma/\lambda \lesssim 300$. The ($\mathcal{R}$-suppressed) perturbative over-estimation in turn can be blamed on a breakdown of the averaging-over-oscillations procedure used to estimate the rate, due the short time-scale here considered. 

For larger coupling, up to $\sigma/\lambda\simeq 5\times 10^3$, the general trend is still of a growing $\rho_{\chi,{\rm max}}$, but this trend is non-monotonic. Instead, we note the appearance of peaks, which correspond to a
maximal energy in $\chi$ that occurs at later times, with $a/a_{\rm end}>5$. Indeed, for larger values of $\sigma$, there is an increase in the number of
unstable bands that each mode can pass through with time. Thus, the larger the
value of $\sigma$, the more possibilities there are to increase the production of particles $\chi$ (and thus achieve larger $\rho_{\chi,{\rm max}}$) over a longer period of time, which means at larger values of the scaling factor $a$. Finally, for $\sigma/\lambda\gtrsim 5\times 10^3$, we enter the backreaction region, which we now proceed to discuss below. We observe that $\rho_{\chi,{\rm max}}$ plateaus to values between $10^{-14}-10^{-15}M_P^4$, equal or (slightly) larger than $\rho_{\phi}$ at the corresponding time (see Fig.~\ref{fig:rhochi2}). 
The stochastic nature of the bosonic resonance, however, makes difficult to pinpoint exactly the moment in time at which this maximum is reached, as demonstrated by the disordered behavior of $a_{\rm max}$ in the back-reaction domain.

%%%%%%%%%%%%%%%%%%%%%%%%%%%%%%%%%%%
\subsubsection{Back-reaction}
\label{sec:nonpert-back}
%%%%%%%%%%%%%%%%%%%%%%%%%%%%%%%%%%%

For all of the values of $\sigma/\lambda$ shown in Fig.~\ref{fig:rhochi}, the energy density of $\chi$ is always small compared to that of the inflaton. Nevertheless, for a sufficiently large coupling $\sigma$, the two become comparable. If this is the case, the effect of the non-perturbative particle production on the background dynamics cannot be ignored. In the homogeneous limit (that is, ignoring mode-mode couplings of the inflaton perturbations and the decay products, i.e.~{\em rescattering}), the effect on the equation of motion of $\phi$ is captured by the Hartree approximation~\cite{preheating2},
\beq\label{eq:phiHart}
\ddot{\phi} + 3H\dot{\phi} + V_{\phi}' + \sigma\langle \chi^2\rangle \phi \;=\;0\,.
\eeq
From Eq.~(\ref{eq:phiHart}), we see the source of the back-reaction processes. Large values of $\sigma$ will explosively generate a large population 
of $\chi$, $\propto \langle \chi^2 \rangle$, which in turn will source in the evolution of $\phi$, and thus $\rho_\phi$.
The UV-finite scalar expectation value can be computed from Eq.(\ref{Eq:X}) writing
$\chi_p =X_p / a$:
\beq
\langle \chi^2\rangle \;=\; \frac{1}{(2\pi)^3a^2} \int d^3\bp \, \left( |X_p|^2 - \frac{1}{2\omega_p}\right)\,.
\eeq

Fig.~\ref{fig:rhochi2} shows the evolution of the energy densities of $\chi$ and $\phi$ in the Hartree approximation for
$\sigma/\lambda=10^4$, $10^5$ and $10^6$. This approximation is shown as the semi-transparent curves. In all three cases, the scalar decay product $\chi$ comprises a non-negligible component of the energy density of the Universe, $\rho_{\chi} > 0.1\,\rho_{\phi}$, at least near the maximum of $\rho_\chi$.
In the top panel, for $\sigma/\lambda=10^4$, backreaction is barely visible
and only affects $\rho_\phi$ at $\rho_{\chi,{\rm max}}$. We observe here, as discussed earlier, that the exponential growth regime for $\rho_{\chi}$ is delayed by a few tens of $\phi$ oscillations, as the resonance builds up. Nevertheless, the inflaton dominates the energy density at all times and the deviation in the decrease of $\rho_{\phi}$ due to  redshift is essentially imperceptible.
In the middle and lower panels, with $\sigma/\lambda=10^5$ and $10^6$, the effect of backreaction is more clearly observed. Here the resonant growth of $\chi$ is so efficient that $\rho_{\chi}\simeq \rho_{\phi}$ around $a/ a_{\rm end}=20$ and 10 respectively.

It is important to note that without
the addition of the Hartree term to the equation of motion (\ref{eq:phiHart}) and without accounting for the contribution of $\rho_{\chi}$ to $H$, the exponential increase in $\rho_{\chi}$ would continue beyond this threshold. This would suggest the completion of the reheating process, albeit suggesting as well a violation of (total) energy conservation. We see here that when we properly account for the effect on the background, the growth of $\rho_{\chi}$ is arrested and $\rho_{\chi}\simeq \rho_{\phi}$ only until $a/a_{\rm end}\simeq 40$ for $\sigma/\lambda=10^5$. 
Subsequently, resonant production shuts off as $m_\chi$ drops below $m_\phi$ (see Appendix~\ref{sec:appA} for details), and the inflaton again dominates the energy density of the Universe since it redshifts as matter, while $\rho_{\chi}$ redshifts as radiation. At this point, we are in the perturbative regime and  unsurprisingly, we observe that this decoupling
happens at the same moment that the slope changes in the perturbative approximation (dashed lines). 
A crude estimate of the time of appearance of the perturbative regime is to estimate $m_\chi|_{\phi\rightarrow \phi_0} <m_\phi$, which gives
\beq
\left(\frac{a}{a_{\rm end}}\right)_{pert.} = \left(\sigma \frac{\phi^2_{\rm end}}{m_\phi^2} \right)^{1/3} \simeq 35
\label{Eqpertubis}
\eeq
for $\sigma=10^5 ~\lambda$.

For the third and last panel, with $\sigma/\lambda=10^6$, the evolution is similar to the previous case, although the momentary domination of $\chi$ over $\phi$ is stronger and lasts longer.  As a consequence, the backreaction regime is reached 
earlier (around $a/a_{\rm max}=8$) through a well-defined single regime of exponential growth and the scalar $\chi$ quickly dominates the energy budget of the Universe. In the Hartree approximation, the inflaton condensate retains its integrity, and the strong coupling results in a pattern of high frequency oscillations as energy is transferred between the inflaton and $\chi$. We note that, despite this exchange, $\rho_{\chi}$ ultimately decreases as $a^{-4}$  i.e.~radiation.\footnote{ Notice that the fact that the radiation bath evolves as $a^{-4}$ as if there was no entropy injection from the annihilation of inflaton is characteristic of scatterings in a quadratic potential. It essentially comes from the fact that the production rate proportional to $\rho_\phi^2$ is more redshifted than $a^{-4}$ \cite{Garcia:2020wiy}.} As a result, the inflaton eventually dominates the energy density of the Universe, 
until it decays at a later time by means of an alternative decay channel.

So far we have discussed the effect of the depletion of $\phi$ into $\chi$ at the homogeneous (Hartree) approximation. This approximation, however, does not take into account the disruption of the homogeneous inflaton condensate, which will be induced by the rescattering of $\chi$ particles into $\phi$, leading to inflaton {\em particle} production, nor the scattering between these particle $\phi$ and $\chi$. At the Fourier level, this manifests as mode-mode couplings of the inflaton perturbations and the decay products. Accounting for these effects in the corresponding equations of motion results in a set of non-linear operator equations in the fields, which pose a challenging mathematical problem.
To get around these issues, it is commonly argued that at large occupation numbers, which are obtained shortly after the beginning of preheating for a sufficiently large coupling,
the quantum fields can be approximated as classical, and hence, their perturbations can be simply studied by solving a classical system of equations.\footnote{Due to this requirement, lattice methods are not suited for the study of reheating for occupation numbers $n_p\lesssim O(1)$, which is the case of $\sigma/\lambda<10^3$, as shown in the top two right panels of Fig.~\ref{fig:rhochi}.} Without the need to track each (quantum) mode function, the resulting set of partial differential equations can be solved with non-spectral methods. A favored integration scheme involves finite-difference techniques on a spatial lattice, in the presence of a spatially homogeneous and isotropic spacetime.\footnote{i.e.~neglecting metric fluctuations, which nevertheless may also be excited by the parametric resonance~\cite{Nambu:1996gf,Bassett:1998wg,Bassett:1999mt,Jedamzik:2010dq,Huang:2011gf,Giblin:2019nuv}.} This is the case of codes such as {\tt LatticeEasy}~\cite{Felder:2000hq}, {\tt Defrost}~\cite{Frolov:2008hy}, {\tt PyCool}~\cite{Sainio:2012mw}, {\tt GFiRe}~\cite{Lozanov:2019jff} and {\tt CosmoLattice}~\cite{Figueroa:2020rrl,Figueroa:2021yhd}, the last of which we use in the present work.

 In Fig.~\ref{boseback} the lattice-computed values of $\rho_{\chi,{\rm max}}$ and $a_{\rm max}$ are shown for discrete values of $\sigma/\lambda$ depicted by red dots. Outside the backreaction regime the resulting values are in agreement with our conventional spectral code. However, for larger values of the coupling, the lattice calculation results in slightly lower values of $\rho_{\chi,{\rm max}}$ compared to the Hartree approximation. Notably though, there is agreement in $a_{\rm max}$.

The effect of re-scattering leads to the disruption of the $\phi$ condensate in favor of particles and suppresses the available amplitude in the inflaton oscillations that feeds the parametric resonance. In addition, energy is transferred from the amplified modes to those lying in the stability regions, resulting in a further slow-down of the resonant particle production after the peak in $\rho_{\chi}$ is reached. Hence, $\rho_{\chi,{\rm max}}$ is lower than in the Hartree approximation as seen in Fig.~\ref{boseback}. This energy exchange process results in a Universe dominated by $\chi$ and a fragmented $\phi$~\cite{Garcia-Bellido:2002fsq,Felder:2006cc,Frolov:2010sz,Amin:2014eta}. 
 
The Universe is momentarily democratic, with $\rho_{\phi}\simeq\rho_{\chi}$, 
with an equation of state intermediate between that of matter and radiation,
until the resonance is fully extinguished in the narrow resonance regime (see above), and $\phi$ and $\chi$ enter their corresponding redshift regimes,  $\rho_\phi \propto a^{-3}$ and $\rho_\chi\propto a^{-4}$ respectively. This can be observed in the solid black and red curves in Fig.~\ref{fig:rhochi2}.

\begin{figure}[!ht]
\centering
    \includegraphics[width=\textwidth]{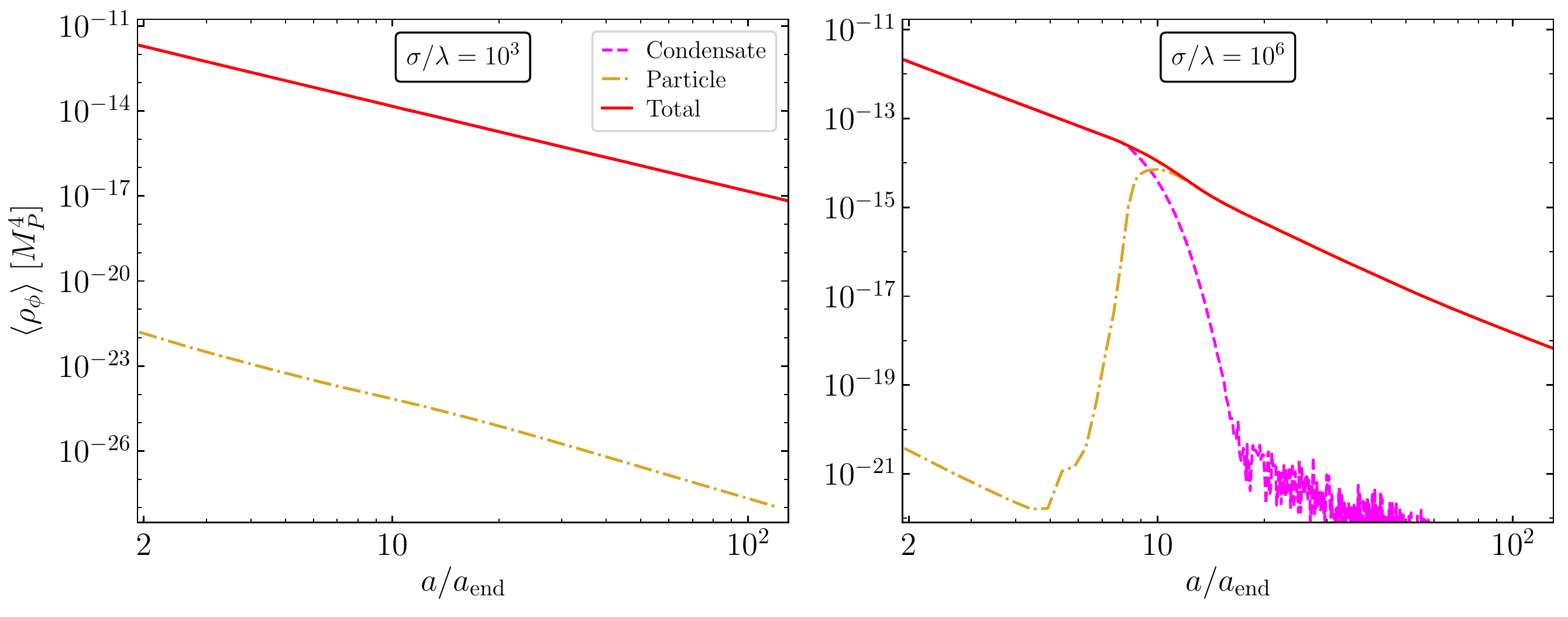}
    \caption{\it Average over inflaton oscillations of the energy density of the inflaton, its condensate component, defined in Eq.~(\ref{eq:rhocond}), and the remainder, which we identify with inflaton particles. Left panel: In the absence of strong backreaction, the integrity of the condensate is preserved during preheating. Right panel: For strong backreaction, the inflaton condensate is quickly replaced by a plasma of inflaton and $\chi$ particles. Without a condensate it is difficult to compute time averages reliably, hence the noise. }
    \label{fig:condpart}%\vspace{-20pt}
\end{figure}

We can estimate the fraction of the energy density in $\rho_{\phi}$ that is in the spatially homogeneous condensate by computing the {\em spatial} average of $\phi$ and its time-derivative, which we denote by $\bar{\phi}$ and $\bar{\dot{\phi}}$, respectively. We then estimate the condensate contribution as
\beq\label{eq:rhocond}
\rho_{\phi,\,{\rm condensate}} \;\simeq\; \frac{1}{2} \bar{\dot{\phi}}^2 + V(\bar{\phi})\,,
\eeq
and the reminder, $\rho_{\phi,\,{\rm particles}} \simeq \rho_{\phi}-\rho_{\phi,\,{\rm condensate}}$. Fig.~\ref{fig:condpart} shows the oscillation-averaged result of this decomposition for two different couplings. We indeed observe in the right-hand panel that for $\sigma/\lambda=10^6$, the backreaction from the strong parametric resonance rapidly replaces the coherent component of $\phi$ for an incoherent one. Hence, at later times the Universe becomes dominated by free particle inflatons instead of an oscillating condensate. Notably, for the set-up chosen in this work 
i.e a quadratic potential, both outcomes lead to the same expansion history, namely a matter dominated Universe.

In the end, the net effect of the destruction of the condensate by 
the scattering of $\chi$ on the homogeneous field $\phi$ is a premature decoupling
of $\rho_\chi$ because of a weakening of the non-perturbative effects generated in
the Mathieu equation. As was the case for the Hartree approximation, this
effect can be effectively parameterized by a change in $\rho_{\rm end}$ with respect
to a pure perturbative approach. This can be seen by the slight deflection in the total energy density in $\rho_\phi$ as the condensate is destroyed in Fig.~\ref{fig:condpart}. Thus as a result of the back scattering, the Universe which enters a phase where it is dominated by $\phi$-particles rather than a $\phi$-condensate.

%%%%%%%%%%%%%%%%%%%%%%%%%%%%%%%%%%%
\subsubsection{Boson decay}\label{sec:unstablechi}
%%%%%%%%%%%%%%%%%%%%%%%%%%%%%%%%%%%

In general, the decay products of the inflaton will be themselves unstable. 
For example, if the inflaton couples to the Higgs boson of the Standard Model, its decays and scatterings will populate the radiation
bath first with massive fermions, and eventually the full set of Standard Model fields.
%When this is the case, their decays will populate both the visible and dark sectors of the Universe. 
Moreover, if the width of $\chi$ is sufficiently large, it will alter the resonant/suppressed particle production that characterizes preheating. 
Let us for definiteness consider the decay of $\chi$ into a fermion pair $\bar{f}f$, with coupling\footnote{ We can imagine for example that 
$\chi$ is the Higgs field, and $y_\chi$ the top-quark Yukawa coupling.} $y_\chi$. 
The decay rate is therefore given by
\beq\label{eq:gammachi}
\Gamma_{\chi} \;=\; \frac{y_\chi^2}{8\pi}m_{\chi}(t)\left(1-\frac{4m_f^2}{m_{\chi}^2(t)}\right)^{3/2}\,.
\eeq
In the absence of strong backreaction, the energy density of the Universe is dominated by $\phi$, and we can easily compare this rate with the instantaneous expansion rate,
\beq
\frac{\Gamma_{\chi}}{H} \;\lesssim\; \frac{y_{\chi}^2 \sqrt{\sigma}\phi_{\rm end}}{8\pi H_{\rm end}} \left(1-\frac{4m_f^2}{m_{\chi}^2(t)}\right)^{3/2} \;\simeq\; 0.06 y_{\chi}^2 \left(\frac{\sigma}{\lambda}\right)^{1/2} \left(1-\frac{4m_f^2}{m_{\chi}^2(t)}\right)^{3/2}\,.
\eeq
We note then that the condition for efficient decay is $y_{\chi}^2 (\sigma/\lambda)^{1/2} \gtrsim 20$ assuming $f$ is massless. This condition scales only with the square root of the inflaton-matter coupling, but with the square of the $\chi$-$f$ coupling. Therefore, even for $\sigma \sim 10^2 \lambda$, dissipation can be efficient for $y_{\chi}\sim \mathcal{O}(1)$. Larger $\sigma$ must of course be explored fully numerically as previously discussed.
We note that if we associate $\chi$ with the Higgs boson, and  $y_\chi= y_t$, we expect decays to be important.

When the dissipation into SM light degrees of freedom is efficient, the equations of motion need to be modified. Assuming that the decay of $\chi$ can be reliably described by the rate (\ref{eq:gammachi}), the modified equations of motion take the form~\cite{Repond:2016sol,Fan:2021otj}\footnote{Here we neglect the gradient contribution to $\rho_f$, as argued in detail in~\cite{Repond:2016sol}. }
\begin{align}
\left(\frac{d^2}{dt^2} - \frac{\nabla^2}{a^2} + (3H+\Gamma_{\chi})\frac{d}{dt} + m^2_{\chi}(t) \right)\chi \;&=\; 0\,,\\
\dot{\rho}_f + 4H\rho_f \;&=\; \Gamma_{\chi}\dot{\chi}^2\,.
\end{align}
Alternatively, in terms of conformal time and the re-scaled $\chi$ mode functions
\beq
Y_p \;\equiv\; a\exp\left(\frac{1}{2}\int a\Gamma_{\chi}\,d\tau\right)\chi_p\,,
\eeq
these equations can be rewritten in a quantization-friendly form as follows,
\begin{align} \label{eq:unstablechi1}
Y_p'' + \Omega_p^2 Y_p \;&=\; 0\,,\\ \label{eq:unstablechi2}
\rho_f' + 4\mathcal{H}\rho_f \;&=\; a\Gamma_{\chi} \dot{\chi}^2\,.
\end{align}
Here\footnote{We disregard conformal time derivatives of $\Gamma_{\chi}$ as the rate~(\ref{eq:gammachi}) is computed  adiabatically in perturbation theory. } 
\beq
\Omega_p^2 \;\equiv\; p^2 + a^2m_{\chi}^2 - \frac{a''}{a} - \frac{1}{4}(a\Gamma_{\chi})^2 - \frac{3}{2}a\mathcal{H}\Gamma_{\chi}\,,
\label{eq:bigomega}
\eeq
with $\mathcal{H}=a'/a$, and the regularized kinetic energy of $\chi$ can be written as
\beq
\dot{\chi}^2 \;=\; \frac{e^{-\int a\Gamma_{\chi}\,d\tau}}{(2\pi)^3a^4} \int d^3\bp \left( |Y_p'|^2 - \frac{1}{4} |Y_p' - i\Omega_p Y_p|^2 \right)\,.
\eeq

We note from Eq.~(\ref{eq:bigomega}) that for an unstable $\chi$ the effective mass is reduced by $\Gamma_{\chi}$-dependent factors. Nevertheless, for the dominant term,
\beq
a^2m_{\chi}^2 - \frac{1}{4}a^2\Gamma_{\chi}^2 \;=\; a^2m_{\chi}^2 \left(1 - \frac{y^4}{256\pi^2}\right)\,,
\eeq
that is, for $y\leq 1$, this reduction is essentially negligible, and the resonance response of the field is maintained. However, the energy density is affected by the exponential decay,
\beq
\rho_{\chi} \;=\; \frac{e^{-\int a\Gamma_{\chi}\,d\tau}}{(2\pi)^3 a^4}\int d^3\bp \,\Omega_p n_p\,, \qquad n_p \;=\; \frac{1}{2\Omega_p} \left[\left|Y_p'  \right|^2 + \Omega_p^2 \left|Y_p\right|^2 \right] -\frac{1}{2}\,.
\eeq
and will therefore be converted into a homogeneous plasma before the end of reheating for a sufficiently large $y_{\chi}$.

The purple dashed lines representing $\rho_R=\rho_{\chi}+\rho_f$ in Figs.~\ref{fig:rhochi} and \ref{fig:rhochi2} show the result of solving the system of equations (\ref{eq:unstablechi1}) and (\ref{eq:unstablechi2}) in the presence of the oscillating inflaton for $y_{\chi}=1$. This choice is inspired by identifying $\chi$ with the Higgs field and $f$ with the top quark. In all cases, during the first oscillation, the production of particles is not limited by expansion, as $\Gamma_{\chi}\gtrsim H$. Thus, unlike the stable $\chi$ scenario, the energy density can be deposited almost immediately into fermions, and the instantaneous maximum in $\rho_R$ is close to the perturbative value. 

Since the ratio $\Gamma_{\chi}/H$ is independent of the scale factor, if the decay dominates over the expansion rate at the beginning of the (p)reheating process,
it will always dominate. In this sense, 
we will not observe the parametric resonance in $\rho_R$ described in the previous section: as soon as a $\chi$-particle is produced, it decays into a radiation bath. This is clearly seen in Figs.~\ref{fig:rhochi} and \ref{fig:rhochi2}. 
Note also that in this case $\rho_R$ evolves as  $\rho_R \propto a^{-4}$,
typical for a bath of relativistic particles (we took $m_f=0$). 
Due to the rapid decay of $\chi$ to radiation, there is no longer a suppression in the energy density due to the delay in production discussed above. 
As a result, the maximum energy density in radiation is in fact similar to the perturbative case with $m_\chi \ne 0$. That is, we can approximate the non-perturbative production in this case using perturbative analysis with the suppression $\mathcal{R}^{-1/2}$ at the maximum. However now, the radiation bath is massless and the energy density falls off as  $a^{-4}$, faster than in the perturbative case with $m_\chi \ne 0$.

At later times, the $\sigma$-dependence of the exponential decay of $\chi$ is noticeable. For $\sigma/\lambda=10$, $\chi$ decays gently throughout the time spanned by the figure, and the radiation reaches energy densities similar to the perturbative result before reaching the redshift regime. For $\sigma/\lambda=10^2$, one observes that after a few oscillations, $\rho_{\chi}$ has decayed sufficiently to stop sourcing $\rho_f$, and the resulting radiation rapidly enters the redshift regime, at around $a/a_{\rm end}=5$. Notably, $\rho_R$ is smaller than the stable $\chi$ cases, perturbative and non-perturbative. For even larger values of $\sigma$, the decay rate of $\chi$ grows accordingly, and very efficiently dissipates $\rho_{\chi}$ after two or less oscillations. The redshift regime is almost immediately reached, and the resulting $\rho_{\chi}$ is parallel to the dotted perturbative line, but smaller by several orders of magnitude. The backreaction regime is therefore never reached.

%%%%%%%%%%%%%%%%%%%%%%%%%%%%%%%%%%%
\subsection{Production of spin-1/2 fermions}
\label{Sec:fermions}
%%%%%%%%%%%%%%%%%%%%%%%%%%%%%%%%%%%

A similar study can be carried out with respect to the production of fermions, paying attention to some subtleties specific to fermionic fields, in particular concerning the treatment of covariant derivatives in a Friedmann-Lemaitre-Robertson-Walker (FLRW) background. From the following action,
\beq
\mathcal{S}_{\psi} \;=\; \int d^4x\, \sqrt{-g} \bar{\psi} \left(i \bar{\gamma}^\mu\nabla_{\mu} - m_{\psi}(t) \right)\psi\,,
\eeq
we can define the field $\Psi\equiv a^{3/2}\psi$, and the equation of motion can be written in a familiar flat-space Dirac equation in terms of the conformal time $\tau$,
\beq\label{eq:dPsi}
\left(i\gamma^{\mu} \partial_{\mu} - am_{\psi}(\tau)\right)\Psi \;=\;0\, ,
\eeq
which is derived in Appendix~\ref{sec:appB}. This choice of variables is convenient since it is straightforward to quantize the field operator $\Psi$ as
we did in the scalar case (see Eq.~(\ref{Eq:X})), and in terms of the usual mode expansion, we find
\beq
\label{modeexpferm}
\Psi(\tau,\bx) \;=\; \sum_{r=\pm}\int \frac{d^3 \bp}{(2\pi)^{3/2}}e^{-i\bp\cdot\bx}\left[ u^{(r)}_p(\tau)\hat{a}^{(r)}_{\bp} + v^{(r)}_p(\tau)\hat{b}^{(r)\,\dagger}_{-\bp}\right]\,,
\eeq
where the commutation relations are given by $\{\hat{a}_{\bp},\hat{a}_{\bp'}^{\dagger}\}=\{\hat{b}_{\bp},\hat{b}_{\bp'}^{\dagger}\}=\delta(\bp-\bp')$, $\{\hat{a}_{\bp},\hat{a}_{\bp'}\}=\{\hat{a}_{\bp}^{\dagger},\hat{a}_{\bp'}^{\dagger}\}=\{\hat{a}_{\bp}^{\dagger},\hat{b}_{\bp'}^{\dagger}\}=0$, and $v^{(r)}_p=\mathcal{C}\bar{u}^{(r)\,T}_{p}$ with $\mathcal{C}=i\gamma^2\gamma^0$ the charge conjugation matrix.

We find that the expectation value corresponding to the energy density in $\psi$ takes the form
\beq
\label{endenferm}
\rho_{\psi} \;=\; \frac{1}{(2\pi)^3 a^4}\int d^3\bp \, \left[ am_{\psi}(|U_2|^2 - |U_1|^2) - p\left( U_1 U_2^* + U_1^* U_2 \right)  + 2\omega_p \right]\, ,
\eeq
where the mode functions $U_{1,2}$ are introduced in Appendix~\ref{sec:appB}, and the angular frequency is given by
\begin{equation}
    \omega_p^2 \;\equiv\; p^2 + (am_{\psi})^2 \, .
\end{equation}

The last term in the previous expression, independent of the field values, corresponds to the adiabatic regulator for the integral. It guarantees the convergence of the integral, and is equivalent to the normal-ordering of the Hamiltonian operator~\cite{Greene:1998nh,Giudice:1999fb,Greene:2000ew,Peloso:2000hy,Nilles:2001fg}. With this regularization, $\rho_{\psi}$ vanishes in the absence of particle production. This can be more explicitly shown by writing the energy density as follows,
\beq\label{eq:rhopsi}
\rho_{\psi} \;=\; \frac{1}{(2\pi)^3 a^4}\int d^3\bp \,\omega_p n_p\,,
\eeq
where the number density per comoving mode can be written as
\beq
n_p \;=\; \frac{1}{2} \left| \left(1+ \frac{am_{\psi}}{\omega_p}\right)^{1/2}U_2 - \left(1-\frac{am_{\psi}}{\omega_p}\right)^{1/2}U_1 \right|^2\,.
\label{Eq:npsibis}
\eeq
In this definition for $n_p$ we have included the multiplicity due to the internal degrees of freedom of $\psi$. Albeit an unconventional choice, it will simplify our discussion of particle production in the next section.

\begin{figure}[!t]
\centering
    \includegraphics[width=0.785\textwidth]{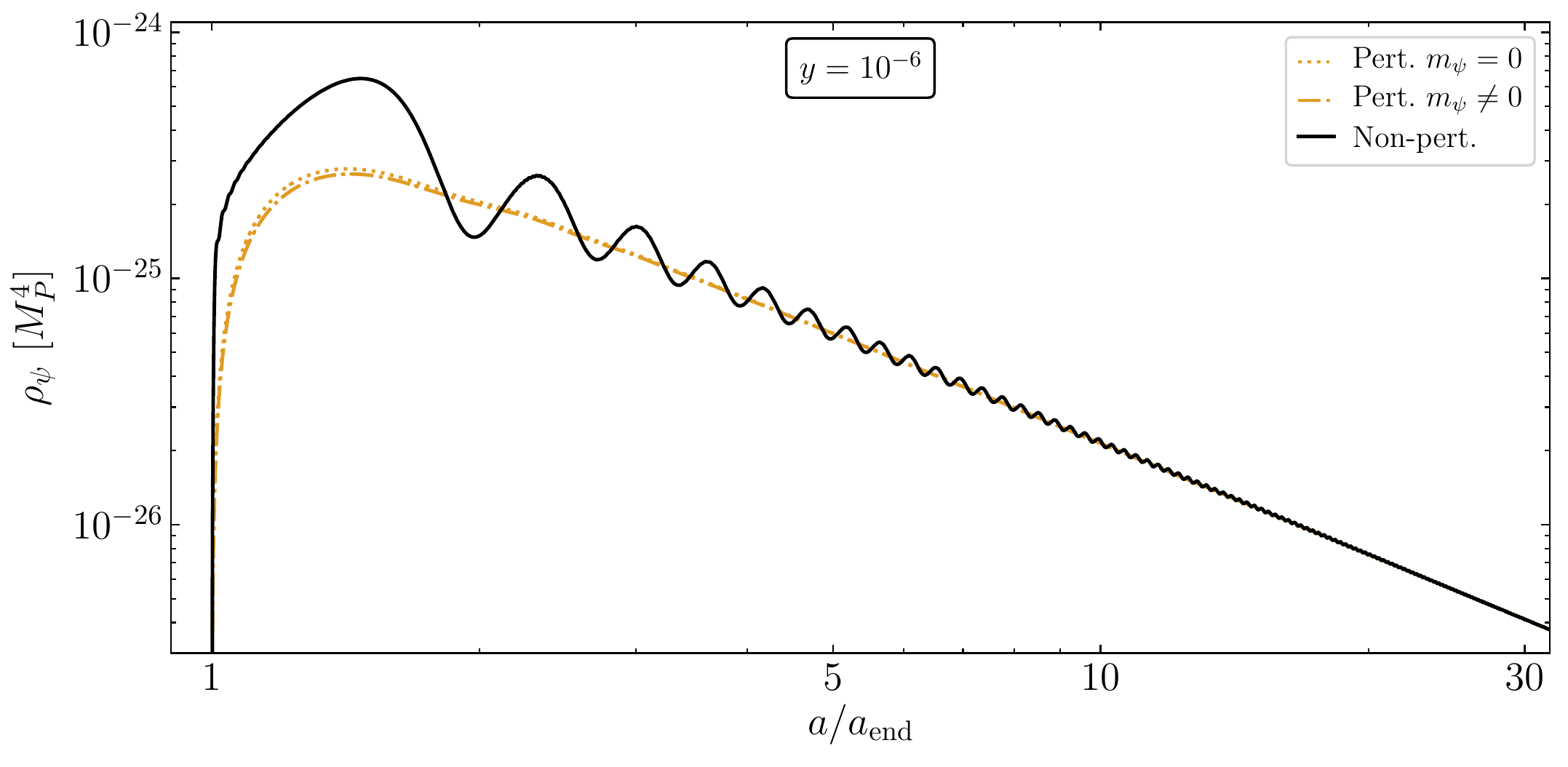}
    \includegraphics[width=0.785\textwidth]{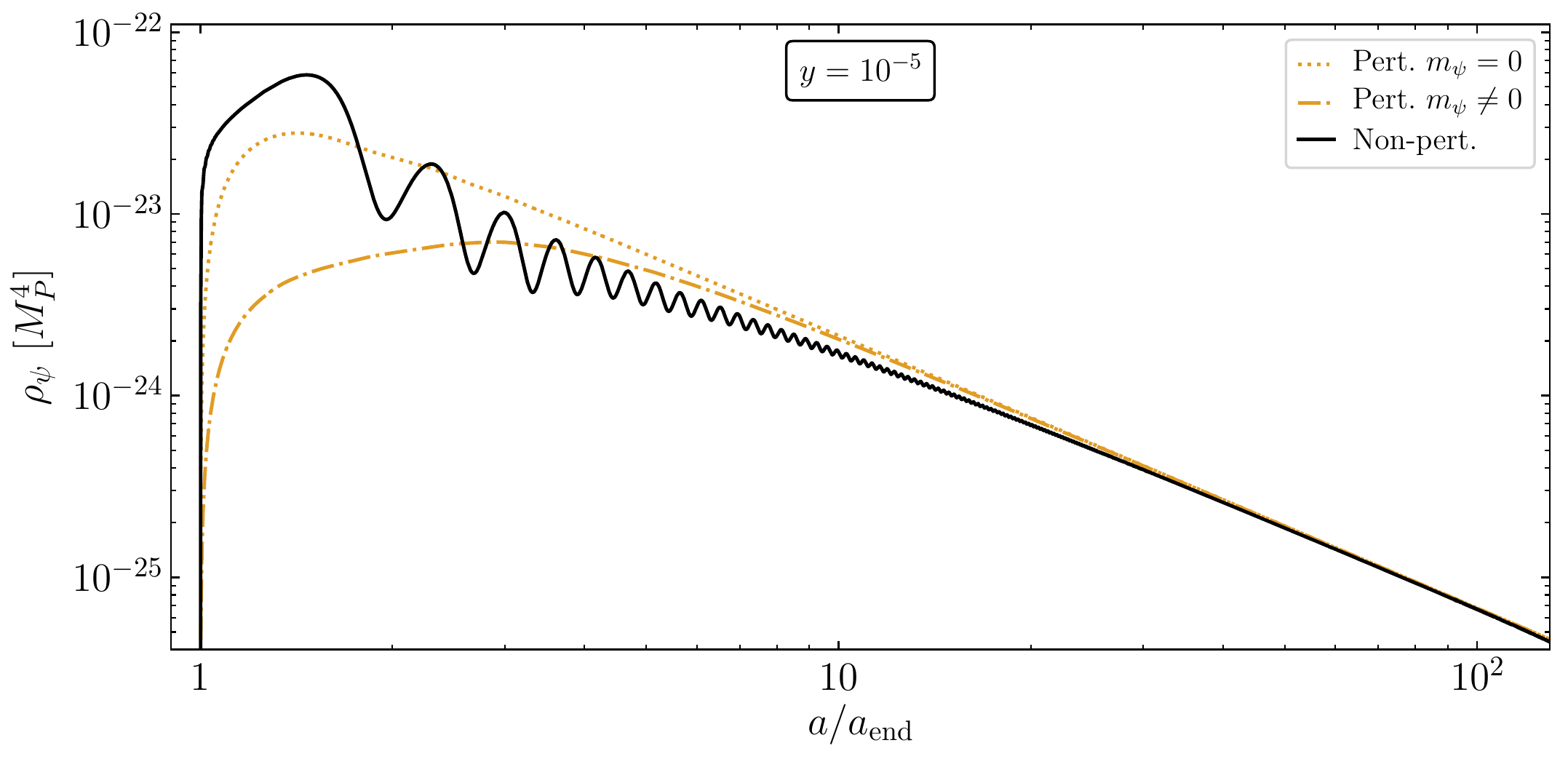}
    \includegraphics[width=0.785\textwidth]{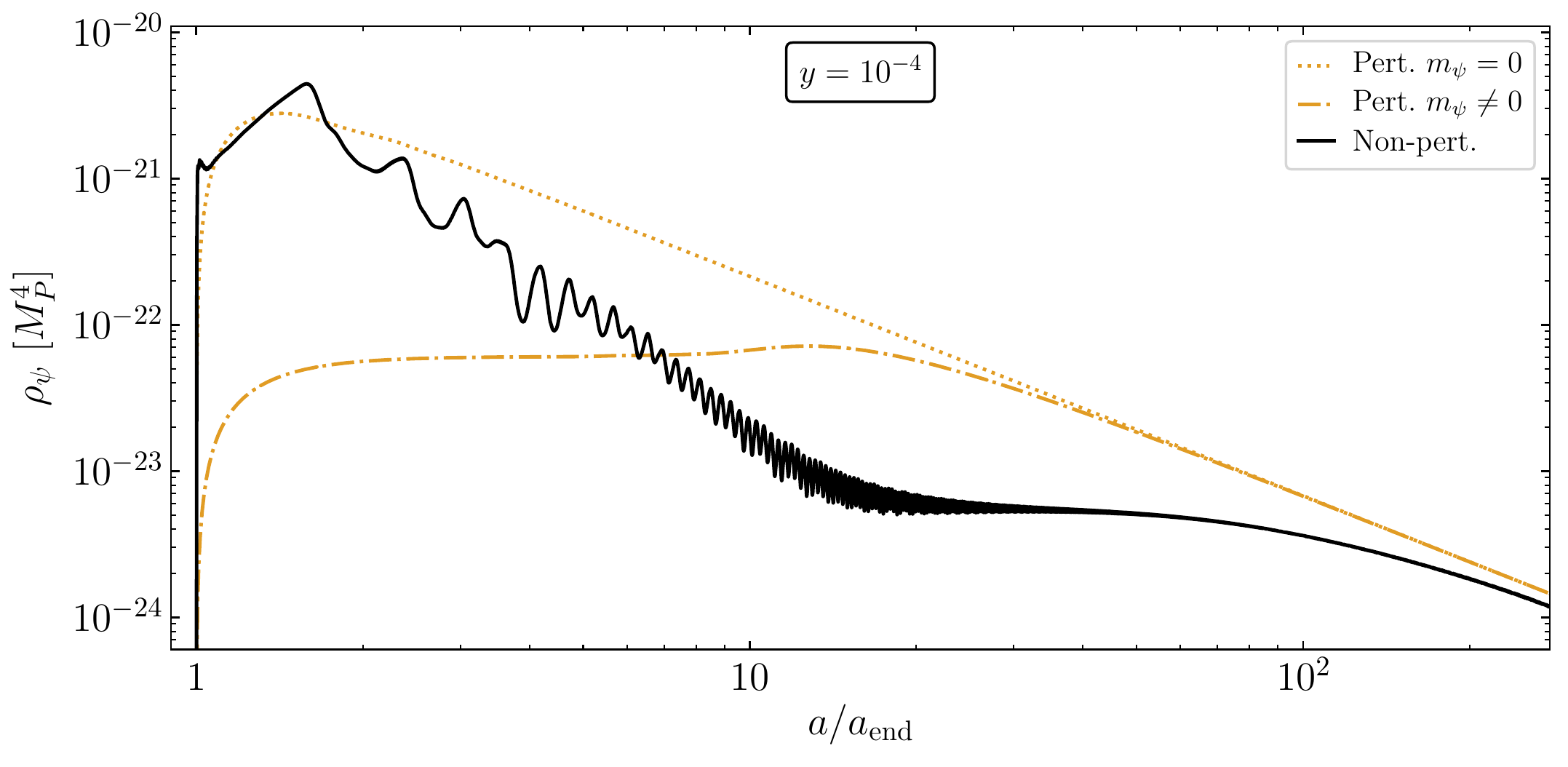}
    \caption{\it Instantaneous energy density in relativistic fermions during reheating sourced via the coupling $y\phi\bar{\psi}\psi$. Shown is the energy density as a function of the scale factor determined perturbatively ignoring the $\phi$-induced mass of $\psi$ (dotted), accounting for this induced mass (dashed-dotted), and computed non-perturbatively (solid). }
    \label{fig:rhopsi}%\vspace{-20pt}
\end{figure}

We summarize our results in Fig.~\ref{fig:rhopsi}, 
which shows the evolution of the energy density of $\psi$ for different Yukawa couplings. In the panels, $\rho_{\psi}$ is shown as a function of the scale factor in the early stages of reheating. We have assumed that the bare mass of $\psi$ can be neglected, $m_{\psi,0}=0$. For comparison, we also show the corresponding energy densities that can be computed by solving the perturbative equations (\ref{eq:FBeqs1})-(\ref{eq:FBeqs3}) with $\rho_R$ associated with $\rho_\psi$. As in Figs.~\ref{fig:rhochi} and \ref{fig:rhochi2}, the ``naïve'' result, which ignores the instantaneous induced mass of $\psi$ due to the non-vanishing VEV of the inflaton, is shown as the dotted curves. The dashed curves correspond to the solution of the perturbative set, where the kinematic effect induced by the oscillation of $\phi$ is kept. The constant evolution of $\rho_\psi$ (in the form of a plateau) for large values of Yukawa coupling $y$, can be easily understood from the expression (\ref{eq:kincond}) and remembering that $\rho_\psi$ evolves na\"ively according to $\rho_\psi\propto a^{-3/2}$ (see e.g., Eq.~(\ref{eq:rho})). Taking into account the kinematic suppression
\beq
{\cal R}^{-1/2}= \frac{\sqrt{\lambda}}{\sqrt{2}y}\frac{M_P}{\phi_{\rm end}}
\left(\frac{a}{a_{\rm end}}\right)^{3/2} \, ,
\eeq
one obtains a $\rho_\psi$ independent of the scale factor $a$.
For a small value of $y$, $\mathcal{R} < 1$ and there is no suppression as one can see by the overlapping dotted and dashed curves in the upper panel of Fig.~\ref{fig:rhopsi}. \\

The effect of Fermi-Dirac statistics is steeper when one performs the non-perturbative analysis. At the beginning of reheating the suppression plays no role, and $\rho_{\psi, {\rm max}}$ is always within $\mathcal{O}(1)$ of the dotted curve showing the na\"ive perturbative result in Fig.~\ref{fig:rhopsi}. This can be verified analytically upon integration of (\ref{numbdens}), valid for the first oscillation also in the fermionic case. For $y<10^{-5}$ the non-perturbative result on average reproduces the perturbative one, although the instantaneous bursts of particle production can leave an impact on dark matter production rates, as we discuss in Section~\ref{sec:DM}. Occupation numbers never saturate the Pauli limit of $n_p=2$ (see Fig.~\ref{fig:npS} in Appendix~\ref{sec:appB}). For $y\gtrsim 10^{-5}$, the decay of the inflaton into $\psi$ is blocked by the saturation of the particle momentum modes (see Fig.~\ref{fig:npL} in Appendix~\ref{sec:appB}). The production of $\psi$ occurs through the population of UV modes, which is less efficient than the perturbative estimate, and the result is a rapidly decreasing $\rho_{\psi}$. This continues until redshift can efficiently counter the Pauli suppression, and a plateau in $\rho_{\psi}$ appears. Eventually the non-perturbative computation asymptotes to the perturbative results  (for the couplings herein explored) implying that the perturbative estimate for the reheating temperature is unchanged~\cite{Garcia:2020eof},
\beq\label{eq:trhfer}
T_{\rm RH} \;\simeq\; \left(\frac{9 \lambda}{20 \pi^4 g_{\rm RH}}\right)^{1/4}y M_P\,.
\eeq 
For even larger couplings, the Pauli suppression and the need to track UV modes makes the numerical analysis difficult, although the expectation is a reduction in $T_{\rm RH}$ relative to the estimate (\ref{eq:trhfer}).

%%%%%%%%%%%%%%%%%%%%%%%%%%%%%%%%%%%%%%%%%
\section{Production of dark matter}
\label{sec:DM}
%%%%%%%%%%%%%%%%%%%%%%%%%%%%%%%%%%%%%%%%%

%%%%%%%%%%%%%%%%%%%%%%%%%%%%%%%%%%%%%%%%%
\subsection{Production rates and particle distributions}
\label{sec:rates}
%%%%%%%%%%%%%%%%%%%%%%%%%%%%%%%%%%%%%%%%%

The decay products of the inflaton may correspond to the initial states necessary for the out-of-equilibrium production of dark relics. If we specialize to the case of dark matter (DM) production via scatterings, then, under the freeze-in approximation, the number density of DM will be determined by the solution of the Boltzmann equation
\beq
\frac{dn_{\rm DM}}{dt} + 3Hn_{\rm DM} \;=\; R(t)\,,
\label{BE}
\eeq
where
%\beq
%R(t) \;\equiv\; 4\int \frac{d^3 \bP_1}{(2\pi)^3 2P_1^0} %\frac{d^3 \bP_2}{(2\pi)^3 2P_2^0}\sqrt{(P_1\cdot P_2)^2 - %m_a^2m_b^2}\,\sigma(s) f_a(P_1)f_b(P_2)\,,
%\eeq
\beq \label{eq:Rdef}
R(t) \;\equiv \; \int \frac{d^3 \bP_1}{(2\pi)^3 2E_1} \frac{d^3 \bP_2}{(2\pi)^3 2E_2} f_1(P_1)f_2(P_2)  \frac{|\mathcal{M}|^2 d\Omega_{13}}{32 \pi^2} \,,
\eeq
denotes the particle production rate for the process $1+2\rightarrow {\rm DM}+3$.\footnote{We denote ${\bf P}= \frac{\bf p}{a}$ as the {\it physical} momentum.}
We can simplify this expression by neglecting 
the initial and final state particles 
\beq
R(t) \simeq \frac{1}{1024 \pi^6} \int f_1 ~f_2~ E_1dE_1 E_2 dE_2 d\cos \theta_{12}
\int |{\cal M}|^2 d\Omega_{13} \, ,
\eeq
though we use the full expression in all our non-perturbative calculations.
If the scattering is due to
light decay products, $i.e$ particles that do not couple directly to the inflaton, 
 (as in the case of the decaying boson discussed in Section
\ref{sec:unstablechi}), $m_{1,2} \ll m_{\phi}$, the dependence on thresholds in the rate will not be important as reheating proceeds.
Let us assume for simplicity that the parent particles are equal, $1=2$, and that the mean, unpolarized squared scattering amplitude can be parametrized in the following way,
\beq
|\mathcal{M}|^2 \;=\; 16\pi \frac{s^{\frac{n}{2}+1}}{\Lambda^{n+2}}\,,
\label{matelem}
\eeq
where $s$ stands for the Mandelstam variable. This generic form of the amplitude is chosen so that, for massless scatterers, the integrated cross section is $\sigma = s^{\frac{n}{2}}/\Lambda^{n+2}$, i.e.~it scales with the $n$-th power of the center-of-mass energy.  In the absence of infrared divergences, for a different combination of $s$, $t$, $u$, our results will generically only differ by numerical factors, which can be absorbed into the value of $\Lambda$. 
In the perturbative case, well after the beginning of reheating, the rate can be approximated by 
%\beq\label{eq:RNPapprox}
%R \;\simeq\; \frac{2^{n+4}}{(2\pi)^4(n+4)\Lambda^{n+2}} \left( \int dE f E^{\frac{n}{2}+2}\right)^2\,.
%\eeq\par\bigskip
\beq\label{eq:RNPapprox}
R(t) \;\simeq\; \frac{2^{n}}{\pi^4 (n+4)\Lambda^{n+2}} \left( \int dE f E^{\frac{n}{2}+2}\right)^2\,.
\eeq
For thermalized parent particles, the distribution function is simply given by
\beq
f_{{\rm eq}} \;=\; \frac{1}{e^{E/T} \pm 1}\,,
\label{Eq:thermaldistribution}
\eeq
where $T$ is the instantaneous temperature, and we have assumed that $\langle E^2\rangle \sim T^2\gg m_i^2$. The rate can then be analytically evaluated \cite{Garcia:2017tuj,Ballesteros:2020adh},
%\beq\label{eq:rthermal}
%R(T) \;=\; \frac{2^{n+4}\Gamma(\frac{n}{2}+3)^2 \zeta(\frac{n}{2}+3)^2 T^{n+6}}{(2\pi)^4(n+4)\Lambda^{n+2}}\times \begin{cases}
%1\,, \quad & 1,2=\chi\,,\\
%\left(1-2^{-(\frac{n}{2}+2)}\right)^2\,, \quad & %1,2=\psi\,.
%\end{cases}
%\eeq
\beq\label{eq:rthermal}
R(T) \;\simeq\; \frac{2^{n}\Gamma(\frac{n}{2}+3)^2 \zeta(\frac{n}{2}+3)^2 T^{n+6}}{\pi^4(n+4)\Lambda^{n+2}}\times \begin{cases}
1\,, \quad & 1,2=\chi\,,\\
\left(1-2^{-(\frac{n}{2}+2)}\right)^2\,, \quad & 1,2=\psi\,.
\end{cases}
\eeq

On the other hand, for not-yet-thermalized parent scatterers, 
assuming $f_a\ll 1$, 
we can also find an analytical approximation for $R$ upon determining the non-thermal distribution functions, $f(E,T)$ \cite{Garcia:2018wtq,Ballesteros:2020adh}.\footnote{Note that the assumption $f_a\ll 1$ is violated when $\mathcal{R}>1$. This occurs for large $\sigma/\lambda$ and initially in the fermionic case. 
 For a related discussion in the scalar case see~\cite{Moroi:2020has,Moroi:2020bkq}.}
At a given time $(t_1)$, these solutions (for both $\chi$ and $\psi$) take the form $f(E) \propto \rho_\phi \Gamma_\phi/H m_\phi^4$. Then using $m_\phi/E \propto a(t)/a(t_1)$, we find that
\begin{align}
f_{\chi}(E,t) \;&\simeq\; \frac{\pi \sigma^2 \rho_{\phi}^2(t) }{8 m_{\phi}^7 H(t) } \left(\frac{m_{\phi}}{E}\right)^{9/2} \theta\left(E- m_{\phi}\left(\frac{a_{\rm end}}{a}\right)\right)\, \theta(m_{\phi}-E)
\,, \label{Eq:nonthermaldistributionchi}\\
f_{\psi}(E,t) \;&\simeq \; \frac{2\pi y^2 \rho_{\phi}(t)}{m_{\phi}^3 H(t)}\left(\frac{m_{\phi}}{2E}\right)^{3/2} \theta(m_{\phi}/2-E)
\,.
\label{Eq:nonthermaldistributionpsi}
\end{align}
For the scalar case, the higher power of $m_\phi/E$
comes from the fact that the distribution is proportional to $\rho_\phi^2/H \sim \rho_\phi^{3/2} \sim (a(t)/a(t_1))^{9/2}\sim (m_\phi/E)^{9/2}$. Note the appearance of a low energy cut-off $E = m_\phi (a_{\rm end}/a)$ since particles produced with energy $m_\phi$ have only a finite time to redshift to lower energy. We also note that the two distribution functions $f_{\chi}$ and $f_{\psi}$ behave quite differently. The density of scalar particles $dn_\chi=d^3P_\chi \tilde f_\chi$ is peaked at low energies, while in the case of fermions, $dn_\psi$ is more peaked towards high energies ($E_\psi \simeq m_\phi / 2$).
This comes directly from the additional redshift in the case of scattering $\phi \phi \rightarrow \chi \chi$ which is proportional to $\rho_\phi^2$ so that the final state scalars are quickly redshifted low energy particles which dominate the distribution.
This is not the case for the decay $\phi \rightarrow \bar{\psi} \psi$ where the more energetic particles are always produced at $E_\psi=m_\phi/2$ over-compensating for the redshifting fermion spectrum. As we will see, this behavior will impact the production of dark matter.
Using these distribution functions, we can again integrate Eq.~(\ref{eq:RNPapprox}). We find
\beq
\label{eq:rchiP}
R(t) \simeq \frac{m_{\phi}^{n-2}}{(n+4)\Lambda^{n+2}} \left(\frac{\rho_{\phi}\Gamma_{\phi}}{H}\right)^2 \begin{cases} \dfrac{2^{n+2}}{(n-3)^2} \left[1 - \left(\dfrac{a_{\rm end}}{a} \right)^{\frac{n-3}{2}} \right]^2 \,, n \ne 3 \quad & 1,2=\chi\,, \\[10pt]
2^n \left[ \log (a/a_{\rm end}) \right]^2 \, , n = 3 \qquad & 1,2=\chi\,, \\[10pt]
\dfrac{16}{(n+3)^2} \,, \quad & 1,2=\psi \,.
\end{cases}
\eeq

Finally, to include the production during the preheating phase, we cannot rely on any analytical
treatment, and one needs to extract numerically the distribution function $f_a$ from the 
number density $n_a$ 
\beq
n_{1,2}(t) \;=\; \frac{1}{(2\pi)^3} \int d^3\bP\, f_{1,2}(P,t)\,,
\eeq
where in this definition the internal number of degrees of freedom is factored into $f_a$.
From Eqs.~(\ref{eq:nchi}) and (\ref{Eq:npsibis}) we can then extract
\beq\label{eq:fchiNP}
f_{\chi}(P,t) \;=\; \frac{1}{2\omega_p} \left|\omega_p X_p - iX'_p \right|^2_{p\rightarrow aP} \,,
\eeq
and
\beq\label{eq:fpsiNP} 
f_{\psi}(P,t)
=\; \frac{1}{2} \left| \left(1+\frac{m_{\psi}}{\omega_p}\right)^{1/2} U_2 - \left(1-\frac{m_{\psi}}{\omega_p}\right)^{1/2} U_1 \right|^2_{p\rightarrow aP}\,.
\eeq

\subsection{Dark matter production from inflaton scattering to bosons}
\label{sec:ndmb}

\begin{figure}[!ht]
\centering
    \includegraphics[width=\textwidth]{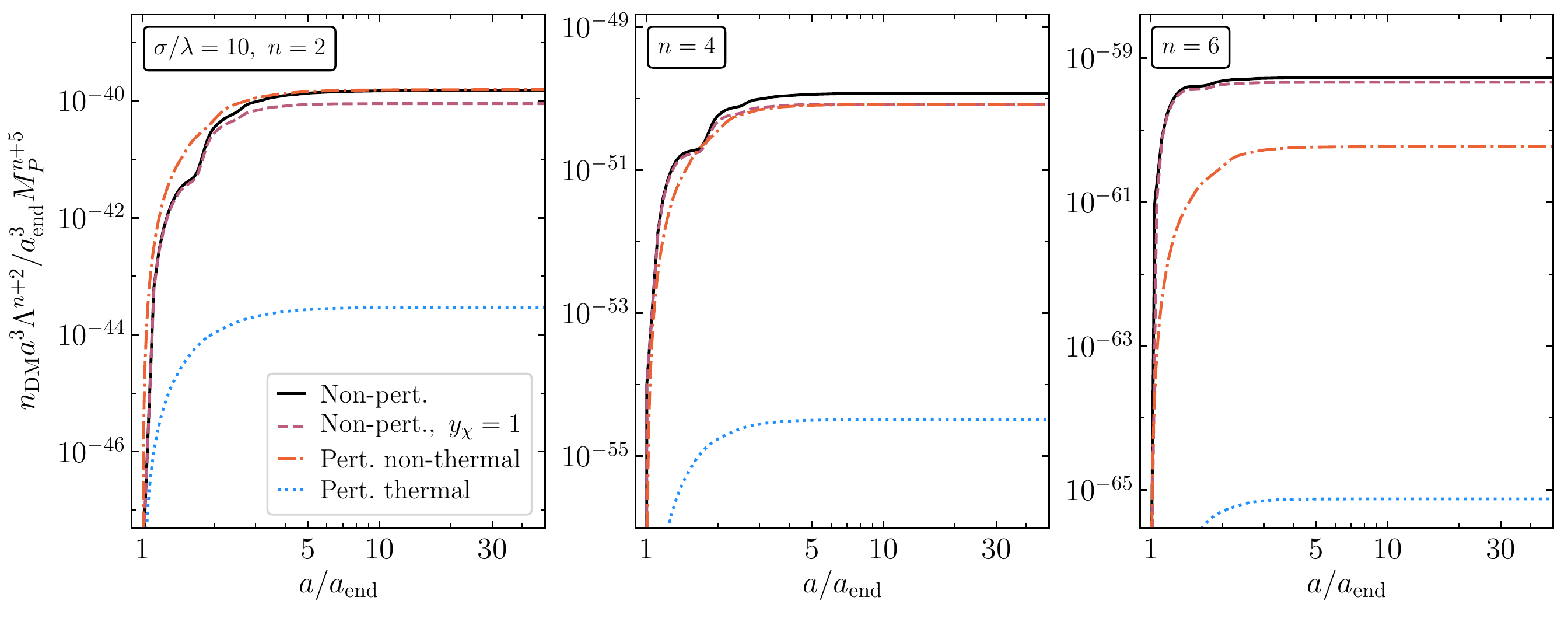}
    \includegraphics[width=\textwidth]{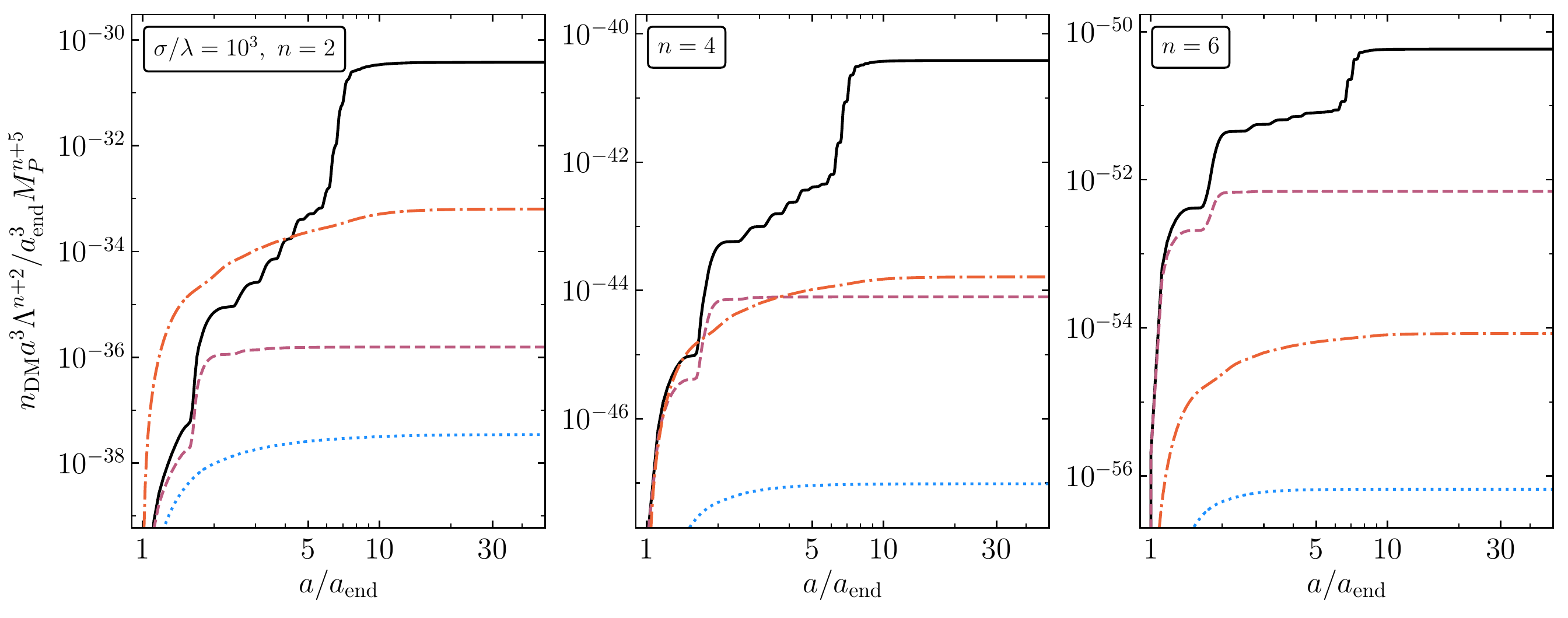}
    \includegraphics[width=\textwidth]{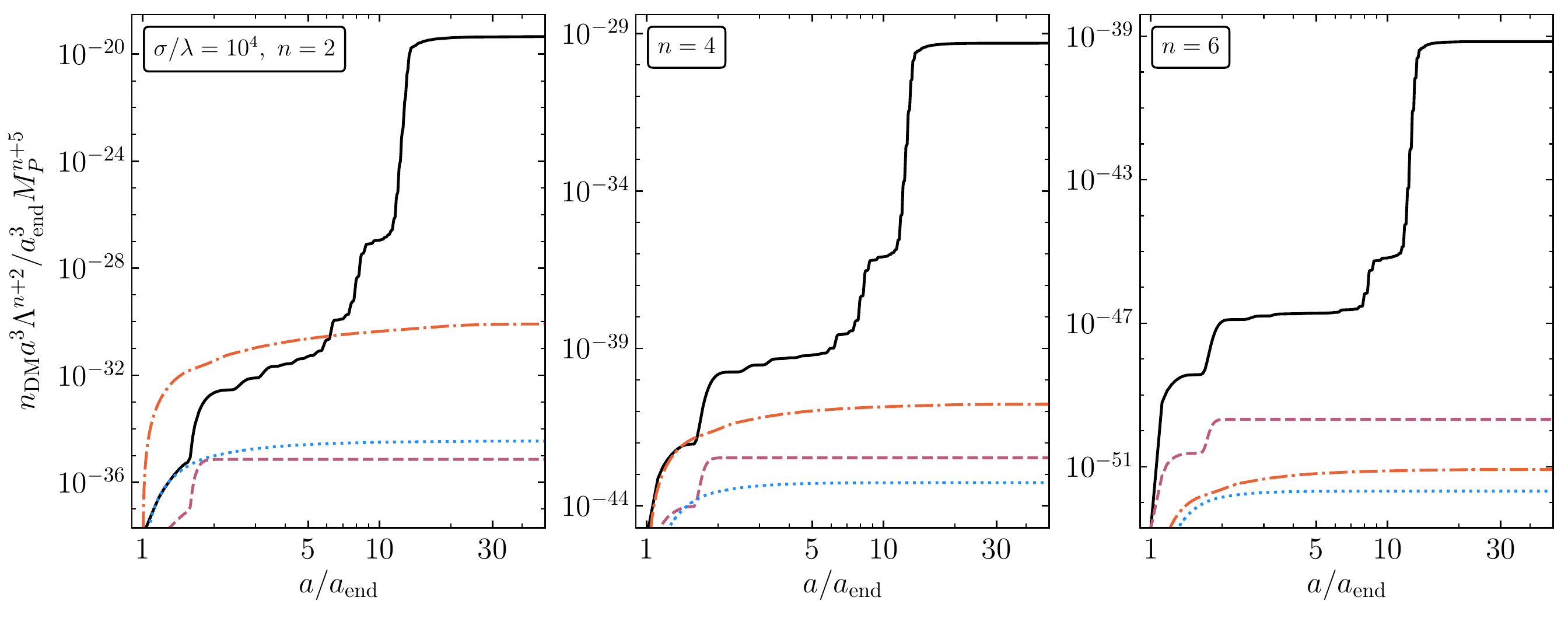}
\caption{\it Instantaneous (scaled) number density of dark matter for $\sigma/\lambda = 10$, $10^3$ and $10^4$, and $n = 2, 4$, 6 in the case of scalar preheating. As in previous figures, the solid black curve is based on the full non-perturbative rate (\ref{eq:Rdef}) to determine $n_{\rm DM}$, while the dashed purple curve corresponds to the case of unstable $\chi$ with $y_{\chi}=1$. The thermal (dotted blue) and non-thermal (dashed-dotted, red) distributions with  $\mathcal{R}$-suppressed perturbative results evaluated from (\ref{eq:rthermal}) and (\ref{eq:rchiP}) are also shown. }\vspace{-20pt}
    \label{fig:FIchiR}
\end{figure}

We show in Fig.~\ref{fig:FIchiR} the evolution of the
dark matter density as function of the scale factor 
when the dark matter is generated by the products of inflaton 
scattering $\phi \phi \rightarrow \chi \chi$ for different values of
$\sigma / \lambda$ (10, $10^3$ and $10^4$). That is, $\chi \chi \to {\rm DM} + \cdots$. Assuming the matrix element given in Eq.~(\ref{matelem}), we compute the number density of dark matter particles by solving the Boltzmann equation (\ref{BE}). 
The resulting relic density of dark matter can be inferred from Fig.~\ref{fig:FIchiR} using
\beq
\Omega_{\rm DM} h^2 = \frac{m_{\rm DM} n_{\rm DM}}{\rho_c} =
\left(\frac{n_{\rm DM}(a_0)}{8 \times 10^{-47}~{\rm GeV^3}}\right)
\frac{m_{\rm DM}}{1~\rm GeV}
=\left(\frac{n_{\rm DM}(a) \times ({a}/{a_0})^3}{8 \times 10^{-47}~{\rm 
GeV^3}}\right)
\frac{m_{\rm DM}}{1~\rm GeV}
\,,
\label{Eq:relic}
\eeq
where $a_0$ is the present scale factor ($a_0=1$). 
For an amplitude of the form in Eq.~(\ref{matelem}), we can
express the yield $n_{\rm DM}(a) a^3 \propto \frac{M_P^{n+5}}{\Lambda^{n+2}}$ such that Eq.~\ref{Eq:relic} becomes
\beq
\frac{\Omega_{\rm DM} h^2}{0.1} = \left[n_{\rm DM}(a) 
\frac{a^3}{a_{\rm end}^3}\frac{\Lambda^{n+2}}{M_P^{n+5}}\right] \left(\frac{10^{18}~\rm GeV}{\Lambda}\right)^{n+2}\times (2.4)^n \times 10^{103}~
 \frac{a_{\rm end}^3}{a_0^3} \left(\frac{m_{\rm DM}}{1~\rm GeV}\right).
\eeq
To evaluate $a_{\rm end}/a_0$, we make the simplification
that the energy density before and after reheating
scales as $a^{-3}$ and $a^{-4}$ respectively.
Then we can obtain $a_{\rm end}/a_0$ in terms of the energy density at reheating, $\rho_{\rm RH} \propto T_{\rm RH}^4$,
\beq
\frac{a_{\rm end}}{a_0}=\left(\frac{a_{\rm end}}{a_{\rm RH}}\right)
\left(\frac{a_{\rm RH}}{a_0}\right)= \left(\frac{\rho_{\rm RH}}{\rho_{\rm end}}\right)^{1/3} \left(\frac{\rho_0}{\rho_{\rm RH}}\right)^{1/4}
~\Rightarrow~ \frac{a_{\rm end}^3}{a_0^3}=\frac{\rho_{\rm RH}^{1/4} \rho_0^{3/4}}{\rho_{\rm end}} \simeq 2.3\times 10^{-91}\left(\frac{\rho_{\rm RH}}{10^{40}~{\rm GeV^4}}\right)^{1/4}
\,,
\eeq
where we took the present density of radiation as  $\rho_0=3.7 \times 10^{-51}~{\rm GeV^4}$ and $\rho_{\rm end}=2 \times 10^{-11}~M_P^4$. The relic density becomes
\beq \label{eq:omegadm}
\frac{\Omega_{\rm DM} h^2}{0.1} \simeq 2.3 \times 10^{12}\left[n_{\rm DM}(a) 
\frac{a^3}{a_{\rm end}^3}\frac{\Lambda^{n+2}}{M_P^{n+5}}\right] \left(\frac{10^{18}~\rm GeV}{\Lambda}\right)^{n+2}\times (2.4)^n 
\frac{\rho_{\rm RH}^{1/4}}{10^{10}~\rm GeV}
\left(\frac{m_{\rm DM}}{1~\rm GeV}\right).
\eeq
Comparing this result with Fig.~\ref{fig:FIchiR}, we see for example that the correct relic abundance can be obtained for a dark matter mass $m_{\rm DM} \sim 100$ EeV assuming $\Lambda=10^{14}$ GeV, $\sigma/\lambda=10$, $\rho_{\rm RH}^{1/4}=10^{10}$ GeV and $n=2$, using the late time value of $n_{\rm DM}(a/a_{\rm end})^3 \Lambda^4/M_P^7 \simeq 10^{-40}$. Alternatively, we obtain the correct relic density for
$m_{\rm DM} \sim 20$ eV with $\sigma/\lambda=10^4$ and $n=6$ in the non-perturbative case (using the solid-black curve).

In addition to the absolute value of the density, we see the effect of the different types of distributions 
discussed in the previous subsection. 
The blue dotted and red dot-dashed curves both use
the perturbative approach with $m_\chi \ne 0$, however
they are distinguished by the choice of distribution functions. For the blue dotted curves, we take the thermal distribution given by Eq.~(\ref{Eq:thermaldistribution}), and for the red dot-dashed curve we use the non-thermal distribution given in Eq.~(\ref{Eq:nonthermaldistributionchi}). 
In the most na\"ive case where the products of the inflaton scattering are instantaneously thermalized (blue dotted curve), the production rate of dark matter is the lowest (see also \cite{Garcia:2020hyo}).
When we assume the non-thermal distribution in Eq.~(\ref{Eq:nonthermaldistributionchi}), 
we observe an enhancement of  $\sim 5$
orders of magnitude.
This difference arises due to the difference in the typical energy of the decay products, $\langle E\rangle_{\rm therm}\sim T$ vs.~$\langle E\rangle_{\rm non\mbox{-}therm}\sim m_{\phi}$. In terms of the rates, 
Eqs.~(\ref{eq:rthermal}) and (\ref{eq:rchiP}), $R_{\rm therm}\sim\frac{T^{n+6}}{\Lambda^{n+2}} $, whereas the non-thermal rate is $R_{\rm non\mbox{-}therm} \sim \sigma^4 \frac{M_P^{6+n}}{\Lambda^{n+2}}10^{6-5n}$ when we take $m_\phi \sim 10^{-5} M_P$ and $\rho_\phi \sim \rho_{\rm end}$.  
If we take $n=6$, the ratio of the non-thermal to thermal rate is roughly $10^{-24} \sigma^4 (M_P/T)^{12}$ and $T\sim T_{\rm max}$ can be estimated from Fig.~\ref{boseback}. The enhancement in the rate is what 
largely contributes to the enhanced production of dark matter seen in Fig.~\ref{fig:FIchiR}.

The non-perturbative result using the distribution in Eq.~(\ref{eq:fchiNP}), is shown by the black curve in Fig.~\ref{fig:FIchiR}. 
For low values of $\sigma / \lambda$,  as we noted in the description of Fig.~\ref{fig:rhochi}, the appearance of 
the radiative bath in the non-perturbative case is delayed relative to that in the perturbative result due to 
the time it takes to pass through the first unstable region. The non-perturbative and perturbative non-thermal dark matter production rates are then comparable up to one order of magnitude. 
For larger values of $\sigma$, we find that the density based on the thermal distribution is suppressed by up to
 13 orders of magnitude relative to the non-perturbative case. 
For example, for $\sigma/\lambda=10^4$ and $n=6$,
with the same value of $\Lambda=10^{14}$ GeV and $\rho_{\rm RH}^{1/4}=10^{10}$ GeV as
discussed previously, 
a 200 TeV 
dark matter candidate would be necessary to fulfill the relic abundance constraint with the thermal hypothesis.
We also note that for larger values $\sigma / \lambda \gtrsim 10^4$, we see a clear jump in the dark matter production in Fig.~\ref{fig:FIchiR} (bottom) for any value of $n$, 
corresponding to the passage through a resonant band at $a/a_{\rm end}\sim 10$. This increases the amount of dark matter by more than 10 orders of magnitude compared to the non-thermal distribution, and 15 orders of magnitude larger than if
one would have considered thermal distribution (for $n=2$).

Shown as the purple dashed curve in all panels of Fig.~\ref{fig:FIchiR} is the re-scaled dark matter number density in the case of unstable $\chi$ with $y_{\chi}=1$, corresponding to the case treated in Section~\ref{sec:unstablechi}. 
Here the contribution to $n_{\rm DM}$ could come from two sources: the $\chi$ particles resonantly produced by $\phi$, or from the plasma of secondary decay products $\rho_f$. Under the assumption that these secondary products quickly thermalize and can produce dark matter with the same amplitude as $\chi$, the more energetic resonantly produced scalars always dominate the production. At small $\sigma$, the difference with respect to the stable $\chi$ scenario is small, at most $\sim 40\%$ in the case when $\sigma/\lambda=10$.
In this case, the bulk of the (stable) non-perturbative production of dark matter occurs early on (there is little delay) and
the effect of thermalization is unimportant as even in the unstable case, the bulk of the particle production occurs early on.
As $\sigma/\lambda$ is increased, as we saw in Figs.~\ref{fig:rhochi} and \ref{fig:rhochi2}, there is a delay in the production of the scalars $\chi$, and hence in the production of dark matter as seen by the solid black curves in Fig.~\ref{fig:FIchiR}. If $\chi$ is unstable, and rapidly thermalizes, the production of dark matter effectively occurs only in the first few oscillations before thermalization is complete. This is seen
by noticing that the purple dashed curve initially tracks the black solid curve and then hits a plateau when thermalization occurs. 
For larger sigma, the decay of $\chi$ is more efficient, and the dark matter relic abundance is suppressed by up to 16 orders of magnitude, for $\sigma/\lambda=10^4$, $n=2$. This corresponds roughly to the 16 orders of magnitude suppression in $\rho_\chi^2$ when 
$y_\chi = 1$ relative to the non-perturbative case with $y_\chi = 0$. In this case, 
the production of dark matter from unstable $\chi$'s is actually similar to that of the perturbative thermal case (blue dotted curves)
except for large $n$ where because of the 
energy dependence in the dark matter production cross-section, there is still some enhanced production as the purple curve again
initally tracks the black curve until thermalization sets in.

As already noted in~\cite{Garcia:2020wiy} in the context of thermal production, we see that in all cases dark matter effectively decouples from the relativistic plasma at the time-scale when $\rho_{\chi,{\rm max}}$ is reached. The subsequent reheating of the Universe due to the eventual decay of $\phi$ only dilutes this early population of dark matter, as shown in Eq.~(\ref{eq:omegadm}).

For small $\sigma$, the (non-thermal) Boltzmann picture adequately mimics the full preheating result, as seen in the first row of Fig.~\ref{fig:FIchiR}. The form of the distribution function (\ref{Eq:nonthermaldistributionchi}) indicates that for $n<3$ low momentum modes, corresponding to the earliest created particles, account for most of the creation of dark matter. For $n>3$, it is the high momentum modes, produced last, that saturate $\Omega_{\rm DM}$. Nevertheless, due to the redshifting number density of $\phi$, the dark relics from $\phi\phi\rightarrow \chi\chi$ are produced within the first few oscillations, i.e.~with minimal redshift, and the difference in qualitative behavior between the different values of $n$ is not significant. Importantly though, in the non-perturbative picture, the excitation of deep sub-horizon modes is suppressed but not forbidden, unlike the result in perturbation theory, Eq.~(\ref{Eq:nonthermaldistributionchi}). These UV modes, although sparsely populated, have a large impact in the dark matter production rate due to its steep energy dependence. Hence, the perturbative approximation is worse as the power $n$ in (\ref{matelem}) is increased. In the presence of broad parametric resonance, this difference only grows.

\subsection{Dark matter production from inflaton decay to fermions}
\label{sec:ndmf}

 \begin{figure}[!ht]
\centering
    \includegraphics[width=\textwidth]{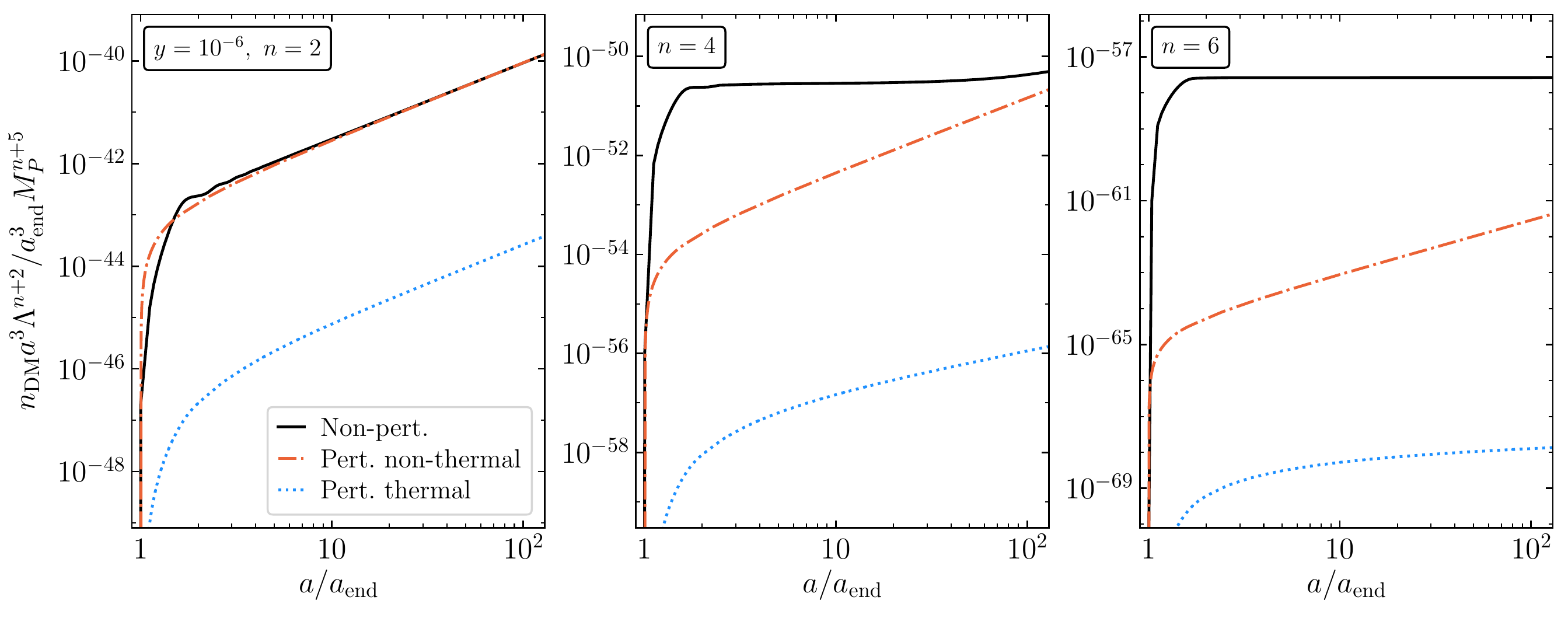}
    \includegraphics[width=\textwidth]{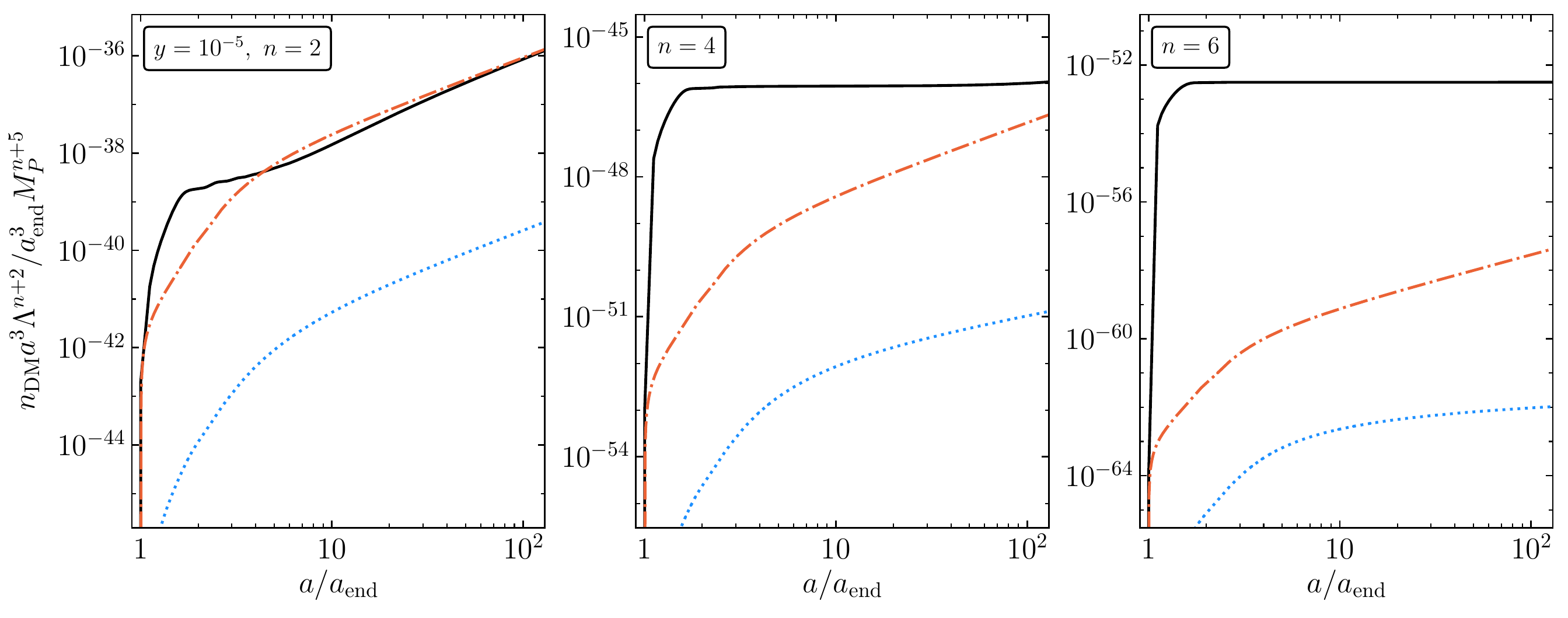}
    \includegraphics[width=\textwidth]{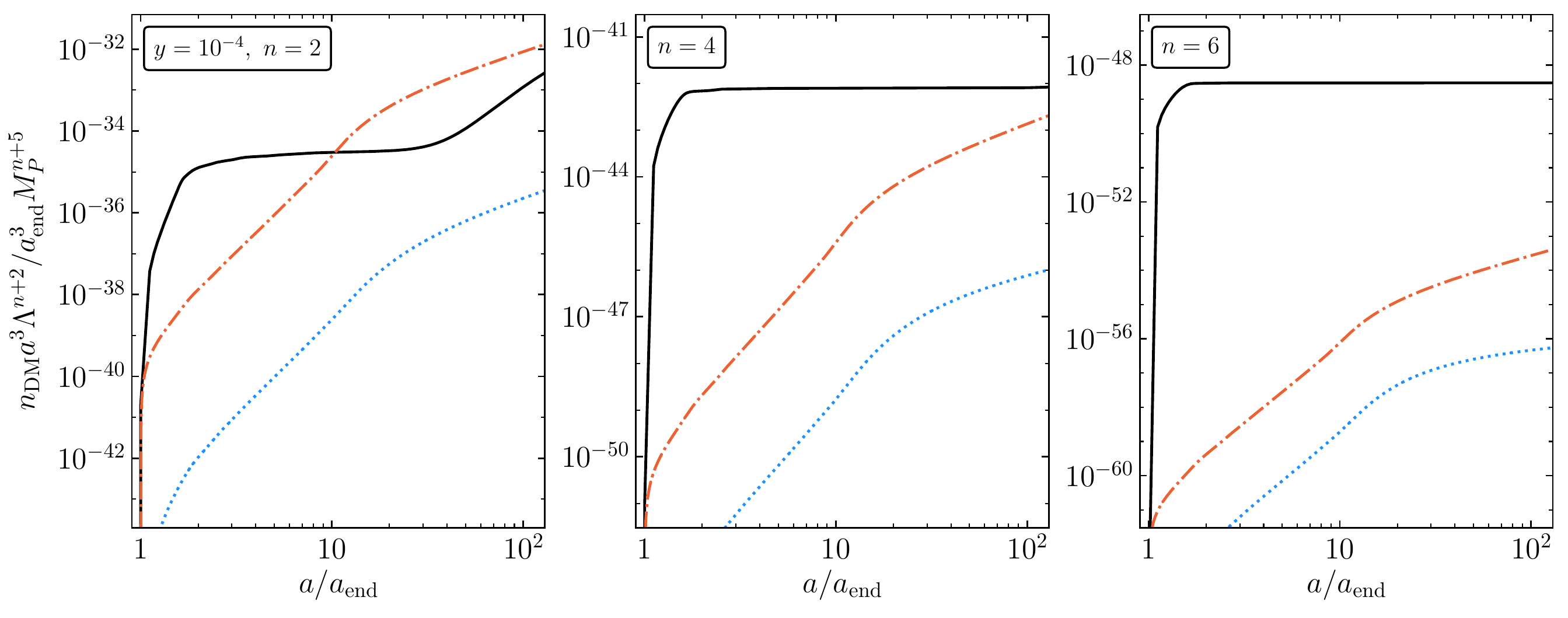}
\caption{\it Instantaneous number density of dark matter for a selection of couplings $y$ and cross-section powers $n$ in the case of fermion preheating. In solid black the full rate (\ref{eq:Rdef}) is used to determine $n_{\rm DM}$. The thermal (dotted blue) and non-thermal (dashed-dotted, red)  $\mathcal{R}$-suppressed perturbative results evaluated from (\ref{eq:rthermal}) and (\ref{eq:rchiP}) are also shown. }
    \label{fig:FIpsiR}%\vspace{-30pt}
\end{figure}

Fig.~\ref{fig:FIpsiR} compares the fermionic thermal perturbative approximation (\ref{eq:rthermal})  to the non-thermal rate  (\ref{eq:rchiP}) and to the non-perturbative rate, computed by numerical integration of (\ref{eq:RNPapprox}) using the distribution in Eq.~(\ref{eq:fpsiNP}). The main difference between these 3 hypotheses lies in the shape of their distribution function. Indeed, by comparing (\ref{Eq:thermaldistribution}) and (\ref{Eq:nonthermaldistributionpsi}), we clearly see that if we consider a non-thermal bath, the majority of the fermions resulting from the disintegration of the inflaton, will have energies of the order of $m_\phi/2$, much larger than the average energy of a thermal bath at temperature $T$. For any $n$, $n_{\rm DM}\sim a^{-3/2}$ for $\mathcal{R}<1$. One consequence is a larger production of dark matter when the production rate depends strongly on the energy, as is the case assuming the amplitude in Eq.~(\ref{matelem}).
Moreover, for larger $n$ the thermal rate decreases faster with $a$ (because $T^{n+6}$
is redshifted faster for larger values of $n$)
 and the density reaches a plateau (logarithmic in the case of $n=6$ and constant for $n>6$).  

In the non-perturbative case, for small values of the Yukawa coupling ($y=10^{-6}$) and $n=2$, the dark matter production is very similar to the perturbative non-thermal case (compare the red-dashed line to the black-full line in Fig.~\ref{fig:FIpsiR}). For larger $y$, the $\mathcal{R}$-suppression initially reduces the perturbative production of dark matter relative to the non-perturbative case. For the latter, the Pauli blocking of the population of momentum modes momentarily reduces the production of $\psi$ from inflaton decay, and $n_{\rm DM}$ plateaus correspondingly. When this blocking is overcome (and $\mathcal{R}<1$) the black and the red-dashed curves converge. Note that for $n=2$ a thermally sourced $n_{\rm DM}$ grows at the same rate as the non-thermal one. When thermalization is finally achieved,\footnote{Perturbatively, $a_{\rm thermalization}\approx 0.33 y^{-4/5} a_{\rm end}$ in the case of Standard Model interactions~\cite{Garcia:2018wtq}.} non-thermal production ceases and the growing thermal fraction of dark matter eventually dominates, saturating $\Omega_{\rm DM}$ at the end of reheating. Hence, for $y\lesssim 10^{-5}$ non-perturbative fermion production with $n\leq 2$ does not make a significant impact on the dark matter relic abundance. On the other hand, for $y>10^{-5}$, the Pauli delay could affect the thermalization rate of the relativistic plasma, and consequently the relative fraction of thermally and non-thermally produced dark matter. 

Non-perturbative effects become more important for larger values of $n$ ($>2$), even for small couplings $y$. In these cases, shown as the two rightmost columns in Fig.~\ref{fig:FIpsiR}, the initial burst of $\psi$ production results in an initial plateau behavior of $n_{\rm DM}\times a^3$. Indeed, in the top panel of Fig.~\ref{fig:rhopsi} we note that, although on average the perturbative and non-perturbative estimates agree, at the maximum of $\rho_{\psi}$ the non-perturbative result is the largest. The same occurs with an even larger difference for larger $y$. This first burst of particle production produces a population of dark matter that dominates over subsequent production bursts even after accounting for its redshift. Hence the plateau for $n=4$ and $n=6$. Nevertheless, at later times, the growing abundance of subsequently produced dark matter eventually overcomes this redshift, and the non-perturbative estimate converges to the perturbative one. Upon thermalization, this non-thermal population of dark matter freezes-in. For $n>2$, thermal production cannot catch up with the non-thermal population.

\section{Summary}
\label{sec:summary}

Two well studied mechanisms for producing dark matter are thermal freeze-out and freeze-in.
In the former, the dark matter candidate is
in thermal equilibrium with the radiation bath. As the temperature falls below the dark matter rest mass, annihilations (or something similar) keep the number density at the equilibrium value which is now dropping exponentially.  Its relic density is typically determined by its equilibrium density once these
interactions freeze-out. That is, when their interaction rate drops below the Hubble expansion rate.  In contrast, in freeze-in production mechanisms, the dark matter may 
never be in thermal equilibrium.
It may be produced thermally from the
existing radiation bath, but at a rate which
is below the expansion rate. Roughly, the abundance of dark matter in this case can be 
approximated by $n_{\rm DM}/n_\gamma \sim \Gamma/H$, where $\Gamma$ is the dark matter production rate.  Both mechanisms as described above rely on a thermal bath created by the reheating process after inflation.

Reheating and thermalization, however, are not instantaneous processes. As the Universe exits from its inflationary expansion, particle production begins as the inflaton begins is series of oscillations about its minimum. 
During these oscillations, inflaton scatterings and/or decays begin to populate the Universe with relativistic particles.
These particles will eventually thermalize and if produced in sufficient abundance will come to dominate the overall energy density and reheating will be complete. But both thermalization and dominance take time.
During this time, scattering may produce dark matter before reheating is complete, and the
abundance of dark matter produced prior to reheating may be the dominant mechanism for dark matter production.

The production of dark matter will obviously depend on both the energy density of the nascent bath of relativistic particles, and their energy distributions (thermal or otherwise). In this paper, we considered a series of assumptions about the state of the 
pre-thermal bath, including both perturbative and non-perturbative effects. To ultimately compute the dark matter relic density through freeze-in, we first concentrated on the production of the radiation bath. We considered both (separately) the production of scalars from inflaton scattering and the production of fermions from inflaton decay.

In case of scalar production, we first assumed
that the scalars are instantly thermalized
and neglected the effects of kinematic suppression 
due to their effective mass from the coupling to the inflaton. With this assumption, the energy density (and temperature) of the radiation peaks early on and falls off 
as $\rho_R \propto a^{-4}$. Including the suppression, we found a lower peak density 
which falls off as $\rho_R \propto a^{-3}$ until the effective mass is sufficiently small and subsequently $\rho_R \propto a^{-4}$.
Neither of computations included non-perturbative effects. When included, 
we found that while initially suppressed,
after a delay of several oscillation periods, the energy density grows through parametric resonance, particularly when the coupling
$\sigma/\lambda$ is large ($\gtrsim 10^3$).
For very large coupling ($\gtrsim 10^5$),
it is necessary to include the effects of backreaction. We also included the case where
the scalars produced from inflaton scattering are unstable. If the lifetime is sufficiently short, the effect of the parametric resonance is obviated and the density falls off as 
$ a^{-4}$ after reaching a quick peak close to the perturbative value (when effective masses are included). If the scalars are associated with the Higgs boson, then the rapid decay scenario is the most realistic. In all of the 
cases considered, reheating does not occur through inflaton scattering as the radiation density is ultimately sub-dominant unless the inflaton finally decays (to either scalars or fermions). We repeated these assumptions
for the production of fermions from inflaton decay (though we did not consider the decay of the fermions). 

In addition to the determination of the energy density, to compute the dark matter abundance,
we must specify the energy distribution functions of the particles in the relativistic bath. We considered three options for the initial distributions: a thermal distribution, Eq.~(\ref{Eq:thermaldistribution}), a non-thermal distribution, Eqs.~(\ref{Eq:nonthermaldistributionchi}) and (\ref{Eq:nonthermaldistributionpsi}), and the exact distribution derived from the non-perturbative approach, Eqs.~(\ref{eq:fchiNP}) and (\ref{eq:fpsiNP}). 
Using these distribution functions,
we computed the dark matter density from scalar and fermion scattering, assuming 
a scattering rate derived from Eq.~(\ref{matelem}). In the scalar case, 
we considered the thermal and non-thermal distributions for the perturbative approach
including the effective scalar mass. We also considered the non-perturbative calculation with and without scalar decay. Generally 
the thermal/perturbative calculation underestimated the dark matter production, sometimes by a very large factor. 
The non-perturbative calculation with stable scalars always produced the most dark matter.
When scalar decays were included in the non-perturbative calculation, the dark matter density initially tracked the stable case briefly (until $a/a_{\rm end} \sim 2$),  
when decays begin and significant dark matter production through the thermal bath of decays production became subdominant. Once again
we repeated this exercise for fermionic inflaton decay products.  

The final dark matter density depends on all of the choices made above. The most complete calculation we considered, includes
the effect of parametric resonance and the rapid decay of the inflaton scattering products. For small coupling ($\sigma/\lambda = 10)$, the effect of decays are minor and the relic density is greatly enhanced when non-thermal distributions are used.
As $\sigma/\lambda$ is increased, we found that decays can greatly suppress the relic density and furthermore, the final abundance becomes sensitive to the energy dependence of the dark matter production cross section.
Finally for $\sigma/\lambda = 10^4$, the final abundance is in fact similar to the abundance
obtained from the thermal perturbative approach. 

\begin{figure}[!ht]
\centering
    \includegraphics[width=\textwidth]{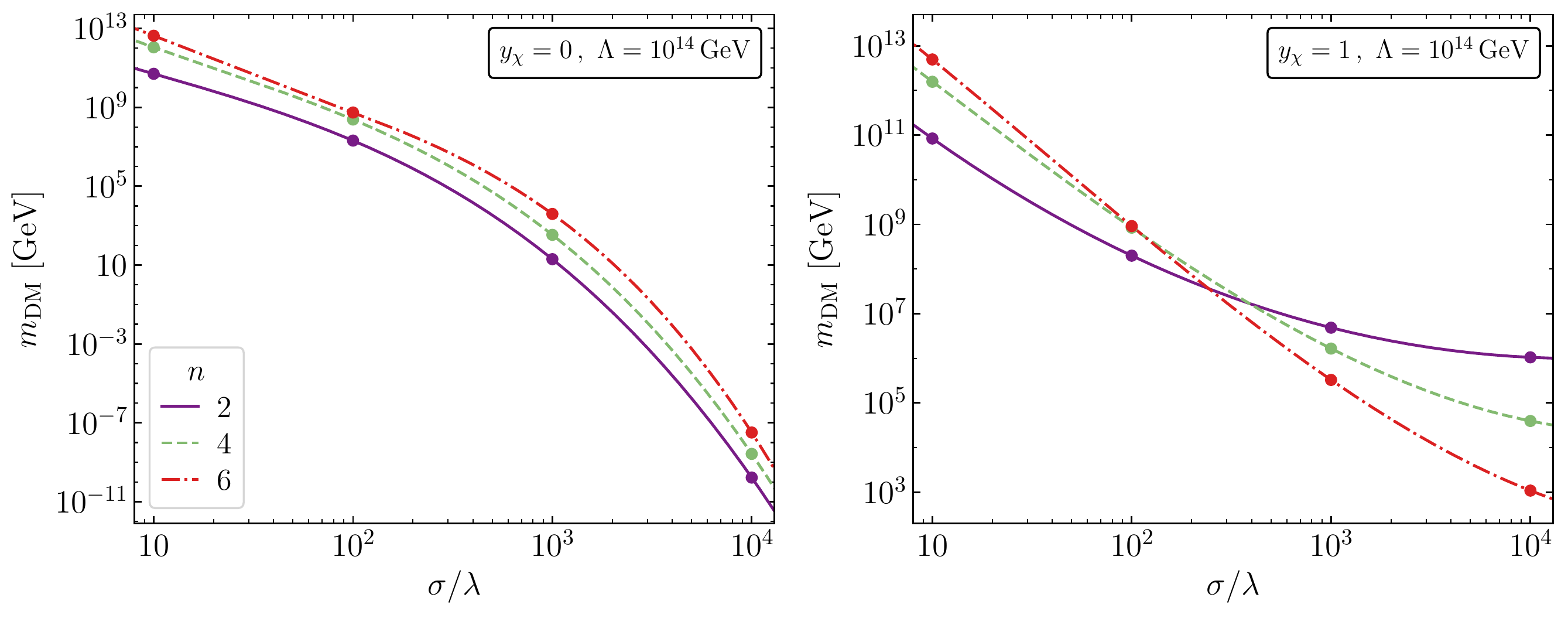}
    \caption{\it Dark matter mass necessary to saturate the measured relic abundance as a function of the coupling $\sigma$ and the power $n$ in (\ref{matelem}), for scalar preheating. Left: stable $\chi$. Right: unstable $\chi$ with $y_{\chi}=1$. 
    }
    \label{fig:masses}
\end{figure}

We have also seen that the correct present-day relic abundance can be obtained for a large range of dark matter masses. 
In Fig.~\ref{fig:masses}, we show the requisite dark matter mass to obtain $\Omega h^2 = 0.1$ in the non-perturbative calculation for both the stable ($y_\chi = 0$) and unstable ($y_\chi  = 1$) scalar products as a function of $\sigma/\lambda$. We use  Eq.~(\ref{eq:omegadm}) and the results for 
$n_{\rm DM}(a) 
\frac{a^3}{a_{\rm end}^3}\frac{\Lambda^{n+2}}{M_P^{n+5}}$ from Fig.~\ref{fig:FIchiR}. Results are plotted assuming $\Lambda = 10^{14}$ GeV and $\rho_{\rm RH}^{1/4}=10^{10}$ GeV, for $n = 2, 4$, and 6. 
For example, for the case of unstable scalar products, 
we find a mass range $m_{\rm DM} = 10^3 -  10^{13}$ GeV and an even wider range in the stable case.
Thus leaving a wide variety of potential dark matter candidates to be produced during the 
reheating process after inflation.

\section*{Acknowledgments}
The authors want to thank M.~Peloso for helpful discussions. This project has received support from the European Union’s Horizon 2020 research and innovation programme under the Marie Sk$\lslash$odowska-Curie grant agreement No 860881-HIDDeN, the CNRS PICS MicroDark and the IN2P3 master project ``UCMN''.
The work of M.G.~was supported by the research grant ``The Dark Universe: A Synergic Multi-messenger Approach'' number 2017X7X85K under the program PRIN 2017 funded by the Ministero dell’Istruzione, Universit\`{a} e della Ricerca (MIUR), and also by Istituto Nazionale di Fisica Nucleare (INFN) through the Theoretical Astroparticle Physics (TAsP) project. The work of M.G. was also supported by the Spanish Agencia Estatal de Investigaci\'{o}n through Grants No.~FPA2015-65929-P (MINECO/FEDER, UE) and No.~PGC2018095161-B-I00, IFT Centro de Excelencia Severo Ochoa Grant No.~SEV-2016-0597, and Red Consolider MultiDark Grant No.~FPA2017-90566-REDC, and by MCIU (Spain) through contract PGC2018-096646-A-I00. 
The work of K.A.O.~was supported in part by DOE grant DE-SC0011842  at the University of
Minnesota.
The work of K.K. was in part supported by JSPS Kakenhi Grant
No.~19H01899. Except where otherwise stated, all numerical results were obtained from a custom Fortran code utilizing the thread-safe arbitrary precision package MPFUN-For written by D.~H.~Bailey~\cite{fortran}.

%%%%%%%%%%%%%%%%%%%%%%%%%%%%%%%%%%%
%\newpage
\appendix
\section{Parametric resonance}
\label{sec:appA}
%%%%%%%%%%%%%%%%%%%%%%%%%%%%%%%%%%%

In this Appendix we discuss the narrow and broad parametric resonance when the inflaton oscillates about a quadratic minimum $V(\phi) \simeq \lambda \phi^2$. We note that here we use cosmic time and Planck units.

\begin{figure}[!t]
\centering
    \includegraphics[width=.785\textwidth]{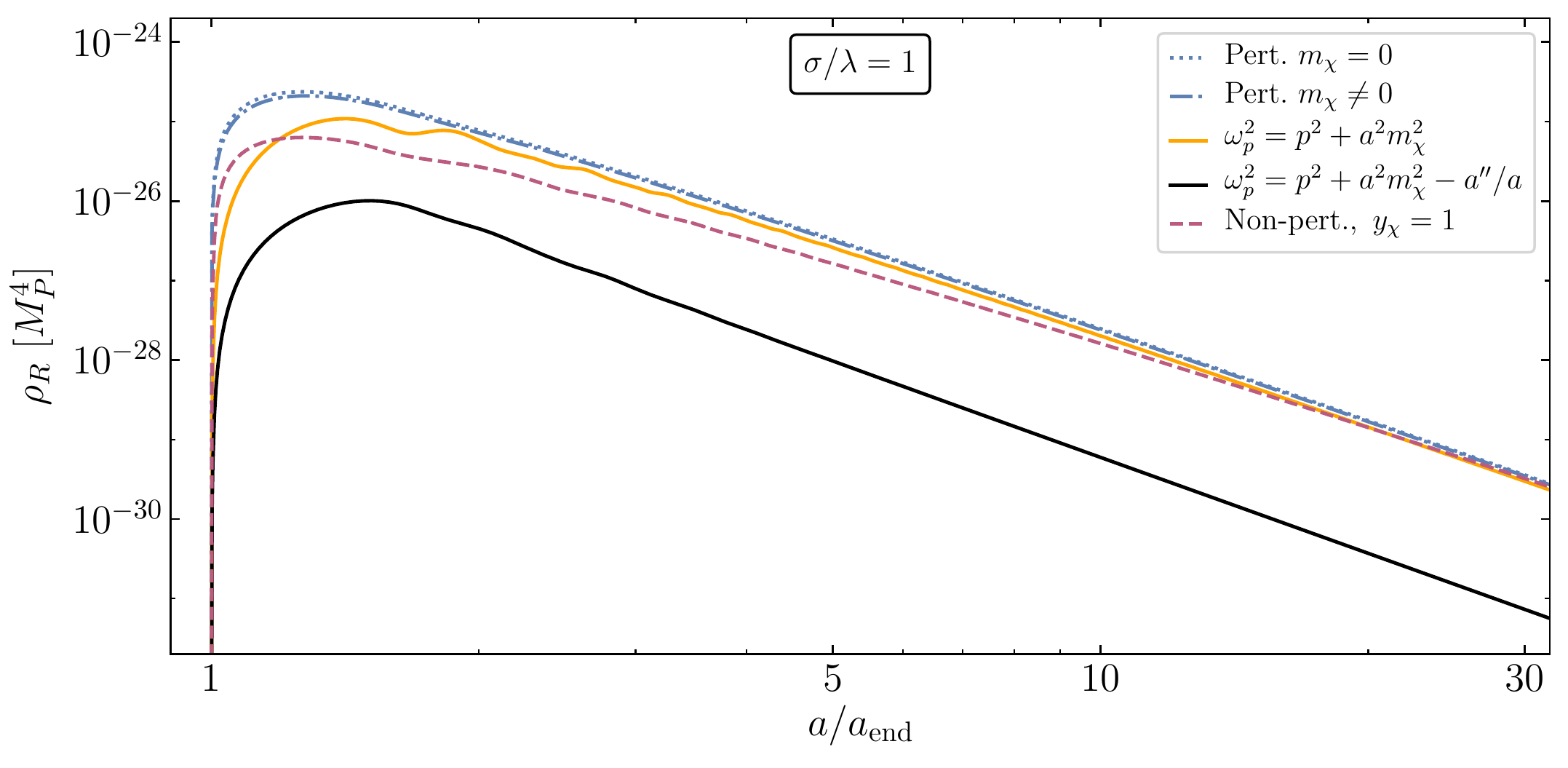}
    \caption{\it Energy density in relativistic bosons during reheating sourced via the coupling $\frac12 \sigma\phi^2\chi^2$ for $\sigma/\lambda = 1$. Shown in blue for $\rho_R=\rho_{\chi}$ are the perturbative estimates ignoring (dotted) or accounting (dashed-dotted) for the $phi$-induced mass of $\chi$. The non-perturbative results accounting (solid, black) or not (solid, orange) for expansion in the effective angular frequency are also shown. We additionally show the energy density $\rho_R=\rho_{\chi}+\rho_f$ when $\chi$ decays rapidly assuming the coupling $y_\chi = 1$ (dashed purple).
    }
    \label{fig:rhochi0}%\vspace{-20pt}
\end{figure}

The equation of motion of the inflaton field $\phi$ is given by
\begin{equation}
    \label{app:eom1}
    \ddot{\phi} + 3H \dot{\phi} + 2 \lambda \phi \; = \; 0 \, .
\end{equation}
After the end of inflation, the inflaton oscillates about a quadratic minimum and we enter a matter-dominated period. The solution to the equation of motion~(\ref{app:eom1}) is given by
\begin{equation}
    \label{app:solphi}
    \phi(t)  = \phi_0(t) \cdot \cos \left(m_{\phi} t \right) \, ,
\end{equation}
where $m_{\phi} =\sqrt{2 \lambda} $, $\phi_0(t) \simeq \frac{\phi_{\rm{end}}}{m_{\phi} t}$, and the scale factor averaged over a few oscillations is given by
\begin{equation}
    a(t) \simeq a_{\rm{end}} \left(\frac{t}{t_{\rm{end}}} \right)^{2/3} \, .
\end{equation}
Using this solution together with the field redefinition, $x_p(t) = a^{3/2}(t) \chi_p(t)$, we can rewrite the equation of motion~(\ref{boson:eom1}) in the Fourier space,
\begin{equation}
    \ddot{x}_p + \omega_p^2 x_p \; = \; 0 \, ,
\end{equation}
where in this case the angular frequency of each mode, $\omega_p$, is given by
\begin{equation}
    \label{app:eom2}
    \omega_p^2 \; = \; \frac{p^2}{a^2} + \sigma \phi_0^2 \cos^{2} \left(m_{\phi} t \right)-\frac{3}{2}\frac{\ddot a}{a} 
- \frac{3}{4}\left(\frac{\dot a}{a}\right)^2 \, ,
\end{equation}
where the last two terms are responsible for gravitational particle production. Using Eq.~(\ref{app:solphi}) we can rewrite the expression for angular frequency as
\beq
\label{app:angfreq}
\omega_p^2 \;=\; \frac{p^2}{a^2} + \sigma\phi(t)^2  + \frac{9}{4}w H^2 \, ,
\eeq
where $w = p/\rho$ is the equation of state parameter. When the Universe is dominated by coherent oscillations of the inflaton field $\phi$, after a few oscillations following the beginning of reheating, $w\rightarrow 0$, and the last term can be safely ignored. However, this approximation cannot be used for the first couple of oscillations right after the end of inflation. When inflation ends, $w_{\rm end} = -1/3$, and if we use the Hubble parameter value at the end of inflation for T-model attractor, 
\begin{equation}
    \label{app:hend}
    H_{\rm end} = \sqrt{3\lambda} \tanh\left(\frac{\phi_{\rm{end}}}{\sqrt{6}}\right) \, ,
\end{equation}
we find that the ratio between the last two terms is
\beq
\frac{\frac{9}{4}wH^2_{\rm end}}{\sigma \phi_{\rm end}^2} \;\simeq\; \frac{9w}{8} \cdot \frac{\lambda}{\sigma}\, , \qquad \phi \lesssim 1 \, .
\eeq
Therefore, the friction term $\frac{9}{4}w H^2$ can only be ignored during the initial stage of reheating if $\sigma \gg \lambda$. This is further illustrated in Fig.~\ref{fig:rhochi0}, where we show for $\sigma/\lambda=1$ the effect of ignoring Hubble-dependent terms in the effective frequency of the $\chi$-modes. Not only is the resulting energy density larger by over an order of magnitude relative to the correct non-perturbative result, but it is in near agreement with the perturbative estimate.\\

For our initial condition of $x_p$, we take the positive-frequency of Bunch-Davies vacuum,
\begin{equation}
    x_p \; \simeq \; \frac{e^{-i \omega_p t}}{\sqrt{2 \omega_p}} \, ,
\end{equation}
and the comoving particle occupation number in the mode $p$ is given by~\cite{preheating2}
\begin{equation}
    \label{app:abund}
    n_p \; = \; \frac{\omega_p}{2} \left(\frac{|\dot{x}_p|^2}{\omega_p} + |x_p|^2 \right) - \frac{1}{2} \, .
\end{equation}
It should be noted that the occupation number $n_p$ does not have a factor of comoving volume $a^3$ in the denominator.

To understand the parametric resonance, it is convenient to introduce the following variables:
\begin{equation}
    z = m_{\phi}t + \frac{\pi}{2}, \qquad A_p = \frac{p^2}{m_{\phi}^2 a^2} + 2q =\frac{E_\chi^2}{m_\phi^2}, \qquad q = \frac{\sigma \phi_0^2}{4m_\phi^2} = \frac{\langle m_{\chi}^2 \rangle}{2m_{\phi}^2} \, ,
\end{equation}
where $\langle m_{\chi}^2 \rangle = \langle \sigma \phi^2 \rangle=\frac{\sigma}{2} \phi_0^2$, $E_{\chi}^2 = p^2/a^2 + \langle m_{\chi}^2 \rangle$,
and $q$ is the resonance parameter, which can be used to rewrite Eq.~(\ref{app:eom2}) in a well-known Mathieu equation form:
\begin{equation}
x_{p}^{\prime \prime}+\left(A_{p}-2 q \cos 2 z\right) x_{p}=0 \, ,
\label{Eq:mathieuequation}
\end{equation}
where in this case prime denotes a derivative with respect to the variable $z$.\footnote{In the following discussion we neglect the backreaction and rescattering effects of created particles.}

\begin{figure}[!t]
\centering
    \includegraphics[width=0.75\textwidth]{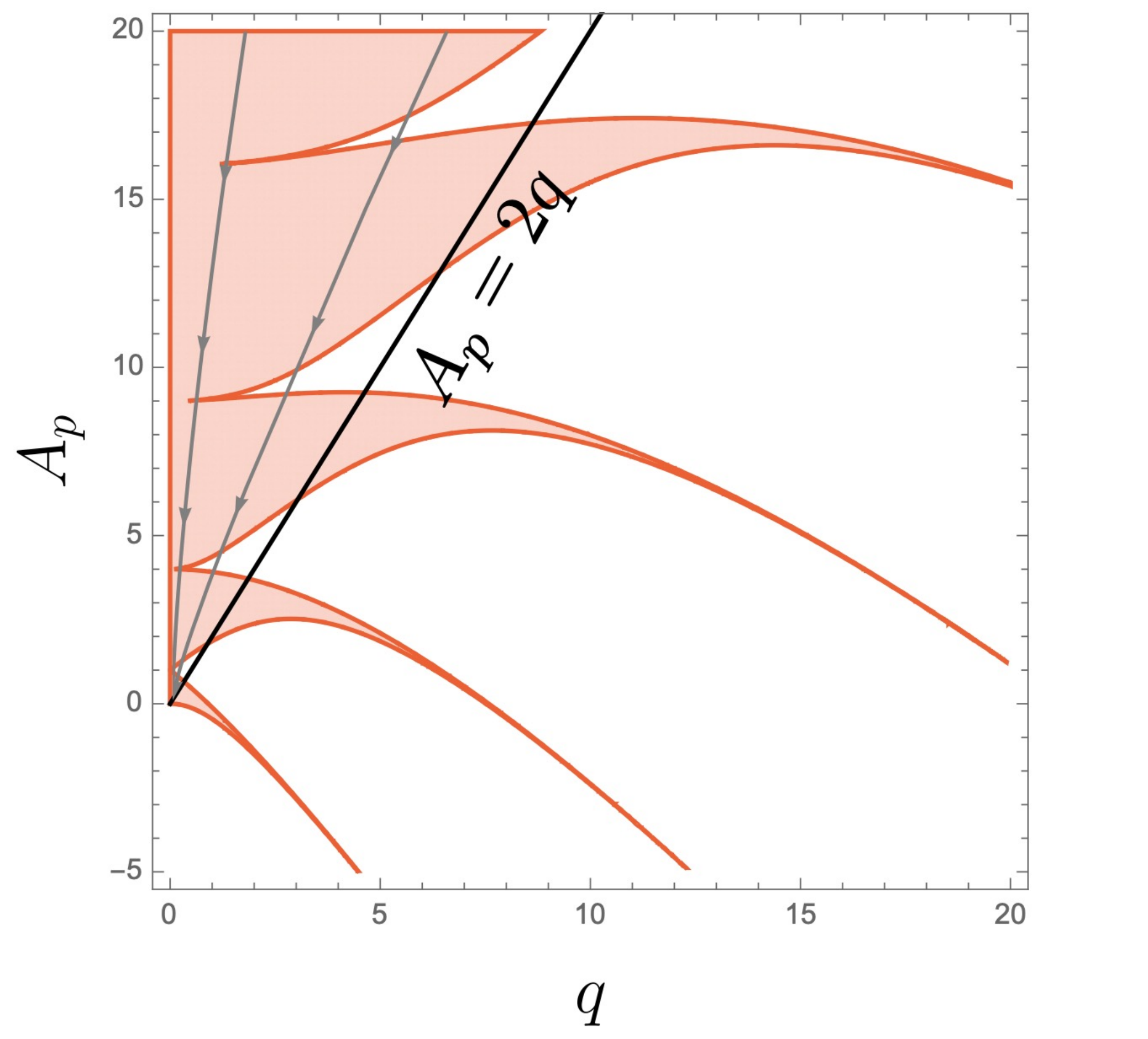}
    \caption{\it Mathieu stability/instability chart. The white regions corresponds to instability bands, where the oscillating mode becomes unstable and grows exponentially, leading to explosive particle production. The red regions correspond to stable regions, where the mode solutions oscillate. We also show two distinct flow lines corresponding to the modes $p=6.4\times 10^{-5}$ (left gray line) and $p=2.6\times 10^{-5}$ (right gray line). The arrows on the gray lines show the time evolution. Below the line $A_p = 2q$, the tachyonic resonance occurs.}
    \label{fig:mathieu}
\end{figure}

We can interpret $A_p$ as the ratio of the mean energy squared of a mode with comoving momentum $p/a$ and the effective mass $\langle m_{\chi} \rangle$ to the inflaton mass squared, whereas $\sqrt{2q}$ represents the mass ratio $\langle m_{\chi} \rangle/m_{\phi}$. The properties of Mathieu equation (\ref{Eq:mathieuequation}) can be understood from its stability/instability chart, which we illustrate in Fig.~\ref{fig:mathieu}. 
The red regions correspond to stable regions, where the mode solutions oscillate, and the white regions correspond to exponentially unstable regions, where the mode solutions grow exponentially as $x_p \propto \exp(\mu_p z)$ \cite{preheating2}. The characteristic exponent typically lies in the range $0 \lesssim \mu_p \lesssim 0.28$~\cite{preheating}. From Eq.~(\ref{app:abund}), we can find that the particle occupation number $n_p(t) \sim |x_p|^2 \sim \exp(2 \mu_p z)$, and it grows exponentially in the unstable regions.

If the resonance parameter is small, $q \ll 1$ or $\langle m_\chi \rangle \ll m_\phi$, the narrow resonance occurs near $A_p \simeq l^2,~l = 1, 2, ...$,  which can be seen from Fig.~\ref{fig:mathieu}, and the strongest resonant particle production will occur near the first instability band, $A_p \sim 1 \pm q$. However, due to expansion of the universe, the resonance parameter $q$ decreases rapidly and the resonance stops. Therefore, significant particle production in an expanding universe occurs only for $q \gg 1$. If we assume that the perturbative inflaton decay during the parametric resonance is inefficient, $\Gamma < H$, then we can estimate the time when a given mode remains in an unstable band by $\Delta t \sim q H^{-1}$. Because the momenta $p$ redshifts, eventually we reach the limit $q \lesssim 1$ and enter the inefficient narrow resonance regime. In the narrow resonance regime, during the time $\Delta t$ the particle occupation number increases by a factor
\beq
e^{2 \mu_p z} \sim e^{q m_{\phi} \Delta t} \sim e^{\frac{q^2 m_{\phi}}{H}} \, ,
\eeq
where in this regime we estimated that the resonance is most efficient and assumes the maximal value when $\mu_p \sim \frac{q}{2}$, and the explosive particle production ends when
\begin{equation}
    q^2 m_{\phi} \lesssim H~~~\Rightarrow ~~~ \frac{\sigma^2 \phi_0^4}{32 \sqrt{2} \lambda^{3/2}} \lesssim H \, .
\end{equation}
If we use the value of Hubble parameter after the end of inflaton $H^2 \simeq \frac{V(\phi_0)}{3}$, where the potential is given by Eq.~(\ref{tmodelpot}), we find the bound
\begin{equation}
    \frac{\sigma^2}{\lambda^2} \lesssim \frac{32 \sqrt{2}}{\sqrt{3} \phi_0^{3}} \, ,
\end{equation}
and if $\phi_0 \sim \mathcal{O}(1)$, we find that when $\sigma \lesssim \lambda$ the non-perturbative particle production is inefficient. A more detailed discussion is presented in~\cite{preheating2}.

When the resonance parameter is large, $q \gg 1$ or $\langle m_\chi\rangle \gg m_\phi$, we are in a highly non-perturbative regime, and the particle production is characterized by the broad parametric resonance. In this case, the resonance occurs above the line $A_p = 2q$, which is illustrated in Fig.~\ref{fig:mathieu}. The band number arising from the Mathieu equation is given by $n = \sqrt{A_p}$, and for the fastest growing modes with $p \sim 0$, we have $A_p \sim 2q \sim n^2$. Because the inflaton amplitude decreases as $\phi_0 \sim 1/t \sim 1/a^{3/2}$, then $q \sim 1/t^2$, and with each oscillation the band number $n$ decreases. Therefore, the modes $x_p$ start evolving from some initial value above the line $A_p = 2q$ 
toward the origin while crossing the Mathieu instability bands, which leads to an increase in the particle occupation number for each mode, $n_p$. 
We illustrate in Fig.~(\ref{fig:mathieu}) the time-evolution of two different momenta modes, $p=6.4 \times 10^{-5}$ (left gray line) and $p=2.6 \times 10^{-5}$ (right gray line). 
To understand the evolution of the comoving modes, it is convenient to express
$A_p$ as
\beq
A_p=\frac{b}{a^2}+\frac{c}{a^3}~~~{\rm and}~~~q = \frac{c}{2a^3} \, ,
\eeq
with
\beq
b=\frac{p^2}{m_\phi^2}, ~~~c =\sigma \frac{\phi_{\rm end}^2 a_{\rm end}^3}{2m_\phi^2}.
\eeq

\noindent
A given mode will follow a line respecting $A_p=\frac{b}{a^2}+2q=b \left(\frac{2}{c}\right)^{2/3}q^{2/3}+ 2q$, converging to the limit $A_p=2q$, 
which corresponds to the fundamental mode $p=0$. From Fig.~(\ref{fig:mathieu}), it can be seen that higher momentum modes will lead to larger values of $A_p$ that will pass through more narrow regions of the resonance bands and the exponential particle production will be significantly smaller. Typically, the parametric resonance ceases when the mode $x_p$ reaches the first stability band, when $q \lesssim 1$ or $\langle m_\chi \rangle \lesssim m_\phi$, and the mode begins entering the perturbative regime.

From Eq.~(\ref{eq:kincond}), we see that we can connect the induced mass parameter, that we discussed in the Introduction, to the resonance parameter using the expression, 
\begin{equation}
    \mathcal{R} = 16 q \, ,
\end{equation}
and if we summarize the discussion of this Appendix, we can identify the following particle production regimes:
\begin{equation}
  \begin{alignedat}{3}
    \mathcal{R} &\gg 1,  \qquad &q &\gg 1, \qquad && {\text{Strong Broad Resonance Regime}} \, , \\
    \mathcal{R} &> 16,  &q &> 1, &&{\text{Weak Broad Resonance Regime}} \, ,\\ 
    \mathcal{R} &> 1,  &q &> \frac{1}{16}, &&{\text{Narrow Resonance Regime}} \, ,\\ 
    \mathcal{R} &\ll 1, &q &\ll \frac{1}{16}, &&{\text{Perturbative Regime}} \, .
 \end{alignedat}
\end{equation}
Therefore, as discussed in~\cite{Garcia:2020wiy}, $\mathcal{R} > 1$ signifies the limit when the non-perturbative effects become important.

%%%%%%%%%%%%%%%%%%%%%%%%%%%%%%%%%%%
%\newpage
\section{Preheating of fermions}
\label{sec:appB}
%%%%%%%%%%%%%%%%%%%%%%%%%%%%%%%%%%%
In this Appendix, we discuss the non-perturbative production of fermions. We begin by considering the excitation of a spin-1/2 fermion due to its coupling to the inflaton during reheating. In order to account for the short time-scale violation of adiabaticity due to the oscillation of the inflaton and the source of particle production one must study the excitation of the field mode-by-mode. At linear order, the equation of motion for a Dirac fermion is obtained from the following action in the Friedmann-Robertson-Walker (FRW) background,

\beq
\mathcal{S}_{\psi} \;=\; \int d^4x\, \sqrt{-g} \bar{\psi} \left(i \bar{\gamma}^\mu\nabla_{\mu} - m_{\psi}(t) \right)\psi\,,
\eeq
where the curved-space gamma matrices are defined in terms of their flat-space counterparts as $\bar{\gamma}^{\mu}=e^{\mu}_a\gamma^a$,\footnote{We use the following convention:\begin{equation}
\gamma^{0}=\left(\begin{array}{cc} \nonumber
\mathbb{1} & 0 \\
0 & -\mathbb{1}
\end{array}\right), \quad \gamma^{i}=\left(\begin{array}{cc}
0 & \sigma_{i} \\
-\sigma_{i} & 0
\end{array}\right) \, ,
\end{equation} 
where $\sigma$ are the Pauli matrices.
} where $e^{\mu}_a$ are the components of the metric tetrad, and the covariant derivative is defined as,
\beq
\nabla_{\mu} \;=\; \partial_{\mu} - \frac{i}{2}\omega_{\mu ab}S^{ab}\,,
\eeq
where $S^{ab}=\frac{i}{4}[\gamma^a,\gamma^b]$, and $\omega_{\mu ab}$ are the components of the spin connection. 
For a Yukawa-like coupling, $\mathcal{L}_{\rm int} = -y\phi \bar{\psi}\psi$, the time-dependent mass of the produced fermion is $m_{\psi}(t)=y\phi(t)+m_{\psi,0}$, which implies that only the contribution of the spatially homogeneous condensate is accounted for the production of fermions. It should be noted that in this analysis we do not consider the fermion $\psi$ couplings to the Standard Model which would affect thermalization.

As in the previous section, for convenience we switch to conformal time, $\tau$.
In an FRW background, one finds that $e^{\mu}_a=a^{-1}\delta^{\mu}_a$ and $- \frac{i}{2}\omega_{\mu ab}S^{ab} = \frac{a'}{4a^2}[\bar{\gamma}_{\mu},\gamma^0]$, and the equation of motion for the spinor $\psi$ takes the form,
\beq
\left[ i\gamma^{\mu}\partial_{\mu} + i \frac{3}{2} \frac{a'}{a} \gamma^0 - am_{\psi}(\tau)\right]\psi \;=\;0\,.
\eeq
We introduce the field redefinition, $\Psi\equiv a^{3/2}\psi$, and the equation of motion can be rewritten in a familiar flat-space Dirac equation with a time-dependent mass form,
\beq\label{eq:dPsi}
\left(i\gamma^{\mu} \partial_{\mu} - am_{\psi}(\tau)\right)\Psi \;=\;0\,.
\eeq
To solve this equation of motion with the field operator $\Psi$ given by Eq.~(\ref{modeexpferm}), it is convenient to introduce the two-spinors $\xi_r$, that are the eigenvectors of the helicity operator, $\frac{\boldsymbol{\sigma}\cdot \bp}{p}\xi_r=r\xi_r$, with $\xi_{1}= \scriptsize \left(\begin{array}{l}1 \\ 0\end{array}\right)$ and $\xi_{2}= \scriptsize \left(\begin{array}{l}0 \\ 1\end{array}\right)$. We introduce the generalized spinors
\beq
u^{(r)}_p(\tau) \;\equiv\; \frac{1}{\sqrt{2}} \left(
\begin{matrix}
U_1(\tau) \xi_r(\bp)\\
U_2(\tau) \xi_r(\bp)
\end{matrix}
\right)\,,
\eeq
where we took the momentum along the $z$-axis, $p \equiv p_z$.
The spinor equation (\ref{eq:dPsi}) is equivalent to the following system of equations,
\begin{align}\label{eq:eomf1}
U_1'(\tau) \;&=\; -i \, p \, U_2(\tau) - i \, a \, m_{\psi} \, U_1(\tau)\,, \\ \label{eq:eomf2}
U_2'(\tau) \;&=\; -i \, p \, U_1(\tau) + i \, a \, m_{\psi} \, U_2(\tau)\,.
\end{align}
The mode functions $U_1$ and $U_2$ are subject to the constraints
\begin{equation}
\begin{aligned}
|U_1|^2 + |U_2|^2 &=2 \, , \\
U_{1} U_{2}^{\dagger} &=U_{2}^{*} U_{1}^{T} \, ,
\end{aligned}
\end{equation}
which are necessary for the consistent quantization of $\Psi$. Their zero-particle initial condition can be chosen as follows,
\beq
U_1(\tau_0) \;=\; \sqrt{1+\frac{am_{\psi}}{\omega_p}} \,, \qquad U_2(\tau_0) \;=\; \sqrt{1-\frac{am_{\psi}}{\omega_p}}\,,
\eeq
where the angular frequency is given by
\beq\label{eq:omF}
\omega_p^2 \;\equiv\; p^2 + (am_{\psi})^2\,.
\eeq
In this case, the energy density in terms of the mode functions $U_{1, 2}$ is given by Eq.~(\ref{endenferm}). A more detailed discussion on non-perturbative fermion production after inflation is presented in~\cite{Giudice:1999fb, Peloso:2000hy, Nilles:2001fg}.\\

\begin{figure}[!t]
\centering
    \includegraphics[width=\textwidth]{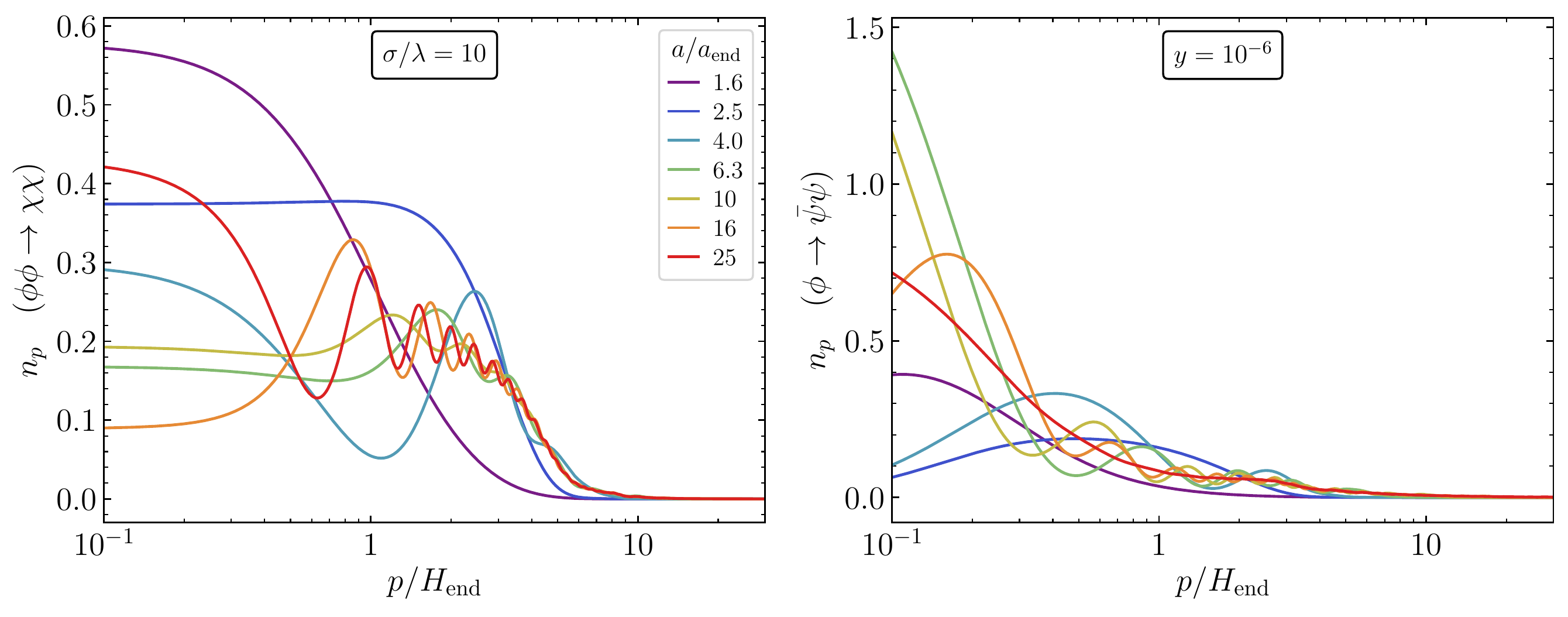}
    \caption{\it Occupation numbers as function of the comoving wavenumber at selected times, for scalar and fermion production from inflaton decay, in the absence of strong Bose enhancement / Pauli blocking (c.f.~Figs.~\ref{fig:rhochi} and \ref{fig:rhopsi}).
    }
    \label{fig:npS}
\end{figure}

\begin{figure}[!t]
\centering
    \includegraphics[width=\textwidth]{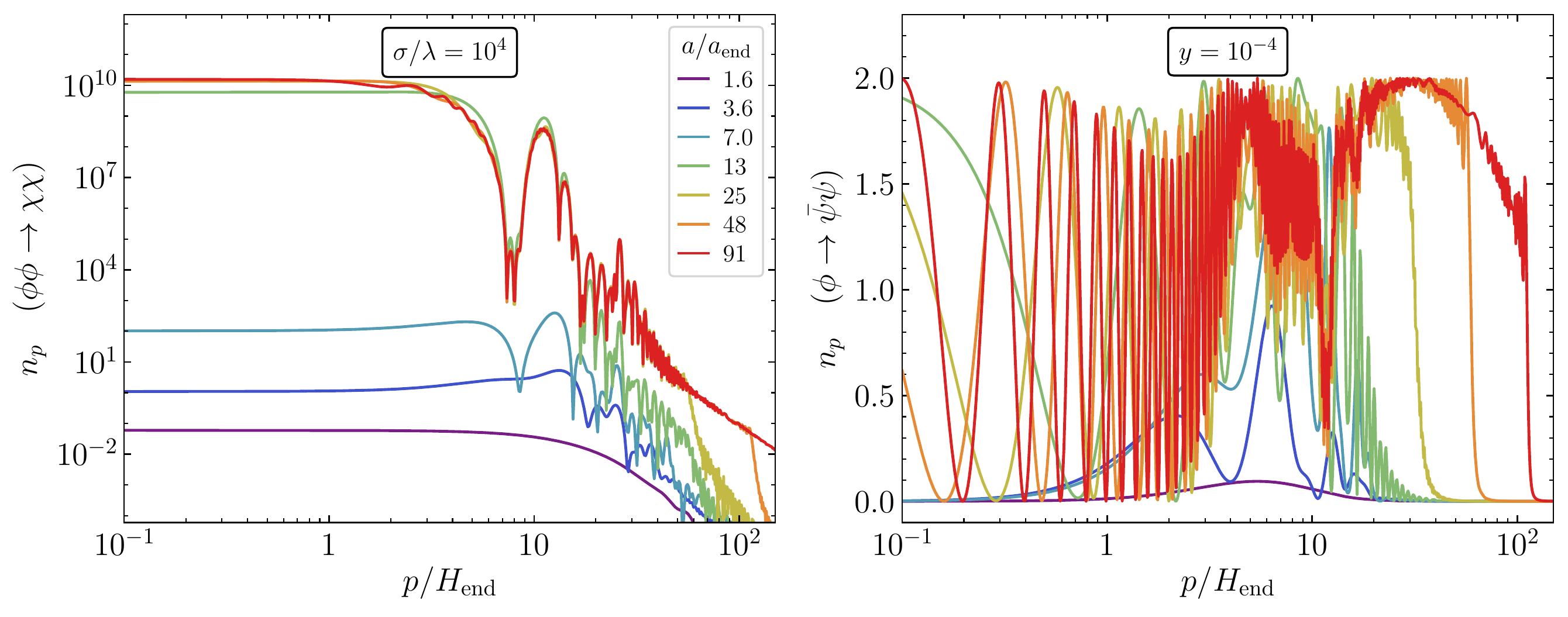}
    \caption{\it Occupation numbers as function of the comoving wavenumber at selected times, for scalar and fermion production from inflaton decay, in the presence of strong Bose enhancement / Pauli blocking (c.f.~Figs.~\ref{fig:rhochi2} and \ref{fig:rhopsi}).
    }
    \label{fig:npL}
\end{figure}

Figs.~\ref{fig:npS} and \ref{fig:npL} highlight the difference in the growth of occupation numbers for a scalar decay channel and a fermionic one. For weak couplings, as shown in Fig.~\ref{fig:npS}, the difference between the two cases is mainly limited to a ``freezing'' of $n_p$ in the scalar case, due to the redshift of the decay rate of $\phi$, while occupation numbers continue to noticeably evolve for $\phi\rightarrow \bar{\psi}\psi$. For large couplings, as demonstrated in Fig.~\ref{fig:npL}, the difference in the quantum statistics becomes evident. For $\chi$, occupation numbers are not bounded from above, and can grow exponentially fast with cosmic time due to the parametric resonance of $\chi$. On the other hand, Fermi-Dirac statistics are manifestly observable: $n_p$ never grows above 2.

%\newpage

%\addcontentsline{toc}{section}{References}
%\bibliographystyle{utphys}
%\bibliography{biblio} 

\begin{thebibliography}{0}



\bibitem{nos}
  D.~V.~Nanopoulos, K.~A.~Olive and M.~Srednicki,
  %``After Primordial Inflation,''
  Phys.\ Lett.\ B {\bf 127}, 30 (1983).
  %%CITATION = PHLTA,B127,30;%%
  
  \bibitem{ehnos}
 J.~R.~Ellis, J.~S.~Hagelin, D.~V.~Nanopoulos, K.~A.~Olive and M.~Srednicki,
%  ``Supersymmetric Relics from the Big Bang,''
  Nucl.\ Phys.\  B {\bf 238}, 453 (1984).

  
   \bibitem{kl}
  M.~Y.~Khlopov and A.~D.~Linde,
  %``Is It Easy to Save the Gravitino?,''
  Phys.\ Lett.\ B {\bf 138}, 265 (1984).
  %%CITATION = PHLTA,B138,265;%%     
  
  \bibitem{oss}
  K.~A.~Olive, D.~N.~Schramm and M.~Srednicki,
  %``Gravitinos as the Cold Dark Matter in an omega = 1 Universe,''
  Nucl.\ Phys.\ B {\bf 255}, 495 (1985).
 % doi:10.1016/0550-3213(85)90149-X
  %%CITATION = doi:10.1016/0550-3213(85)90149-X;%%
  
  \bibitem{fimp}
  L.~J.~Hall, K.~Jedamzik, J.~March-Russell and S.~M.~West,
  %``Freeze-In Production of FIMP Dark Matter,''
  JHEP {\bf 1003} (2010) 080
%  doi:10.1007/JHEP03(2010)080
  [arXiv:0911.1120 [hep-ph]];
  %%CITATION = doi:10.1007/JHEP03(2010)080;%%
  %182 citations counted in INSPIRE as of 25 Oct 2016
  X.~Chu, T.~Hambye and M.~H.~G.~Tytgat,
  %``The Four Basic Ways of Creating Dark Matter Through a Portal,''
  JCAP {\bf 1205} (2012) 034
%  doi:10.1088/1475-7516/2012/05/034
  [arXiv:1112.0493 [hep-ph]];
  Y.~Mambrini, K.~A.~Olive, J.~Quevillon and B.~Zaldivar,
%``Gauge Coupling Unification and Nonequilibrium Thermal Dark Matter,''
Phys. Rev. Lett. \textbf{110} (2013) no.24, 241306
[arXiv:1302.4438 [hep-ph]];
  X.~Chu, Y.~Mambrini, J.~Quevillon and B.~Zaldivar,
  %``Thermal and non-thermal production of dark matter via Z'-portal(s),''
  JCAP {\bf 1401} (2014) 034
%  doi:10.1088/1475-7516/2014/01/034
  [arXiv:1306.4677 [hep-ph]];
  A.~Biswas, D.~Borah and A.~Dasgupta,
%``UV complete framework of freeze-in massive particle dark matter,''
Phys.\ Rev.\ D \textbf{99}, no.1, 015033 (2019)
% doi:10.1103/PhysRevD.99.015033
[arXiv:1805.06903 [hep-ph]].

\bibitem{brv}          
N.~Bernal, J.~Rubio and H.~Veerm\"ae,
%``UV Freeze-in in Starobinsky Inflation,''
JCAP \textbf{10} (2020), 021
%doi:10.1088/1475-7516/2020/10/021
[arXiv:2006.02442 [hep-ph]].
%21 citations counted in INSPIRE as of 24 Sep 2021

  
  \bibitem{Bernal:2017kxu}
  N.~Bernal, M.~Heikinheimo, T.~Tenkanen, K.~Tuominen and V.~Vaskonen,
  %``The Dawn of FIMP Dark Matter: A Review of Models and Constraints,''
  Int.\ J.\ Mod.\ Phys.\ A {\bf 32} (2017) no.27,  1730023
 % doi:10.1142/S0217751X1730023X
  [arXiv:1706.07442 [hep-ph]].
  %%CITATION = doi:10.1142/S0217751X1730023X;%%
  %29 citations counted in INSPIRE as of 17 Feb 2018

  \bibitem{Bernal:2019mhf}
N.~Bernal, F.~Elahi, C.~Maldonado and J.~Unwin,
%``Ultraviolet Freeze-in and Non-Standard Cosmologies,''
JCAP \textbf{11} (2019), 026
%doi:10.1088/1475-7516/2019/11/026
[arXiv:1909.07992 [hep-ph]].
%22 citations counted in INSPIRE as of 01 Nov 2020

%\cite{Bernal:2020gzm}
\bibitem{Bernal:2020gzm}
N.~Bernal,
%``Boosting Freeze-in through Thermalization,''
JCAP \textbf{10}, 006 (2020)
%doi:10.1088/1475-7516/2020/10/006
[arXiv:2005.08988 [hep-ph]].
%12 citations counted in INSPIRE as of 04 Dec 2020    

 \bibitem{dg}
   A.~D.~Dolgov and A.~D.~Linde,
  %``Baryon Asymmetry in Inflationary Universe,''
  Phys.\ Lett.\  {\bf 116B}, 329 (1982);
%  doi:10.1016/0370-2693(82)90292-1
  %%CITATION = doi:10.1016/0370-2693(82)90292-1;%%
  L.~F.~Abbott, E.~Farhi and M.~B.~Wise,
  %``Particle Production in the New Inflationary Cosmology,''
  Phys.\ Lett.\  {\bf 117B}, 29 (1982).
 % doi:10.1016/0370-2693(82)90867-X
  %%CITATION = doi:10.1016/0370-2693(82)90867-X;%%
  
  
   \bibitem{Davidson:2000er} 
  S.~Davidson and S.~Sarkar,
  %``Thermalization after inflation,''
  JHEP {\bf 0011}, 012 (2000)
  [hep-ph/0009078].
  %%CITATION = HEP-PH/0009078;%%

  \bibitem{Harigaya:2013vwa}
  K.~Harigaya, K.~Mukaida and M.~Yamada,
%``Dark Matter Production during the Thermalization Era,''
JHEP \textbf{07} (2019), 059
%doi:10.1007/JHEP07(2019)059
[arXiv:1901.11027 [hep-ph]];
%9 citations counted in INSPIRE as of 09 Jun 2020
K.~Harigaya, M.~Kawasaki, K.~Mukaida and M.~Yamada,
%``Dark Matter Production in Late Time Reheating,''
Phys. Rev. D \textbf{89} (2014) no.8, 083532
%doi:10.1103/PhysRevD.89.083532
[arXiv:1402.2846 [hep-ph]];
%54 citations counted in INSPIRE as of 09 Jun 2020 
K.~Harigaya and K.~Mukaida,
%``Thermalization after/during Reheating,''
JHEP \textbf{05}, 006 (2014)
%doi:10.1007/JHEP05(2014)006
[arXiv:1312.3097 [hep-ph]].
%77 citations counted in INSPIRE as of 03 Dec 2020

%\cite{Mukaida:2015ria}
\bibitem{Mukaida:2015ria}
K.~Mukaida and M.~Yamada,
%``Thermalization Process after Inflation and Effective Potential of Scalar Field,''
JCAP \textbf{02}, 003 (2016)
%doi:10.1088/1475-7516/2016/02/003
[arXiv:1506.07661 [hep-ph]].
%48 citations counted in INSPIRE as of 03 Dec 2020

 \bibitem{Garcia:2018wtq}
  M.~A.~G.~Garcia and M.~A.~Amin,
  %``Prethermalization production of dark matter,''
  Phys.\ Rev.\ D {\bf 98}, no. 10, 103504 (2018)
 % doi:10.1103/PhysRevD.98.103504
  [arXiv:1806.01865 [hep-ph]].
  %%CITATION = doi:10.1103/PhysRevD.98.103504;%%
  
  \bibitem{Giudice:2000ex}
  G.~F.~Giudice, E.~W.~Kolb and A.~Riotto,
  %``Largest temperature of the radiation era and its cosmological implications,''
  Phys.\ Rev.\ D {\bf 64} (2001) 023508
 % doi:10.1103/PhysRevD.64.023508
  [hep-ph/0005123];
  %%CITATION = doi:10.1103/PhysRevD.64.023508;%%
  %319 citations counted in INSPIRE as of 04 Apr 2018
   D.~J.~H.~Chung, E.~W.~Kolb and A.~Riotto,
  %``Production of massive particles during reheating,''
  Phys.\ Rev.\ D {\bf 60} (1999) 063504
%  doi:10.1103/PhysRevD.60.063504
  [hep-ph/9809453].
  %%CITATION = doi:10.1103/PhysRevD.60.063504;%%
  %270 citations counted in INSPIRE as of 04 Apr 2018
  
  \bibitem{egnop}
  J.~Ellis, M.~A.~G.~Garcia, D.~V.~Nanopoulos, K.~A.~Olive and M.~Peloso,
  %``Post-Inflationary Gravitino Production Revisited,''
  JCAP {\bf 1603}, no. 03, 008 (2016)
%  doi:10.1088/1475-7516/2016/03/008
  [arXiv:1512.05701 [astro-ph.CO]].
  %%CITATION = doi:10.1088/1475-7516/2016/03/008;%%
  
  \bibitem{Garcia:2017tuj}
  M.~A.~G.~Garcia, Y.~Mambrini, K.~A.~Olive and M.~Peloso,
  %``Enhancement of the Dark Matter Abundance Before Reheating: Applications to Gravitino Dark Matter,''
  Phys.\ Rev.\ D {\bf 96} (2017) no.10,  103510
%  doi:10.1103/PhysRevD.96.103510
  [arXiv:1709.01549 [hep-ph]].
  %%CITATION = doi:10.1103/PhysRevD.96.103510;%%
  %4 citations counted in INSPIRE as of 17 Feb 2018
  
    %\cite{Chen:2017kvz}
\bibitem{Chen:2017kvz}
S.~L.~Chen and Z.~Kang,
%``On UltraViolet Freeze-in Dark Matter during Reheating,''
JCAP \textbf{05}, 036 (2018)
%doi:10.1088/1475-7516/2018/05/036
[arXiv:1711.02556 [hep-ph]].
%20 citations counted in INSPIRE as of 03 Dec 2020




 \bibitem{Garcia:2020eof}
M.~A.~Garcia, K.~Kaneta, Y.~Mambrini and K.~A.~Olive,
%``Reheating and Post-inflationary Production of Dark Matter,''
Phys. Rev. D \textbf{101} (2020) no.12, 123507
%doi:10.1103/PhysRevD.101.123507
[arXiv:2004.08404 [hep-ph]].
%3 citations counted in INSPIRE as of 20 May 2020
  


%\cite{Co:2020xaf}
\bibitem{Co:2020xaf}
R.~T.~Co, E.~Gonzalez and K.~Harigaya,
%``Increasing Temperature toward the Completion of Reheating,''
JCAP \textbf{11}, 038 (2020)
%doi:10.1088/1475-7516/2020/11/038
[arXiv:2007.04328 [astro-ph.CO]].
%2 citations counted in INSPIRE as of 04 Dec 2020

%\cite{Garcia:2020wiy}
\bibitem{Garcia:2020wiy}
M.~A.~G.~Garcia, K.~Kaneta, Y.~Mambrini and K.~A.~Olive,
%``Inflaton Oscillations and Post-Inflationary Reheating,''
JCAP \textbf{04}, 012 (2021)
%doi:10.1088/1475-7516/2021/04/012
[arXiv:2012.10756 [hep-ph]].
%7 citations counted in INSPIRE as of 16 Jun 2021

%\cite{Turner:1983he}
\bibitem{Turner:1983he}
M.~S.~Turner,
%``Coherent Scalar Field Oscillations in an Expanding Universe,''
Phys. Rev. D \textbf{28}, 1243 (1983)
%doi:10.1103/PhysRevD.28.1243
%659 citations counted in INSPIRE as of 14 Aug 2021



%\cite{Garcia:2020hyo}
\bibitem{Garcia:2020hyo}
M.~A.~G.~Garcia, Y.~Mambrini, K.~A.~Olive and S.~Verner,
%``Case for decaying spin- 3/2 dark matter,''
Phys. Rev. D \textbf{102}, no.8, 083533 (2020)
%doi:10.1103/PhysRevD.102.083533
[arXiv:2006.03325 [hep-ph]].
%13 citations counted in INSPIRE as of 27 Jul 2021

%\cite{Anastasopoulos:2020gbu}
\bibitem{Anastasopoulos:2020gbu}
P.~Anastasopoulos, K.~Kaneta, Y.~Mambrini and M.~Pierre,
%``Energy-momentum portal to dark matter and emergent gravity,''
Phys. Rev. D \textbf{102}, no.5, 055019 (2020)
%doi:10.1103/PhysRevD.102.055019
[arXiv:2007.06534 [hep-ph]].
%10 citations counted in INSPIRE as of 27 Jul 2021


%\cite{Brax:2020gqg}
\bibitem{Brax:2020gqg}
P.~Brax, K.~Kaneta, Y.~Mambrini and M.~Pierre,
%``Disformal dark matter,''
Phys. Rev. D \textbf{103}, no.1, 015028 (2021)
%doi:10.1103/PhysRevD.103.015028
[arXiv:2011.11647 [hep-ph]].
%8 citations counted in INSPIRE as of 27 Jul 2021

\bibitem{Kaneta:2019zgw}
  K.~Kaneta, Y.~Mambrini and K.~A.~Olive,
  %``Radiative production of nonthermal dark matter,''
  Phys.\ Rev.\ D {\bf 99} (2019) no.6,  063508
%  doi:10.1103/PhysRevD.99.063508
  [arXiv:1901.04449 [hep-ph]].
  %%CITATION = doi:10.1103/PhysRevD.99.063508;%%
  %4 citations counted in INSPIRE as of 21 May 2019
  
  \bibitem{Mambrini:2021zpp}
Y.~Mambrini and K.~A.~Olive,
%``Gravitational Production of Dark Matter during Reheating,''
Phys. Rev. D \textbf{103} (2021) no.11, 115009
%doi:10.1103/PhysRevD.103.115009
[arXiv:2102.06214 [hep-ph]].

%\cite{Bernal:2021kaj}
\bibitem{Bernal:2021kaj}
N.~Bernal and C.~S.~Fong,
%``Dark matter and leptogenesis from gravitational production,''
JCAP \textbf{06}, 028 (2021)
%doi:10.1088/1475-7516/2021/06/028
[arXiv:2103.06896 [hep-ph]].
%2 citations counted in INSPIRE as of 27 Jul 2021

%\cite{Kaneta:2021pyx}
\bibitem{Kaneta:2021pyx}
K.~Kaneta, P.~Ko and W.~I.~Park,
%``Conformal Portal to Dark Matter,''
[arXiv:2106.01923 [hep-ph]].

\bibitem{Bhattacharyya:2018evo}
G.~Bhattacharyya, M.~Dutra, Y.~Mambrini and M.~Pierre,
%``Freezing-in dark matter through a heavy invisible Z',''
Phys. Rev. D \textbf{98} (2018) no.3, 035038
%doi:10.1103/PhysRevD.98.035038
[arXiv:1806.00016 [hep-ph]].

\bibitem{Chowdhury:2018tzw}
D.~Chowdhury, E.~Dudas, M.~Dutra and Y.~Mambrini,
%``Moduli Portal Dark Matter,''
Phys. Rev. D \textbf{99} (2019) no.9, 095028
%doi:10.1103/PhysRevD.99.095028
[arXiv:1811.01947 [hep-ph]].

%\cite{Garcia:2021gsy}
\bibitem{Garcia:2021gsy}
M.~A.~G.~Garcia, Y.~Mambrini, K.~A.~Olive and S.~Verner,
%``On the Realization of WIMPflation,''
[arXiv:2107.07472 [hep-ph]].
%0 citations counted in INSPIRE as of 27 Jul 2021

\bibitem{book} Y.~Mambrini, ``Particles in the dark Universe", 
Springer Ed., ISBN 978-3-030-78139-2


  \bibitem{Planck}
  N.~Aghanim \textit{et al.} [Planck],
%``Planck 2018 results. VI. Cosmological parameters,''
Astron. Astrophys. \textbf{641}, A6 (2020)
%doi:10.1051/0004-6361/201833910
[arXiv:1807.06209 [astro-ph.CO]];
%\cite{Akrami:2018odb}
Y.~Akrami \textit{et al.} [Planck],
%``Planck 2018 results. X. Constraints on inflation,''
Astron. Astrophys. \textbf{641}, A10 (2020)
%doi:10.1051/0004-6361/201833887
[arXiv:1807.06211 [astro-ph.CO]].
%1360 citations counted in INSPIRE as of 17 Jun 2021

%\cite{Dufaux:2006ee}
\bibitem{Dufaux:2006ee}
J.~F.~Dufaux, G.~N.~Felder, L.~Kofman, M.~Peloso and D.~Podolsky,
%``Preheating with trilinear interactions: Tachyonic resonance,''
JCAP \textbf{07}, 006 (2006)
%doi:10.1088/1475-7516/2006/07/006
[arXiv:hep-ph/0602144 [hep-ph]].
%154 citations counted in INSPIRE as of 14 Aug 2021

\bibitem{Kainulainen:2016vzv}
K.~Kainulainen, S.~Nurmi, T.~Tenkanen, K.~Tuominen and V.~Vaskonen,
%``Isocurvature Constraints on Portal Couplings,''
JCAP \textbf{06} (2016), 022
%doi:10.1088/1475-7516/2016/06/022
[arXiv:1601.07733 [astro-ph.CO]].

\bibitem{Ichikawa:2008ne}
K.~Ichikawa, T.~Suyama, T.~Takahashi and M.~Yamaguchi,
%``Primordial Curvature Fluctuation and Its Non-Gaussianity in Models with Modulated Reheating,''
Phys. Rev. D \textbf{78} (2008), 063545
%doi:10.1103/PhysRevD.78.063545
[arXiv:0807.3988 [astro-ph]].


\bibitem{Kallosh:2013hoa}
R.~Kallosh and A.~Linde,
%``Universality Class in Conformal Inflation,''
JCAP \textbf{07} (2013), 002
%doi:10.1088/1475-7516/2013/07/002
[arXiv:1306.5220 [hep-th]].


\bibitem{gravprod}
L.~Parker,
%``Particle creation in expanding universes,''
Phys. Rev. Lett. \textbf{21}, 562-564 (1968);
%doi:10.1103/PhysRevLett.21.562
%561 citations counted in INSPIRE as of 28 Jun 2020
L.~Parker,
%``Quantized fields and particle creation in expanding universes. 1.,''
Phys. Rev. \textbf{183}, 1057-1068 (1969);
%doi:10.1103/PhysRev.183.1057
%935 citations counted in INSPIRE as of 28 Jun 2020
L.~Parker,
%``Quantized fields and particle creation in expanding universes. 2.,''
Phys. Rev. D \textbf{3}, 346-356 (1971);
%doi:10.1103/PhysRevD.3.346
%400 citations counted in INSPIRE as of 28 Jun 2020
V.~Frolov, S.~Mamaev and V.~Mostepanenko,
%``On the Difference in Creation of Particles with Spin 0 and 1/2 in Isotropic Cosmologies,''
Phys. Lett. A \textbf{55}, 389-390 (1976);
%doi:10.1016/0375-9601(75)90555-1
%19 citations counted in INSPIRE as of 28 Jun 2020
A.~Grib, B.~Levitskii and V.~Mostepanenko,
%``Particle creation from vacuum by non-stationary gravitational field in the canonical formalism,''
Teor. Mat. Fiz. \textbf{19}, 59-75 (1974);
%10 citations counted in INSPIRE as of 28 Jun 2020
A.~Grib, S.~Mamaev and V.~Mostepanenko,
%``Particle Creation from Vacuum in Homogeneous Isotropic Models of the Universe,''
Gen. Rel. Grav. \textbf{7}, 535-547 (1976);
%doi:10.1007/BF00766413
%82 citations counted in INSPIRE as of 28 Jun 2020
Y.~Zel'dovich and A.~Starobinsky,
%``Rate of particle production in gravitational fields,''
JETP Lett. \textbf{26}, no.5, 252 (1977);
%95 citations counted in INSPIRE as of 28 Jun 2020
L.~Ford,
%``Gravitational Particle Creation and Inflation,''
Phys. Rev. D \textbf{35}, 2955 (1987);
%doi:10.1103/PhysRevD.35.2955
%371 citations counted in INSPIRE as of 28 Jun 2020
%\bibitem{Hashiba:2018iff}
S.~Hashiba and J.~Yokoyama,
%``Gravitational reheating through conformally coupled superheavy scalar particles,''
JCAP \textbf{01}, 028 (2019)
%doi:10.1088/1475-7516/2019/01/028
[arXiv:1809.05410 [gr-qc]];
%22 citations counted in INSPIRE as of 25 Aug 2020
%\cite{Haro:2018zdb}
%\bibitem{Haro:2018zdb}
J.~Haro, W.~Yang and S.~Pan,
%``Reheating in quintessential inflation via gravitational production of heavy massive particles: A detailed analysis,''
JCAP \textbf{01}, 023 (2019)
%doi:10.1088/1475-7516/2019/01/023
[arXiv:1811.07371 [gr-qc]].
%19 citations counted in INSPIRE as of 25 Aug 2020

\bibitem{ema}
Y.~Ema, R.~Jinno, K.~Mukaida and K.~Nakayama,
%``Gravitational Effects on Inflaton Decay,''
JCAP \textbf{05}, 038 (2015)
%doi:10.1088/1475-7516/2015/05/038
[arXiv:1502.02475 [hep-ph]];
%29 citations counted in INSPIRE as of 27 Jan 2021
 %\cite{Ema:2016hlw}
%\bibitem{Ema:2016hlw}
Y.~Ema, R.~Jinno, K.~Mukaida and K.~Nakayama,
%``Gravitational particle production in oscillating backgrounds and its cosmological implications,''
Phys. Rev. D \textbf{94}, no.6, 063517 (2016)
%doi:10.1103/PhysRevD.94.063517
[arXiv:1604.08898 [hep-ph]];
%45 citations counted in INSPIRE as of 27 Jan 2021
%\cite{Ema:2018ucl}
%\bibitem{Ema:2018ucl}
Y.~Ema, K.~Nakayama and Y.~Tang,
%``Production of Purely Gravitational Dark Matter,''
JHEP \textbf{09}, 135 (2018)
%doi:10.1007/JHEP09(2018)135
[arXiv:1804.07471 [hep-ph]].
%49 citations counted in INSPIRE as of 27 Jan 2021


\bibitem{preheating2}
L.~Kofman, A.~D.~Linde and A.~A.~Starobinsky,
%``Towards the theory of reheating after inflation,''
Phys. Rev. D \textbf{56}, 3258-3295 (1997)
%doi:10.1103/PhysRevD.56.3258
[arXiv:hep-ph/9704452 [hep-ph]].
%1448 citations counted in INSPIRE as of 28 Jun 2020

%\cite{Felder:1999pv}
\bibitem{Felder:1999pv}
G.~N.~Felder, L.~Kofman and A.~D.~Linde,
%``Inflation and preheating in NO models,''
Phys. Rev. D \textbf{60}, 103505 (1999)
%doi:10.1103/PhysRevD.60.103505
[arXiv:hep-ph/9903350 [hep-ph]].
%171 citations counted in INSPIRE as of 27 Jul 2021

%\cite{Chung:1999ve}
\bibitem{Chung:1999ve}
D.~J.~H.~Chung, E.~W.~Kolb, A.~Riotto and I.~I.~Tkachev,
%``Probing Planckian physics: Resonant production of particles during inflation and features in the primordial power spectrum,''
Phys. Rev. D \textbf{62}, 043508 (2000)
%doi:10.1103/PhysRevD.62.043508
[arXiv:hep-ph/9910437 [hep-ph]].
%225 citations counted in INSPIRE as of 27 Jul 2021

\bibitem{Felder:1998vq}
G.~N.~Felder, L.~Kofman and A.~D.~Linde,
%``Instant preheating,''
Phys. Rev. D \textbf{59} (1999), 123523
%doi:10.1103/PhysRevD.59.123523
[arXiv:hep-ph/9812289 [hep-ph]].

\bibitem{Nambu:1996gf}
Y.~Nambu and A.~Taruya,
%``Evolution of cosmological perturbation in reheating phase of the universe,''
Prog. Theor. Phys. \textbf{97}, 83-89 (1997)
%doi:10.1143/PTP.97.83
[arXiv:gr-qc/9609029 [gr-qc]].

\bibitem{Bassett:1998wg}
B.~A.~Bassett, D.~I.~Kaiser and R.~Maartens,
%``General relativistic preheating after inflation,''
Phys. Lett. B \textbf{455}, 84-89 (1999)
%doi:10.1016/S0370-2693(99)00478-5
[arXiv:hep-ph/9808404 [hep-ph]].

\bibitem{Bassett:1999mt}
B.~A.~Bassett, F.~Tamburini, D.~I.~Kaiser and R.~Maartens,
%``Metric preheating and limitations of linearized gravity. 2.,''
Nucl. Phys. B \textbf{561}, 188-240 (1999)
%doi:10.1016/S0550-3213(99)00495-2
[arXiv:hep-ph/9901319 [hep-ph]].

\bibitem{Jedamzik:2010dq}
K.~Jedamzik, M.~Lemoine and J.~Martin,
%``Collapse of Small-Scale Density Perturbations during Preheating in Single Field Inflation,''
JCAP \textbf{09}, 034 (2010)
%doi:10.1088/1475-7516/2010/09/034
[arXiv:1002.3039 [astro-ph.CO]].

\bibitem{Huang:2011gf}
Z.~Huang,
%``The Art of Lattice and Gravity Waves from Preheating,''
Phys. Rev. D \textbf{83}, 123509 (2011)
%doi:10.1103/PhysRevD.83.123509
[arXiv:1102.0227 [astro-ph.CO]].

\bibitem{Giblin:2019nuv}
J.~T.~Giblin and A.~J.~Tishue,
%``Preheating in Full General Relativity,''
Phys. Rev. D \textbf{100}, no.6, 063543 (2019)
%doi:10.1103/PhysRevD.100.063543
[arXiv:1907.10601 [gr-qc]].

\bibitem{Felder:2000hq}
G.~N.~Felder and I.~Tkachev,
%``LATTICEEASY: A Program for lattice simulations of scalar fields in an expanding universe,''
Comput. Phys. Commun. \textbf{178}, 929-932 (2008)
%doi:10.1016/j.cpc.2008.02.009
[arXiv:hep-ph/0011159 [hep-ph]].

\bibitem{Frolov:2008hy}
A.~V.~Frolov,
%``DEFROST: A New Code for Simulating Preheating after Inflation,''
JCAP \textbf{11}, 009 (2008)
%doi:10.1088/1475-7516/2008/11/009
[arXiv:0809.4904 [hep-ph]].

\bibitem{Sainio:2012mw}
J.~Sainio,
%``PyCOOL - a Cosmological Object-Oriented Lattice code written in Python,''
JCAP \textbf{04}, 038 (2012)
%doi:10.1088/1475-7516/2012/04/038
[arXiv:1201.5029 [astro-ph.IM]].

\bibitem{Lozanov:2019jff}
K.~D.~Lozanov and M.~A.~Amin,
%``GFiRe\textemdash{}Gauge Field integrator for Reheating,''
JCAP \textbf{04}, 058 (2020)
%doi:10.1088/1475-7516/2020/04/058
[arXiv:1911.06827 [astro-ph.CO]].

%\cite{Figueroa:2020rrl}
\bibitem{Figueroa:2020rrl}
D.~G.~Figueroa, A.~Florio, F.~Torrenti and W.~Valkenburg,
%``The art of simulating the early Universe -- Part I,''
JCAP \textbf{04}, 035 (2021)
%doi:10.1088/1475-7516/2021/04/035
[arXiv:2006.15122 [astro-ph.CO]].
%11 citations counted in INSPIRE as of 27 Jul 2021

%\cite{Figueroa:2021yhd}
\bibitem{Figueroa:2021yhd}
D.~G.~Figueroa, A.~Florio, F.~Torrenti and W.~Valkenburg,
%``CosmoLattice,''
[arXiv:2102.01031 [astro-ph.CO]].
%4 citations counted in INSPIRE as of 27 Jul 2021


\bibitem{Garcia-Bellido:2002fsq}
J.~Garcia-Bellido, M.~Garcia Perez and A.~Gonzalez-Arroyo,
%``Symmetry breaking and false vacuum decay after hybrid inflation,''
Phys. Rev. D \textbf{67}, 103501 (2003)
%doi:10.1103/PhysRevD.67.103501
[arXiv:hep-ph/0208228 [hep-ph]].

\bibitem{Felder:2006cc}
G.~N.~Felder and L.~Kofman,
%``Nonlinear inflaton fragmentation after preheating,''
Phys. Rev. D \textbf{75}, 043518 (2007)
%doi:10.1103/PhysRevD.75.043518
[arXiv:hep-ph/0606256 [hep-ph]].

\bibitem{Frolov:2010sz}
A.~V.~Frolov,
%``Non-linear Dynamics and Primordial Curvature Perturbations from Preheating,''
Class. Quant. Grav. \textbf{27}, 124006 (2010)
%doi:10.1088/0264-9381/27/12/124006
[arXiv:1004.3559 [gr-qc]].

\bibitem{Amin:2014eta}
M.~A.~Amin, M.~P.~Hertzberg, D.~I.~Kaiser and J.~Karouby,
%``Nonperturbative Dynamics Of Reheating After Inflation: A Review,''
Int. J. Mod. Phys. D \textbf{24}, 1530003 (2014)
%doi:10.1142/S0218271815300037
[arXiv:1410.3808 [hep-ph]].






\bibitem{Repond:2016sol}
J.~Repond and J.~Rubio,
%``Combined Preheating on the lattice with applications to Higgs inflation,''
JCAP \textbf{07}, 043 (2016)
%doi:10.1088/1475-7516/2016/07/043
[arXiv:1604.08238 [astro-ph.CO]].

\bibitem{Fan:2021otj}
J.~Fan, K.~D.~Lozanov and Q.~Lu,
%``Spillway Preheating,''
JHEP \textbf{05}, 069 (2021)
%doi:10.1007/JHEP05(2021)069
[arXiv:2101.11008 [hep-ph]].

%\cite{Shapiro:2016pfm}
%\bibitem{Shapiro:2016pfm}
%I.~L.~Shapiro,
%``Covariant derivative of fermions and all that,''
%[arXiv:1611.02263 [gr-qc]].
%4 citations counted in INSPIRE as of 27 Jul 2021






%\cite{Bernal:2018hjm}
%\bibitem{Bernal:2018hjm}
%N.~Bernal, A.~Chatterjee and A.~Paul,
%``Non-thermal production of Dark Matter after Inflation,''
%JCAP \textbf{12}, 020 (2018)
%doi:10.1088/1475-7516/2018/12/020
%[arXiv:1809.02338 [hep-ph]].
%14 citations counted in INSPIRE as of 27 Jul 2021




\bibitem{Greene:1998nh}
P.~B.~Greene and L.~Kofman,
%``Preheating of fermions,''
Phys. Lett. B \textbf{448}, 6-12 (1999)
%doi:10.1016/S0370-2693(99)00020-9
[arXiv:hep-ph/9807339 [hep-ph]].

%\cite{Giudice:1999fb}
\bibitem{Giudice:1999fb}
G.~F.~Giudice, M.~Peloso, A.~Riotto and I.~Tkachev,
%``Production of massive fermions at preheating and leptogenesis,''
JHEP \textbf{08}, 014 (1999)
%doi:10.1088/1126-6708/1999/08/014
[arXiv:hep-ph/9905242 [hep-ph]].
%295 citations counted in INSPIRE as of 01 Sep 2021

\bibitem{Greene:2000ew}
P.~B.~Greene and L.~Kofman,
%``On the theory of fermionic preheating,''
Phys. Rev. D \textbf{62}, 123516 (2000)
%doi:10.1103/PhysRevD.62.123516
[arXiv:hep-ph/0003018 [hep-ph]].

%\cite{Peloso:2000hy}
\bibitem{Peloso:2000hy}
M.~Peloso and L.~Sorbo,
%``Preheating of massive fermions after inflation: Analytical results,''
JHEP \textbf{05}, 016 (2000)
%doi:10.1088/1126-6708/2000/05/016
[arXiv:hep-ph/0003045 [hep-ph]].
%95 citations counted in INSPIRE as of 01 Sep 2021


%\cite{Nilles:2001fg}
\bibitem{Nilles:2001fg}
H.~P.~Nilles, M.~Peloso and L.~Sorbo,
%``Coupled fields in external background with application to nonthermal production of gravitinos,''
JHEP \textbf{04}, 004 (2001)
%doi:10.1088/1126-6708/2001/04/004
[arXiv:hep-th/0103202 [hep-th]].
%115 citations counted in INSPIRE as of 01 Sep 2021

%\cite{Ballesteros:2020adh}
\bibitem{Ballesteros:2020adh}
G.~Ballesteros, M.~A.~G.~Garcia and M.~Pierre,
%``How warm are non-thermal relics? Lyman-$\alpha$ bounds on out-of-equilibrium dark matter,''
JCAP \textbf{03}, 101 (2021)
%doi:10.1088/1475-7516/2021/03/101
[arXiv:2011.13458 [hep-ph]].
%12 citations counted in INSPIRE as of 27 Jul 2021

\bibitem{Moroi:2020has}
T.~Moroi and W.~Yin,
%``Light Dark Matter from Inflaton Decay,''
JHEP \textbf{03}, 301 (2021)
%doi:10.1007/JHEP03(2021)301
[arXiv:2011.09475 [hep-ph]].

\bibitem{Moroi:2020bkq}
T.~Moroi and W.~Yin,
%``Particle Production from Oscillating Scalar Field and Consistency of Boltzmann Equation,''
JHEP \textbf{03}, 296 (2021)
%doi:10.1007/JHEP03(2021)296
[arXiv:2011.12285 [hep-ph]].

\bibitem{fortran}
D.H. Bailey, ``MPFUN2015: A thread-safe arbitrary precision package'', http://www.davidhbailey.com/dhbpapers/mpfun2015.pdf.

  \bibitem{preheating} 
L.~Kofman, A.~D.~Linde and A.~A.~Starobinsky,
%``Reheating after inflation,''
Phys. Rev. Lett. \textbf{73}, 3195-3198 (1994)
%doi:10.1103/PhysRevLett.73.3195
[arXiv:hep-th/9405187 [hep-th]].
%1403 citations counted in INSPIRE as of 28 Jun 2020

\end{thebibliography}

\end{document}